\documentclass[superscriptaddress, twocolumn, prx, longbibliography, nofootinbib]{revtex4-1}
\usepackage{amsmath, amssymb, color, graphicx, stmaryrd, wasysym, esint}
\definecolor{linkcolor}{rgb}{0,0,0.6} 
\usepackage[pdftex,colorlinks=true,
	pdfstartview = FitV,
	linkcolor    = linkcolor,
	citecolor    = linkcolor,
	urlcolor     = linkcolor,	
	hyperindex   = true,
	hyperfigures = false]{hyperref}

\newcommand{\dd}{\text{d}}
\newcommand{\ee}{\text{e}}
\newcommand{\ii}{\text{i}}

\begin{document}

\title{How dissipation constrains fluctuations in nonequilibrium liquids:\\Diffusion, structure and biased interactions}
\author{Laura Tociu}
\affiliation{James Franck Institute, University of Chicago, Chicago, IL 60637}
\affiliation{Department of Chemistry, University of Chicago, Chicago, IL 60637}

\author{\'Etienne Fodor}
\affiliation{DAMTP, Centre for Mathematical Sciences, University of Cambridge, Wilberforce Road, Cambridge CB3 0WA, UK}

\author{Takahiro Nemoto}
\affiliation{Philippe Meyer Institute for Theoretical Physics, Physics Department, \'Ecole Normale Sup\'erieure \& PSL Research University, 24, rue Lhomond, 75231 Paris Cedex 05, France}

\author{Suriyanarayanan Vaikuntanathan}
\affiliation{James Franck Institute, University of Chicago, Chicago, IL 60637}
\affiliation{Department of Chemistry, University of Chicago, Chicago, IL 60637}

\begin{abstract}

The dynamics and structure of nonequilibrium liquids, driven by non-conservative forces which can be either external or internal, generically hold the signature of the net dissipation of energy in the thermostat. Yet, disentangling precisely how dissipation changes collective effects remains challenging in many-body systems due to the complex interplay between driving and particle interactions. First, we combine explicit coarse-graining and stochastic calculus to obtain simple relations between diffusion, density correlations and dissipation in nonequilibrium liquids. Based on these results, we consider large-deviation biased ensembles where trajectories mimic the effect of an external drive. The choice of the biasing function is informed by the connection between dissipation and structure derived in the first part. Using analytical and computational techniques, we show that biasing trajectories effectively renormalizes interactions in a controlled manner, thus providing intuition on how driving forces can lead to spatial organization and collective dynamics. Altogether, our results show how tuning dissipation provides a route to alter the structure and dynamics of liquids and soft materials.

\end{abstract}

\maketitle


\section{Introduction}

Nonequilibrium forces can drive novel and specific pathways to modulate phase transitions and self-assembly in materials. The close connection between the net dissipation of energy, powered by these forces, internal transport and spatial organization is especially apparent in living systems~\cite{Toyabe2010, Ahmed2016, Battle604, Mura2018}. As an example, the flagella motors of {\it E. Coli} exhibit a unique phenomenology combining ultra-sensitive response, adaptation, and motor restructuring as a function of the applied load~\cite{Lele2013, Lan2012, Wang2017}. Moreover, {\it in vivo} studies of the cellular cytoskeleton, as well as {\it in vitro} experiments on reconstituted systems, have also shown that motor-induced forces control a large variety of functionality in the cell~\cite{Silva2011, Sanchez2012, Blanchoin2014, Murrell2015, Decamp2015}.

To elucidate the role of nonequilibrium forces in materials, it is crucial to examine how dissipation affects the emerging dynamics and structure. While equilibrium features are well established, progress in controlling systems with sustained dissipation has been hampered by a lack of general principles~\cite{Cates2015, Solon2015a, Nguyen2016, Fodor2016, Murugan2017, Nardini2017, Nguyen2018}. {In this context}, minimal models of active and driven systems provide analytically and numerically tractable test beds to investigate the interplay between dissipation and material properties far from equilibrium~\cite{Marchetti2013, Han2016, Bechinger2016, delJunco2018, Marchetti2018}. They have illustrated, for instance, how nonequilibrium driving can induce phase transitions and excite novel collective responses in soft media~\cite{Vicsek1995, Tailleur2008, Han2016, Nguyen2016, VanZuiden2016}. Recent theoretical work has proposed extending equilibrium concepts to active media, such as the definition of pressure~\cite{Takatori2015, Solon2015a}, to rationalize their phenomenology~\cite{Solon2018, Solon2018b}. Others have striven to obtain stationary properties of active matter through perturbation close to equilibrium~\cite{Farage2015, Fodor2016, Brader2017}, inspired by other approaches on driven systems~\cite{McLennan1959, Komatsu2008, Maes2009, Maes2010}.

To investigate how dissipation controls emerging behavior, yet another approach has focused on introducing a bias in dynamical ensembles. Using large deviation techniques, trajectories are conditioned to promote atypical realizations of the dynamics~\cite{Touchette2009, Jack2010}. Such techniques have been used, for instance, to investigate the role of dynamical heterogeneities in glassy systems~\cite{garrahan2007, Hedges2009, Pitard2011, Speck2012, Bodineau2012a, Limmer2014, Jack2017} and soliton solutions in high-dimensional chaotic chains~\cite{tailleur2007probing, laffargue2013}. More recently, it has been shown that changing dissipation, through a dynamical bias, strongly affects the internal transport and the density fluctuations of nonequilibrium liquids~\cite{Cagnetta2017, Nemoto2019}, thus confirming that controlling dissipation is indeed a fruitful route to tailoring material properties. In spite of these advances, anticipating the emergent dynamics and structure of biased nonequilibrium systems is still challenging in the presence of many-body interactions~\cite{Chetrite2013, Jack2010}, so that precise control has remained elusive so far in this context. Consequently, any generic principle rationalizing spatial organization in terms of dissipation is still lacking.

In this paper, we explore how dissipation affects the dynamics and structure of many-body diffusive systems. First, we consider in Sec.~\ref{sec:method} two types of assemblies of Brownian particles: one in which only a subset is driven by an external force, and one in which a subset of the particles experience an internal active force. We first focus on instances where the fraction of driven particles is less than the fraction of undriven particles, so that driven and undriven particles are respectively referred to as {\it tracer} and {\it bath} particles. Using the diffusion coefficient of a tagged tracer particle and the density correlations between tracer and bath particles, we connect dissipation to liquid properties. In contrast with~\cite{delJunco2018}, our prediction for diffusion follows from a systematic coarse-graining with explicit dependence in terms of microscopic details~\cite{Dean1996, Demery2011, Demery2014}.

Next, importantly, we put forward a generic relation between density correlations and dissipation valid for an arbitrary driving force: this is our first main result. We demonstrate that this result holds both for fluids in which a fraction of the particles are driven by a fixed external drive and for fluids in which either a fraction of the liquid or the entire liquid is driven by an internal noise, analogous to the driving used in model active matter systems. This result opens the door to estimating dissipation directly from the liquid structure, in contrast with previous approaches based either on perturbing the system~\cite{Harada2005, Mizuno2007, Visco2015, Turlier2016, Ahmed2018} or on analyzing trajectories and currents in phase space~\cite{Battle604, Gingrich2017, Roldan2018, Parrondo2018, Li2018}. We illustrate this with numerical simulations for which dissipation is quantified by the deviation from equilibrium tracer-bath correlations. Using these results as a basis, we also show how various aspects of the pair correlation function of a nonequilibrium liquid are effectively constrained by the energy dissipation. Altogether, this set of results clarify how nonequilibrium forces affect the transport and structure of the liquid, thus showing how liquid properties can be modified at the cost of energy dissipation.

Motivated by these results, and to provide concrete intuition for how particular configurations can be stabilized by nonequilibrium forces, we next investigate in Sec.~\ref{sec:bias} the emerging structure of Brownian particles subject to a dynamical bias. The explicit form of the bias is inspired by the results of Sec.~\ref{sec:method} connecting dissipation to many-body interactions. Using analytical calculations and numerical simulations based on the cloning algorithm~\cite{Giadina2006, tailleur2007probing, Hurtado2009, Nemoto2016, Ray2018, Klymko2018, Brewer2018}, we show that biased sampling trajectories can be used to renormalize any specific interparticle interaction in a multi-component liquid. The rare noise fluctuations sampled with dynamical bias effectively drive the system away from typical behavior~\cite{garrahan2007, Hedges2009, Jack2010, Pitard2011, Speck2012, Bodineau2012a, Chetrite2013, Limmer2014, Jack2017}. Such noise realizations can then serve as proxies of how to control the dynamics by applying an external force with complex protocols. We also illustrate the generality of our ideas by considering an assembly of aligning self-propelled particles~\cite{Farrell2012}. Specifically, we show how biased energy flows can renormalize interactions between particles and stabilize a flocking transition. Overall, our results lay the groundwork for precise control of the emerging structure and collective dynamics in many-body diffusive nonequilibrium systems.


\section{Dissipation and liquid properties}\label{sec:method}

In this Section, we provide a series of relations between energy dissipation and liquid properties in nonequilibrium liquids. Specifically, we consider interacting Brownian particles where a specific set of particles $\Omega$ is driven by a non-conservative force ${\bf F}_{{\rm d},i}$:
\begin{equation}\label{eq:dyn}
	\gamma\dot{\bf r}_i = \delta_{i\in\Omega}{\bf F}_{{\rm d},i} - \nabla_i \sum_j v({\bf r}_i-{\bf r}_j) + {\boldsymbol\xi}_i ,
\end{equation}
where $\delta_{i\in\Omega}=1$ if $i\in\Omega$ and $\delta_{i\in\Omega}=0$ otherwise. The driven particles belonging to the set $\Omega$ are referred to as {\it tracers}, and others as {\it bath} particles. The fluctuating term ${\boldsymbol\xi}_i$ is a zero-mean Gaussian white noise with correlations $\langle\xi_{i\alpha}(t)\xi_{j\beta}(0)\rangle=2\gamma T\delta_{ij}\delta_{\alpha\beta}\delta(t)$, where $\gamma$ and $T$ respectively denote the damping coefficient and the bath temperature, with the Boltzmann constant set to unity ($k_{\rm B}=1$).


\subsection{Deterministic vs. active drive}\label{sec:map}

In what follows, we consider two types of drive: (i)~an external force following the same deterministic protocol for all driven particles, and (ii)~an internal force given by a noise term independent for each driven particle. Building on recent work~\cite{Han2016, delJunco2018}, we take for drive (i) a time-periodic protocol given in two dimensions by
\begin{equation}\label{eq:theta}
	{\bf F}_{\rm d}(t) = f \big[\sin(\omega t) \hat{\bf e}_x + \cos(\omega t)\hat{\bf e}_y \big] ,
\end{equation}
where $f$ and $\omega$ are respectively the amplitude and the frequency of the drive, so that the drive persistence reads $\tau=2\pi/\omega$. The relative strength of the drive is given by the P\'eclet number $\text{Pe} = \sigma f/T$, where $\sigma$ is the typical particle size~\cite{Han2016, delJunco2018}. In the absence of interactions ($v=0$), the average position of driven tracers follows a periodic orbit, describing a circle in two dimensions. In contrast, drive (ii) corresponds to a random self-propulsion as is often considered in active liquids~\cite{Fily2012, Redner2013, Maggi2015}. Specifically, we use a set of zero-mean Gaussian noises with correlations
\begin{equation}\label{eq:theta_ac}
	\langle F_{{\rm d},i\alpha}(t) F_{{\rm d},j\beta}(0) \rangle = \delta_{ij}\delta_{\alpha\beta} \frac{f^2}{d} {\rm e}^{-|t|/\tau} ,
\end{equation}
where $d$ is the spatial dimension. The parameters $f$ and $\tau$ respectively control the amplitude and the persistence of fluctuations.

Interestingly, the active force with correlations~\eqref{eq:theta_ac} can be obtained from a generalized version of the deterministic force~\eqref{eq:theta} where each particle $i$ is now subjected to an independent drive. The period of the orbit is determined by a series of $n$ oscillators with identical frequencies for all particles, yet independent amplitudes:
\begin{equation}\label{eq:theta_dis}
	{\bf F}_{{\rm d}, i}(t) = \frac{f}{\sqrt{nd}} \sum_{a=1}^n \big[ {\bf A}_{ai} \cos(\omega_a t) + {\bf B}_{ai} \sin(\omega_a t) \big] .
\end{equation}
The essential ingredient of the mapping into active force is to implement disorder in the drive. This is done by taking the oscillator amplitudes as uncorrelated zero-mean Gaussian variables with unit variance:
\begin{equation}
	\langle A_{ai\alpha}A_{bj\beta}\rangle_{\rm d} = \delta_{ab}\delta_{ij}\delta_{\alpha\beta} = \langle B_{ai\alpha}B_{bj\beta}\rangle_{\rm d} ,
\end{equation}
where $\langle\cdot\rangle_{\rm d}$ denotes an average over the disorder. It follows that ${\bf F}_{{\rm d},i}$ is also a Gaussian process with zero mean and correlations given by:
\begin{equation}
	\langle F_{{\rm d},i\alpha} (t) F_{{\rm d},j\beta}(0) \rangle_{\rm d} = \delta_{ij} \delta_{\alpha\beta} \frac{f^2}{nd} \sum_{a=1}^n \cos (\omega_a t). 
\end{equation}
In the limit of a large number of oscillators ($n\gg1$), we express these correlations in terms of the density of driving frequencies $\phi$ as
\begin{equation}\label{eq:drive_corr}
	\langle F_{{\rm d},i\alpha} (t) F_{{\rm d},j\beta}(0) \rangle_\text{d} \underset{n\gg1}{=} \delta_{ij} \delta_{\alpha\beta} \frac{f^2}{d} \int \phi(\omega') \ee^{\ii\omega'|t|} \frac{\dd\omega'}{2\pi} .
\end{equation}
This establishes that, in the limit of many oscillators, the deterministic drive~\eqref{eq:theta_dis} with disordered amplitude is equivalent to a noise term with spectrum $\phi$. In particular, by choosing $\phi(\omega') = 2\tau/ \big[1+(\omega'\tau)^2\big]$, the drive correlations~\eqref{eq:drive_corr} reproduce exactly the ones of the random force in~\eqref{eq:theta_ac}.

\begin{figure}
	\centering
	\includegraphics[width=.8\columnwidth]{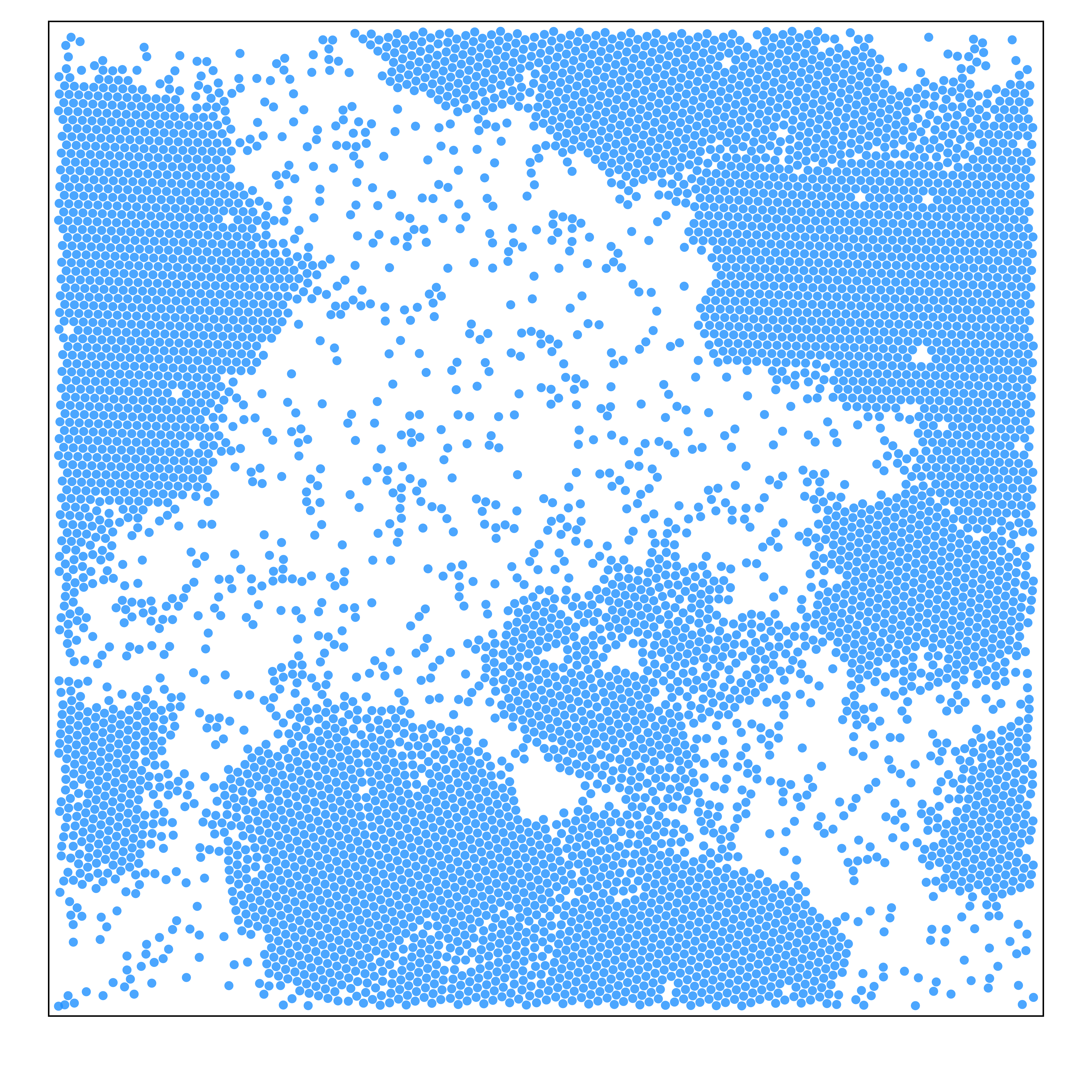}
	\caption{\label{fig0}
		Snapshot of particles subjected to a disordered drive. A phase separation emerges which is analogous to the motility-induced phase separation of active particles~\cite{Tailleur2008, Cates2015}. Simulation details in Appendix~\ref{app:simu} and movie in~\cite{movie}.
	}
\end{figure}

To illustrate the relevance of this mapping, we simulate numerically the many-body dynamics~\eqref{eq:dyn} where every particle is subjected to a disordered drive of the form~\eqref{eq:theta_dis}. We use the potential $v({\bf r})=v_0(1-|{\bf r}|/\sigma)^2\Theta(\sigma-|{\bf r}|)$, where $\Theta$ denotes the Heaviside step function, which sets purely repulsive interactions. To implement numerically the disorder in driving, it is sufficient to sample the amplitudes $\{{\bf A}_{ai}, {\bf B}_{ai}\}$ and frequencies $\{\omega_i\}$ at initial time. In the regime of high persistence $\tau$ and large average density $\rho_0$, we observe the spontaneous formation of clusters up to a complete phase separation at large time, see Fig.~\ref{fig0}. This is analogous to the motility-induced phase separation commonly reported in standard models of active particles~\cite{Tailleur2008, Cates2015}. Interestingly, it appears in our case even in the absence of fluctuations ($T=0$), namely for a purely deterministic set of equations.

In short, we thus demonstrate that the disordered drive alone reproduces the emerging physics of active systems. This important result bridges the gap between two main classes of nonequilibrium liquids, where the driving force stems from either a deterministic protocol or a random noise. In what follows, we obtain analytic and numerical results for both drives to illustrate the broad applicability of our framework, ranging from systems driven by deterministic fields to active matter systems.

\subsection{Dissipation controls tracer diffusion}\label{sec:diff}

To connect tracer diffusion with dissipation, we first describe the dynamics of undriven particles in terms of a coarse-grained variable. Using standard techniques, the dynamics of the density field $\rho({\bf r},t) = \sum_{i\not\in\Omega}\delta[{\bf r}-{\bf r}_i(t)]$ can be written as a non-linear Langevin equation~\cite{Dean1996}. In the regime of weak interactions, the density fluctuations $\delta \rho({\bf r},t ) = \rho({\bf r}, t) - \rho_0$ around the average density $\rho_0$ are Gaussian and captured by the following Hamiltonian~\cite{Chandler1993, Demery2014, Kruger2017}:
\begin{equation}
	\begin{aligned}
		{\cal H} &= \frac{T}{2} \int \delta \rho ({\bf r}) K({\bf r} - {\bf r}') \delta \rho ({\bf r}') \dd{\bf r}\dd{\bf r}'
		\\
		&\quad + \int \sum_{i\in\Omega} v({\bf r}-{\bf r}_i) \rho({\bf r}) \dd{\bf r} ,
	\end{aligned}
\end{equation}
where $K({\bf r})= \delta({\bf r})/\rho_0 + v({\bf r})/T$. Note that density fluctuations remain generally Gaussian even for a homogeneous active liquid~\cite{Fily2012}. The conserved density dynamics reads
\begin{equation}\label{eq:EvolutionField}
	\begin{aligned}
		\frac{\partial \delta \rho({\bf r}, t)}{\partial t} &= D_{\rm G} \nabla^2 \int K({\bf r}-{\bf r}') \delta \rho({\bf r}', t) \dd{\bf r}'
		\\
		&\quad + \frac{1}{\gamma_{\rm G}} \nabla^2 \sum_{i\in\Omega}v({\bf r}-{\bf r}_i(t)) + \nabla \cdot {\boldsymbol\Lambda}({\bf r},t) ,
	\end{aligned}
\end{equation}
where $D_{\rm G} = \rho_0 T/\gamma$ and $\gamma_{\rm G}=\gamma/\rho_0$ are respectively the field diffusion coefficient and the field damping coefficient. The term $\boldsymbol\Lambda$ is a zero-mean Gaussian white noise with correlations $\langle \Lambda_\alpha({\bf r}, t) \Lambda_\beta({\bf r}', t') \rangle = 2 D_{\rm G} \delta_{\alpha\beta}\delta({\bf r} - {\bf r}')\delta(t-t')$.

Owing to the linearity of the density dynamics~\eqref{eq:EvolutionField}, it can be readily written in Fourier space $\delta\rho({\bf q},t)=\int\rho({\bf r},t){\rm e}^{-{\rm i}{\bf q}\cdot{\bf r}} {\rm d}{\bf r}$ as
\begin{equation}\label{eq:FourierEvolutionField}
	\begin{aligned}
		\frac{\partial \delta \rho({\bf q},t)}{\partial t} &= - |{\bf q}|^2 D_{\rm G} K({\bf q}) \delta \rho({\bf q},t)
		\\
		&\quad - |{\bf q}|^2 \frac{v({\bf q})}{\gamma_{\rm G}} \sum_{j\in\Omega}\ee^{-\ii {\bf q} \cdot {\bf r}_j(t)} + \ii{\bf q}\cdot{\boldsymbol\Lambda} ({\bf q},t) ,
	\end{aligned}
\end{equation}
so that the field dynamics can be directly solved as
\begin{equation}\label{eq:rho}
	\begin{aligned}
		\delta\rho({\bf q},t) &= \int_{-\infty}^t \dd s \ee^{-D_{\rm G} |{\bf q}|^2 K({\bf q})(t-s)}
		\\
		&\quad\times \bigg[ \ii{\bf q} \cdot {\boldsymbol\Lambda}({\bf q},s) - |{\bf q}|^2 \frac{v({\bf q})}{\gamma_{\rm G}} \sum_{j\in\Omega}\ee^{-\ii {\bf q}\cdot {\bf r}_j(s)}\bigg] .
	\end{aligned}
\end{equation}
Considering the limit of dilute driven tracers, where interactions among them are negligible, their dynamics reads
\begin{equation}\label{eq:EvolutionTracer}
	\gamma\dot{\bf r}_j = {\bf F}_\text{d} + {\boldsymbol\xi_j} - \int_{\bf q} \ii{\bf q} v(-{\bf q})\ee^{\ii{\bf q} \cdot {\bf r}_j(t)} \delta \rho({\bf q}, t) ,
\end{equation}
with $\int_{\bf q}=\int\dd{\bf q}/(2\pi)^d$. As a result,~\eqref{eq:rho} and~\eqref{eq:EvolutionTracer} provide closed time-evolution equations for tracers only. It should only be valid for weak interactions {\it a priori}, yet previous works have shown that it remains qualitatively relevant even beyond this regime in practice~\cite{Demery2015, Martin2018, Demery2019}. Indeed, Gaussian field theories for density fluctuations provide a very good description of simple liquids~\cite{Chandler1993}.

To characterize the transport properties of the liquid in the presence of driving forces, our first goal is to obtain an explicit expression, in terms of microscopic details, for the tracer diffusion coefficient:
\begin{equation}
	D = \underset{t\to\infty}{\lim} \frac{1}{2dt} \big\langle \big[\langle{\bf r}_i(t)\rangle-{\bf r}_i(t)\big]^2 \big\rangle .
\end{equation}
We aim to explore connections between $D$ and dissipation, which is defined from stochastic thermodynamics as the power of the forces exerted by all tracers on solvent: ${\cal J} = \sum_i\langle\dot{\bf r}_i\cdot(\gamma\dot{\bf r}_i-{\boldsymbol\xi}_i)\rangle$, where $\cdot$ denotes a Stratonovich product~\cite{Sekimoto1998, Seifert2012}. Dissipation is directly related to entropy production, as a measure of irreversibility, both when the drive is deterministic~\cite{Sekimoto1998, Seifert2012} and when it is a correlated noise~\cite{Mandal2017, Seifert2018, Shankar2018, Bo2019}. Substituting the dynamics~\eqref{eq:dyn}, the dissipation coincides with the power of driving forces: ${\cal J} = \sum_{i\in\Omega}\langle\dot{\bf r}_i\cdot{\bf F}_{{\rm d},i}\rangle$. Besides, replacing $\dot{\bf r}_i$ by its expression in~\eqref{eq:dyn}, and using the fact that ${\boldsymbol\xi}_i$ and ${\bf F}_{{\rm d},i}$ are uncorrelated, we deduce that the dissipation can be further separated into free-motion and interaction contributions as ${\cal J} = N f^2/\gamma - \dot w$, where the {\it rate of work} reads
\begin{equation}\label{eq:work}
	\dot w = \frac{1}{\gamma} \sum_{i\in\Omega,j}\big\langle {\bf F}_{{\rm d},i}\cdot\nabla_iv({\bf r}_i-{\bf r}_j)\big\rangle.
\end{equation}
Given that $\dot w$ is the only non-trivial contribution to dissipation, connecting diffusion and dissipation simply amounts to expressing $D$ in terms of $\dot w$.

Deriving transport coefficients in nonequilibrium many-body systems, whose collective effects result from the complex interplay between interaction and driving forces, is a notoriously difficult task~\cite{Voituriez2016, Brady2017, Stenhammar2017, Tailleur2017, Illien2018}. We set up a proper perturbation scheme by scaling the pair potential $v$ with a small dimensionless parameter $h\ll1$ which controls the coupling between tracer and bath. In Appendix~\ref{app:diff}, we obtain some explicit expressions for $D$ and $\dot w$ to quadratic order in $h$ and in the scaled driving amplitude $\rm Pe$.

First, we discuss the case of the deterministic drive~\eqref{eq:theta}, and we focus on the limits of small and large driving frequency, respectively $\omega\tau_{\rm r}\ll 1$ and $\omega\tau_{\rm r}\gg 1$, where the relaxation time scale $\tau_{\rm r}=(D_{\rm G}/\sigma^2)K(|{\bf q}|=1/\sigma)$ is set by density diffusion over the tracer size $\sigma$. First, at high frequencies $\omega\tau_{\rm r}\gg 1$, the rate of work per particle $\dot w/N$ and the deviation from equilibrium diffusion $D-D_{\rm eq}$, where $D_{\rm eq}$ is the diffusion coefficient for ${\rm Pe}=0$, are given by
\begin{equation}\label{eq:large}
	\begin{aligned}
		\frac{\dot w}{N} &= \Big(\frac{h{\rm Pe}}{\omega}\Big)^2 \cdot \frac{(T/\sigma)^2}{d\gamma^3} \int_{\bf q} |{\bf q}|^4 |v({\bf q})|^2 \frac{1 + \rho_0K({\bf q})}{K({\bf q})} ,
		\\
		D-D_{\rm eq} &= \Big(\frac{h{\rm Pe}}{\omega}\Big)^2 \cdot \frac{T /\sigma^2}{d\gamma^3} \int_{\bf q}\frac{|{\bf q}|^2 |v({\bf q})|^2 } {K({\bf q})\big[1 + \rho_0 K({\bf q})\big]} .
	\end{aligned}
\end{equation}
In the opposite limit of low frequencies $\omega\tau_{\rm r}\ll 1$, we get
\begin{equation}\label{eq:small}
	\begin{aligned}
		\frac{\dot w}{N} &= \frac{(h{\rm Pe})^2}{d\gamma\sigma^2} \int_{\bf q} \frac{|v({\bf q})|^2}{K({\bf q})\big[1 + \rho_0K({\bf q})\big]} ,
		\\
		D - D_{\rm eq} &= \frac{5(h{\rm Pe})^2}{d\gamma T\sigma^2} \int_{\bf q} \frac{|v({\bf q})|^2}{|{\bf q}|^2K({\bf q})\big[1 + \rho_0K({\bf q})\big]^3} .
	\end{aligned}
\end{equation}
Both $\dot w/N$ and $D-D_{\rm eq}$ are now independent of the driving frequency $\omega$. As a result, our perturbation theory shows that the scalings of $\dot w$ and $D-D_{\rm eq}$ are identical, both in terms of the drive amplitude $\rm Pe$ and of its frequency $\omega$, in asymptotic frequency regimes. Note that the scaled rate of work $\gamma\dot w/(Nf^2)$ coincides with the reduced equilibrium diffusion $\gamma D_{\rm eq}/T-1$ to this order~\cite{Demery2011, Demery2014}, as expected from linear response.

The case of the active drive with correlations~\eqref{eq:theta_ac} follows by using the mapping between disordered drive and active forcing in Sec.~\ref{sec:map}. In practice, we first derive the diffusion coefficient $D$ and the rate of work $\dot w$ for the driving force~\eqref{eq:theta_dis} at fixed disorder, as a straightforward generalization of the deterministic driving case, and we then average over the disorder. At small persistence $\tau\ll\tau_{\rm r}$, we get
\begin{equation}\label{eq:small_tau}
	\begin{aligned}
		\frac{\dot w}{N} &=  \frac{\tau T(h{\rm Pe})^2}{d(\sigma\gamma)^2} \int_{\bf q} \frac{|{\bf q}|^2 |v({\bf q})|^2}{K({\bf q})} ,
		\\
		D-D_{\rm eq} &= \frac{3\tau(h{\rm Pe})^2}{d(\sigma\gamma)^2} \int_{\bf q}\frac{|v({\bf q})|^2 } {K({\bf q})\big[1 + \rho_0 K({\bf q})\big]^2} .
	\end{aligned}
\end{equation}
In contrast, the large persistence limit $\tau\gg\tau_{\rm r}$ yields the same results as for the low frequencies regime of deterministic drive, namely the expressions~\eqref{eq:small}. Indeed, the force ${\bf F}_{{\rm d},i}$ has a constant direction in such a limit, and the difference between deterministic and active drives, which respectively correspond to independent or similar directions for each tracer, is irrelevant in the limit of dilute tracers.

When the size $a$ of the bath particles is significantly smaller than the tracer size $\sigma\gg a$, which amounts to setting different pair potential $v$ for bath-bath and for bath-tracer interactions, one can safely neglect the variation of $K({\bf q})$ in~(\ref{eq:large}-\ref{eq:small_tau}), so that $K({\bf q}) \simeq K(|{\bf q}|=1/a)$. Then, in both regimes $\omega\tau_{\rm r}\gg 1$ ($\tau\ll\tau_{\rm r}$) and $\omega\tau_{\rm r}\ll 1$ ($\tau\gg\tau_{\rm r}$), the renormalization of the diffusion coefficient $D-D_{\rm eq}$ can be simply written in terms of the rate of work per particle $\dot w/N$ for ${\rm Pe}\ll1$ as
\begin{equation}\label{eq:work_D}
	\frac{D-D_{\rm eq}}{\sigma^2} \sim \frac{\dot w}{NT} .
\end{equation}
Thus, the excess rate at which tracers move over their own size compared to equilibrium, set by the lhs of~\eqref{eq:work_D}, is controlled by the rate at which work is applied on tracers by nonequilibrium forces, set by the rhs of~\eqref{eq:work_D}. The proportionality factor depends on the details of interactions and of density fluctuations. Interestingly, this result is valid both for deterministic and active drives. It corroborates numerical observations obtained previously in a system where composition-dependent diffusion constants can lead to phase transitions~\cite{delJunco2018}.


\subsection{Dissipation sets density correlations}\label{sec:struc}

We now explore how dissipation relates with static density correlations of the liquid. To this end, we treat undriven bath particles without any approximation in what follows, instead of relying on the Gaussian density field theory for $\delta\rho$ as in Sec.~\ref{sec:diff}, and we consider an arbitrary set of driving forces ${\bf F}_{{\rm d},i}$. In equilibrium, the liquid structure can be derived from a hierarchy of equations for density correlations, whose explicit form reflects the steady-state condition on the many-body distribution function~\cite{Hansen2013}. In our settings, steady-state conditions should now provide modified equations for density correlations, which can potentially make apparent the connection with dissipation.

This motivates us to consider the average rate at which the potential $U = \sum_{i\in\Omega,j} v({\bf r}_i-{\bf r}_j)$ changes, which can be written using It\^o calculus as
\begin{equation}
	\gamma\langle\dot U \rangle = \sum_{i\in\Omega,j} \big\langle\big[\gamma(\dot{\bf r}_i - \dot{\bf r}_j) + 2 T \nabla_i\big] \cdot \nabla_i v({\bf r}_i-{\bf r}_j)\big\rangle .
\end{equation}
Substituting the dynamics~\eqref{eq:dyn} and using $\langle{\boldsymbol\xi}_i\cdot\nabla_i v\rangle=0$ within It\^o convention, we get
\begin{equation}\label{eq:dotU}
	\begin{aligned}
		&\gamma\langle\dot U \rangle = \sum_{i\in\Omega,j} (1+\delta_{j\in\Omega}) \big\langle{\bf F}_{{\rm d},i} \cdot \nabla_i v({\bf r}_i-{\bf r}_j)\big\rangle
		\\
		&\,+ \sum_{i\in\Omega,j,k} \big\langle \big[\nabla_i v({\bf r}_i-{\bf r}_j)\big] \cdot \nabla_k\big[ v({\bf r}_i-{\bf r}_k) - v({\bf r}_j-{\bf r}_k) \big] \big\rangle
		\\
		&\,+ \sum_{i\in\Omega,j} 2T\big\langle\nabla_i^2 v({\bf r}_i-{\bf r}_j) \big\rangle .
    \end{aligned}
\end{equation}
In the first line of~\eqref{eq:dotU}, we recognize the rate of work $\dot w$ as defined in~\eqref{eq:work}, and the term $\gamma\dot w_{\rm act} = \sum_{\{i,j\}\in\Omega} \big\langle{\bf F}_{{\rm d},i} \cdot \nabla_i v({\bf r}_i-{\bf r}_j)\big\rangle$ which quantifies the contribution of interactions among driven particles to dissipation. The latter vanishes exactly when the drive is identical for all particles, since $\sum_{\{i,j\}\in\Omega}\nabla_iv({\bf r}_i-{\bf r}_j) = 0$ by symmetry, and it can be neglected for active drive when the fraction of driven particles is small. Then, using the steady-state condition $\langle\dot U \rangle=0$, we deduce
\begin{equation}\label{eq:balance}
	\begin{aligned}
		&\dot w + \dot w_{\rm act} = \frac{2\rho_0}{\gamma} \int g({\bf r}) \big[(\nabla v({\bf r}))^2 - T \nabla^2 v({\bf r})\big] \dd{\bf r}
		\\
		&\, + \frac{\rho_0^2}{\gamma} \iint\big[ g_{3a}({\bf r}, {\bf r}') + g_{3b}({\bf r}, {\bf r}')\big]  \big[\nabla v({\bf r})\big] \cdot \big[\nabla v({\bf r}')\big] \dd{\bf r} \dd{\bf r}' ,
	\end{aligned}
\end{equation}
where
\begin{equation}
	\begin{aligned}
 		g({\bf r}) &= \frac{1}{N} \underset{i\in\Omega,j}{{\sum}'} \big\langle \delta({\bf r}-{\bf r}_i+{\bf r}_j)\big\rangle ,
		\\
 		g_{3a}({\bf r},{\bf r}') &= \frac{1}{N^2} \underset{i\in\Omega,j,k}{{\sum}'} \big\langle \delta({\bf r}-{\bf r}_i+{\bf r}_j) \delta({\bf r}'-{\bf r}_i+{\bf r}_k)\big\rangle ,
		\\
 		g_{3b}({\bf r},{\bf r}') &= \frac{1}{N^2} \underset{i\in\Omega,j,k}{{\sum}'} \big\langle \delta({\bf r}-{\bf r}_i+{\bf r}_j) \delta({\bf r}'-{\bf r}_j+{\bf r}_k)\big\rangle ,
	\end{aligned}
\end{equation}
and $\sum'$ denotes a sum without the overlap of indices: $i\neq j$, $k\neq i$ and $k\neq j$. The power balance~\eqref{eq:balance}, valid for an arbitrary driving, either deterministic or active, is our first main result. Importantly, it holds for generic interactions and for any number of driven particles, namely not only in the limit of dilute tracers, in contrast with the results in Sec.~\ref{sec:diff}.

In practice, it reflects how density correlations adapt to the presence of nonequilibrium forces. For a vanishing rate of work ($\dot w = 0 = \dot w_{\rm act}$), one recovers the first order of the equilibrium Yvon-Born-Green (YBG) hierarchy, in its integral form, for two-component fluids~\cite{Hansen2013}. At finite rate of work ($\dot w\neq0$), the relation between the two-body correlation $g$ and the three-body terms $\{g_{3a},g_{3b}\}$ is now implicitly constrained by dissipation. A direct implication is that the rate of work can now be inferred simply by measuring static density correlations, provided that the pair-wise interaction potential is known, for a given driven liquid. Importantly, such an approach does not require any invasive methods based on comparing fluctuations and response~\cite{Harada2005, Mizuno2007, Visco2015, Turlier2016, Ahmed2018}, and it does not rely on a detailed analysis of particle trajectories~\cite{Roldan2018, Parrondo2018} or currents in phase space~\cite{Gingrich2017, Li2018}, whose experimental implementation can require elaborate techniques~\cite{Battle604, Mura2018}.

\begin{figure}
	\centering
	\includegraphics[width=\linewidth]{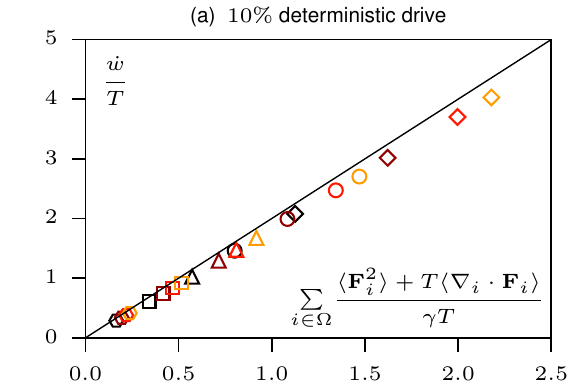}
	\vskip.3cm
	\includegraphics[width=\linewidth]{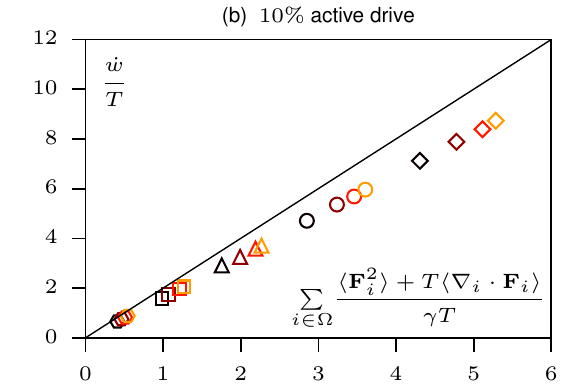}
	\caption{\label{fig:fig1}
		Parametric plot of the rate of work $\dot w/T$ and of the statistics of bath-tracer forces $\sum_{i\in\Omega}\big[\langle{\bf F}_i^2\rangle + T\langle\nabla_i\cdot{\bf F}_i\rangle\big]/(\gamma T)$ when $10\%$ of particles are driven by either (a)~a deterministic force, or (b)~an active force. The solid line with slope $2$ refers to the approximate relation~\eqref{eq:approx}. The satisfying agreement with numerical data indicates that the rate of work can be estimated by only measuring bath-tracer forces.
		The simulations were performed with $N=4500$ particles using the procedure described in Appendix~\ref{app:simu}. Parameters: ${\rm Pe}=12$ ($\hexagon$), $18$ ($\boxempty$), $24$ ({\large$\vartriangle$}), $30$ ({$\Circle$}), $36$ (${\Diamond}$); (a)~$\tau T/(\gamma\sigma^2)=2\times10^{-1}$ (black), $3\times10^{-1}$ (brown), $4\times10^{-1}$ (red), $5\times10^{-1}$ (orange); (b)~$\tau T/(\gamma\sigma^2)=2\times10^{-2}$ (black), $3\times10^{-2}$ (brown), $4\times10^{-2}$ (red), $5\times10^{-2}$ (orange).
	}
\end{figure}

\begin{figure*}
	\centering
	\includegraphics[width=\linewidth]{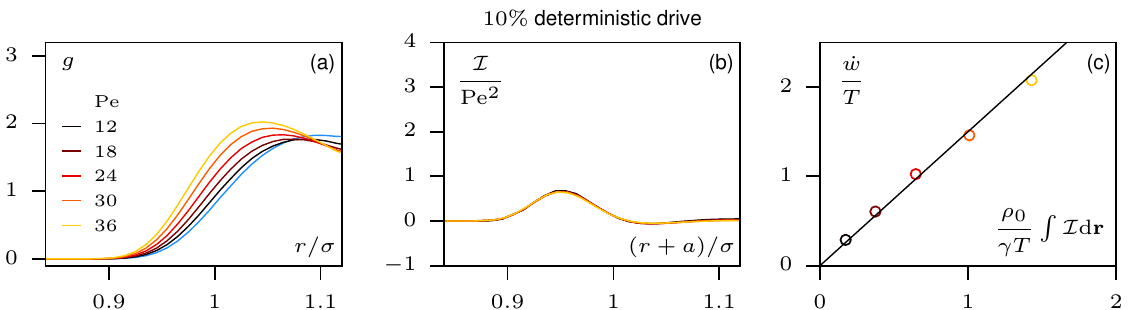}
	\includegraphics[width=\linewidth]{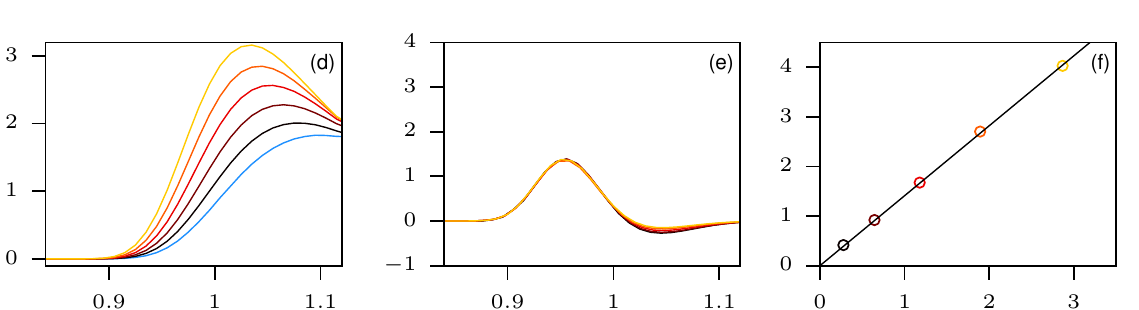}
	\vskip.5cm
	\includegraphics[width=\linewidth]{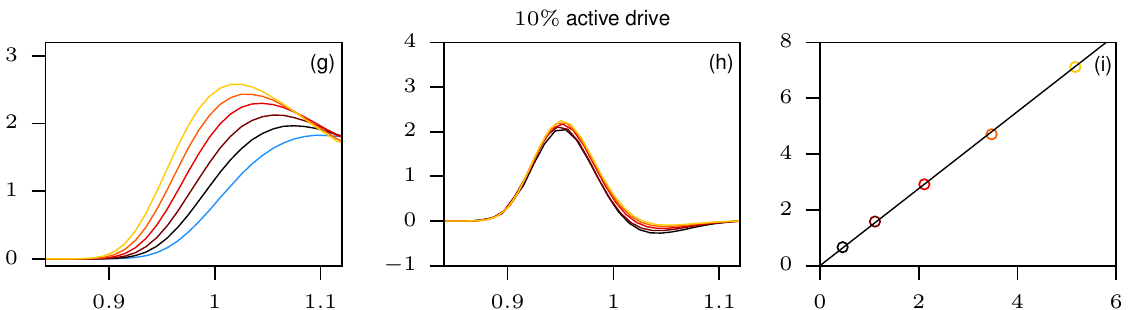}
	\includegraphics[width=\linewidth]{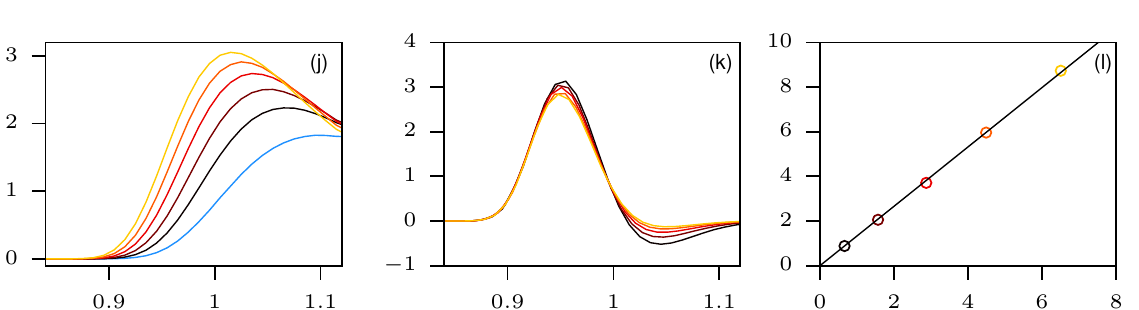}
	\caption{\label{fig:fig2}
		Connecting dissipation and structure for a liquid where $10\%$ of particles are driven by either deterministic or active forces.
		(Left)~Bath-tracer density correlation $g$ as a function of interparticle distance $r/\sigma$. The blue solid line corresponds to the equilibrium correlation function $g_{\rm eq}$ for ${\rm Pe}=0$.
		(Middle)~Deviation from equilibrium correlations ${\cal I} = \big[(\nabla v)^2-T\nabla^2v\big](g - g_{\rm eq})$ scaled by ${\rm Pe}^2$ as a function of $(r+a)/\sigma$ where $a({\rm Pe})$ is a fitting parameter. The data almost collapse into a master curve for each row, namely at a given $\tau$.
		(Right)~Parametric plot of the rate of work $\dot w/T$ and the integrated deviation from equilibrium correlations $\rho_0\int{\cal I}({\bf r}){\rm d}{\bf r}/(\gamma T)$ showing a linear relation. The black solid line with slope $\alpha$ is the best linear fit, and the marker colors refer to the same P\'eclet values as in the left and right columns.
		Simulation details in Appendix~\ref{app:simu}. Parameters: $\{\tau T/(\gamma\sigma^2), \alpha\} = \{2\times10^{-1}, 1.50\}$ (a-c), $\{5\times10^{-1}, 1.41\}$ (d-f), $\{2\times10^{-2}, 1.38\}$ (g-i), $\{5\times10^{-2}, 1.33\}$ (j-l).
	}
\end{figure*}

\begin{figure*}
	\centering
	\includegraphics[width=\linewidth]{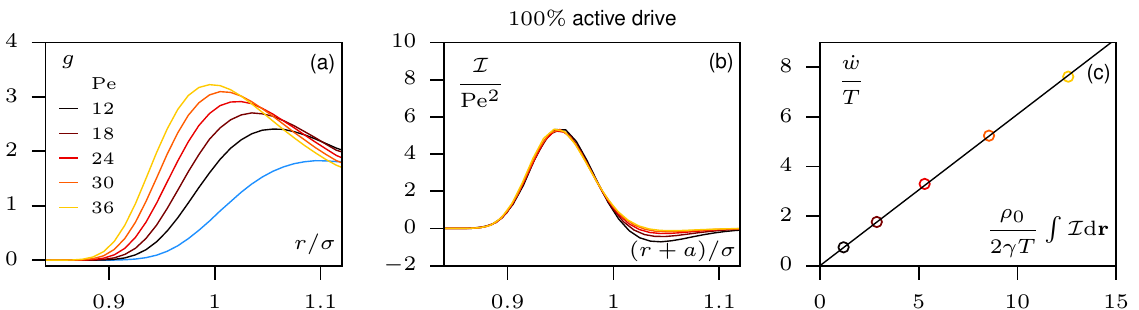}
	\includegraphics[width=\linewidth]{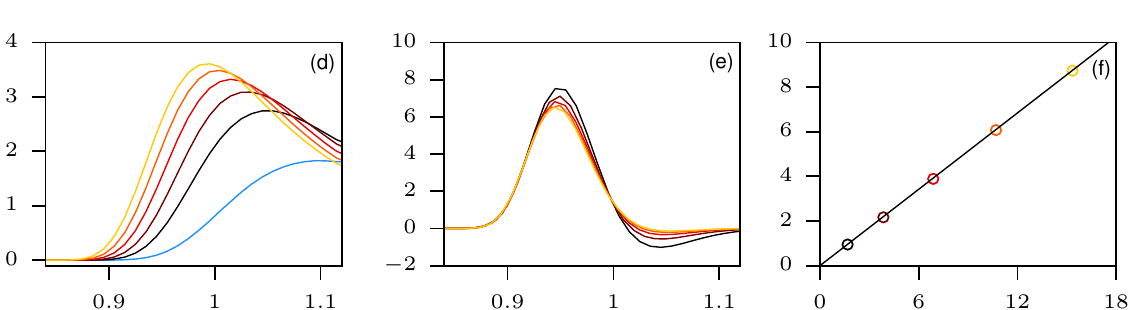}
	\caption{\label{fig:fig2_ac}
		Connecting dissipation and structure for a liquid where $100\%$ of particles are driven by an active force.
		(Left)~Density correlation $g$ as a function of interparticle distance $r/\sigma$. The blue solid line corresponds to the equilibrium correlation function $g_{\rm eq}$ for ${\rm Pe}=0$.
		(Middle)~Deviation from equilibrium correlations ${\cal I} = \big[(\nabla v)^2-T\nabla^2v\big](g - g_{\rm eq})$ scaled by ${\rm Pe}^2$ as a function of $(r+a)/\sigma$ where $a({\rm Pe})$ is a fitting parameter. The data almost collapse into a master curve for each row, namely at a given $\tau$.
		(Right)~Parametric plot of the rate of work $\dot w/T$ and the integrated deviation from equilibrium correlations $\rho_0\int{\cal I}({\bf r}){\rm d}{\bf r}/(2\gamma T)$ showing a linear relation. The black solid line with slope $\alpha$ is the best linear fit, and the marker colors refer to the same P\'eclet values as in the left and right columns.
		Simulation details in Appendix~\ref{app:simu}. Parameters: $\{\tau T/(\gamma\sigma^2), \alpha\} = \{2\times10^{-2}, 1.22\}$ (a-c), $\{5\times10^{-2}, 1.14\}$ (d-f).
	}
\end{figure*}

However, the power balance~\eqref{eq:balance} is not straightforward to test, either numerically or experimentally, due to the three-body correlations. In equilibrium, where tracer and bath particles are indistinguishable, we get $g_{3a}=g_{3b}$. Assuming that this remains approximately valid in the driven case for a small fraction of tracers, the rate of work can simply be written in terms of the force exerted on a tracer ${\bf F}_i =- \sum_j\nabla_iv({\bf r}_i-{\bf r}_j)$ as
\begin{equation}\label{eq:approx}
	\dot w \simeq \frac{2}{\gamma}\sum_{i\in\Omega} \big[ \big\langle{\bf F}_i^2\big\rangle + T \big\langle\nabla_i\cdot{\bf F}_i\big\rangle \big] .
\end{equation}
To probe the validity of this result, we simulate the dynamics~\eqref{eq:dyn} where $10\%$ of particles are subject to the driving force, considering either the deterministic periodic drive~\eqref{eq:theta} or the active noise drive~\eqref{eq:theta_ac}. Interactions are set by the WCA potential $v({\bf r})=4v_0\big[(\sigma/|{\bf r}|)^{12}-(\sigma/|{\bf r}|)^6\big]\Theta(2^{1/6}\sigma-|{\bf r}|)$~\cite{WCA1971}. Our measurements in Fig.~\ref{fig:fig1} show that~\eqref{eq:approx} is indeed a good approximation at small $\rm Pe$ and small $\tau$, namely when the drive only weakly perturbs the liquid. The discrepancy is higher for the active case compared with the deterministic one, since $\dot w_{\rm act} = 0$ in the latter without any approximation. In contrast with previous approaches~\cite{Harada2005, Lander2012, Battle604}, which rely on prospecting the whole system, our results demonstrate that the rate of work can actually be evaluated with only a small error by considering solely forces acting on tracer: the contribution of forces on other particles is negligible for a small fraction of driven tracers.

To evaluate further the change in liquid structure induced by dissipation, we measure the deviation from equilibrium pair correlations $g-g_{\rm eq}$ due to the driving forces (left column in Fig.~\ref{fig:fig2}). In particular, inspired by the two-body contribution in the power balance~\eqref{eq:balance}, we focus on the observable ${\cal I} = \big[(\nabla v)^2-T\nabla^2v\big](g - g_{\rm eq})$. At a given $\tau$, scaling ${\cal I}$ by ${\rm Pe}^2$ reveals that all curves almost collapse into a master curve for our numerical range ${\rm Pe}\in[12,36]$, as reported in the middle column of Fig.~\ref{fig:fig2}. In practice, particles overlap more for a stronger drive, so that $g$ departs from zero at smaller interparticle distance. To correct for this, we introduce a shift of the curves ${\cal I}(|{\bf r}|)$ as $|{\bf r}|\to |{\bf r}|+a$, where $a(\rm Pe)$ is a fitting parameter. Given that the rate of work also scales like ${\rm Pe}^2$, it suggests the existence of an underlying relation between $\int{\cal I}({\bf r}){\rm d}{\bf r}$ and $\dot w$. In practice, a linear fitting provides a satisfactory agreement between them, as shown in the right column of Fig.~\ref{fig:fig2}:
\begin{equation}\label{eq:emp}
	\dot w = \frac{\alpha\rho_0}{\gamma} \int \big[ g({\bf r}) - g_{\rm eq}({\bf r}) \big]\big[(\nabla v({\bf r}))^2-T\nabla^2v({\bf r})\big] {\rm d}{\bf r} ,
\end{equation}
where $\alpha$ is a fitting parameter independent of the P\'eclet number. This empirical relation demonstrates that, in the limit of dilute tracers, the rate of work can actually be directly estimated by comparing driven and equilibrium pair correlations for both deterministic and active drives. Comparing~\eqref{eq:balance} and~\eqref{eq:emp}, we deduce the following integral relation between density correlations
\begin{equation}\label{eq:emp_b}
	\begin{aligned}
		&\int \big[ (2-\alpha) g({\bf r}) + \alpha \,g_{\rm eq}({\bf r}) \big] \big[T \nabla^2 v({\bf r}) - (\nabla v({\bf r}))^2\big] \dd{\bf r}
		\\
		&= \rho_0 \iint\big[ g_{3a}({\bf r}, {\bf r}') + g_{3b}({\bf r}, {\bf r}')\big]  \big[\nabla v({\bf r})\big] \cdot \big[\nabla v({\bf r}')\big] \dd{\bf r} \dd{\bf r}' .
	\end{aligned}
\end{equation}
Interestingly, it is reminiscent again of the connection between density correlations provided by the YBG hierarchy at equilibrium~\cite{Hansen2013}. Similarly, the relation~\eqref{eq:emp_b} amounts to a constraint on density correlations, now valid for nonequilibrium liquids, which could guide the search for explicit predictions on the emerging structure. Importantly, it does not rely on any equilibrium mapping, in contrast with previous works~\cite{Maggi2015b, Rein2016, Wittmann2017}, since it remains valid for non-negligible dissipation.

The power balance~\eqref{eq:balance} can actually be extended to the case where all particles in the liquid are driven as
\begin{equation}\label{eq:balance_b}
	\begin{aligned}
		\dot w &= \frac{\rho_0}{\gamma} \int g({\bf r}) \big[(\nabla v({\bf r}))^2 - T \nabla^2 v({\bf r})\big] \dd{\bf r} ,
		\\
		&\quad + \frac{\rho_0^2}{\gamma} \iint g_3({\bf r}, {\bf r}') \big[\nabla v({\bf r})\big] \cdot \big[\nabla v({\bf r}')\big] \dd{\bf r} \dd{\bf r}' ,
	\end{aligned}
\end{equation}
where $g$ and $g_3$ now refer respectively to the two-body and three-body density correlations among all particles. This leads to an exact relation between the rate of work and the forces applied to particles ${\bf F}_i$ as
\begin{equation}\label{eq:balance_c}
	\dot w = \frac{1}{\gamma}\sum_i \big[ \big\langle{\bf F}_i^2\big\rangle + T \big\langle\nabla_i\cdot{\bf F}_i\big\rangle \big] ,
\end{equation}
which differs from the relation~\eqref{eq:approx} for driven tracers by an overall factor of $2$. A result analogous to~\eqref{eq:balance_c} was found previously for deterministic drive~\cite{Maes2010, Maes2015}. The main difference is that~\eqref{eq:balance_c} only features interaction forces ${\bf F}_i$ in the rhs, thus allowing one to evaluate the rate of work without any prior knowledge on the driving force. Besides, it is valid for both deterministic and active drives. Moreover, conducting the same analysis of density correlations as for driven tracers, ${\cal I}$ exhibits again a scaling with ${\rm Pe}^2$, as reported in Fig.~\ref{fig:fig2_ac}. We show that $\dot w$ and $\int{\cal I}({\bf r}){\rm d}{\bf r}$ are also linearly related. Introducing the linear coefficient as $\alpha\rho_0/(2\gamma)$ is consistent with substituting~\eqref{eq:emp_b} into~\eqref{eq:balance_b}, where $g_{3,a}+g_{3,b}$ is now replaced by $2g_3$. Hence, it demonstrates that the rate of work is also accessible from the nonequilibrium deviation of pair correlations in fully driven liquids. 

Overall, the results of this Section illustrate how dissipation affects the transport and structural properties of driven liquids, measured in terms of diffusion coefficient and density fluctuations. These findings motivate the following question: can nonequilibrium forces be tuned to reliably stabilize target configurations? To explore this, we rely in what follows on the framework of large deviation theory. In practice, our strategy amounts to biasing trajectories in terms of dissipation, related to many-body interactions by~\eqref{eq:balance}, to mimic the effect of an external drive. Following this route, our analytical and numerical results provide some concrete intuition for how interactions in a multicomponent system can be controllably renormalized by nonequilibrium forces. Hence, we demonstrate the ability to nucleate structures different from those characteristic of the equilibrium Boltzmann distribution to help guide self-assembly~\cite{Bisker2018} and collective motion~\cite{Vicsek1995} far from equilibrium. These results further illustrate the interplay between energy dissipation and organization in nonequilibrium many-body settings.


\section{Interactions in biased ensembles}\label{sec:bias}

To investigate how target structures and dynamics can be promoted by means of a dynamical bias, we begin by considering a system of interacting Brownian particles without any driving force
\begin{equation}\label{eq:dyn_eq}
	\gamma\dot{\bf r}_i = - \nabla_i \sum_j v({\bf r}_i-{\bf r}_j) + {\boldsymbol\xi}_i ,
\end{equation}
where the statistics of the noise term ${\boldsymbol\xi}_i$ is the same as the one in~\eqref{eq:dyn}. The rate of work $\dot w$ defined in~\eqref{eq:work} is zero because of the absence of driving. In Sec.~\ref{sec:method}, to obtain a non-zero rate of energy flow through the system, we consider an explicit driving force ${\bf F}_{{\rm d},i}$ and we explore its effects on the transport and structural properties of the liquid. In practice, different types of driving can lead to the same dissipation. In this section, using the framework of 
large deviation theory, 
we take an alternative approach where the dynamics is now conditioned by enforcing a required energy flow without any explicit driving. Thus, exploring how the system adapts to this requirement provides a new insight into the relation between dissipation and organization in driven systems, which is distinct from yet complements the approach in Sec.~\ref{sec:method}.

To this end, we focus on the subset of noise realizations that are conditioned on a non-zero rate of work. In particular, these realizations no longer have zero average, so that one can re-define the noise term in~\eqref{eq:dyn_eq} as ${\boldsymbol \xi}_i \to {\boldsymbol \xi}_i + {\bf F}_{{\rm aux},i}$ by introducing an {\it auxiliary force} ${\bf F}_{{\rm aux},i}$~\cite{Jack2010, Chetrite2013}. Hence, the stochastic dynamics given by~\eqref{eq:dyn_eq} with added force ${\bf F}_{{\rm aux},i}$ provides an explicit case which ensures a non-zero energy flow rate. In practice, this dynamics can be drastically different from the original one, thus opening the door to stabilizing unexpected structure and to promoting novel collective effects. Interestingly, such a dynamics can actually be regarded as the optimal strategy to effectively enforce a target condition on rate of work~\cite{Evans2004PRL}.

Formally, to study the dynamics conditioned by dissipation, we bias the probability of trajectories. This is done by introducing an exponential weighting factor $\exp\big[\kappa\int_0^t{\cal E}(s){\rm d}s\big]$ where ${\cal E}$ is the observable which conditions the dynamics, {\it e.g.} energy flow rate, and $\kappa$ is a conjugate field. In practice, the relative importance of biasing in the dynamics is controlled by $\kappa$, which in turn controls the average value of $\cal E$~\cite{Touchette2009}. Before deriving the central results of this section, namely relations between biased energy flow rates and organization, we first introduce a simple example in which the connection between auxiliary forces and exponentially biased ensemble can be clearly seen.


\subsection{Dynamical bias and external forces}\label{sec:biasexternal}

To introduce pedagogically our methods, we first show how biasing trajectories can lead to effectively introduce a driving force. Inspired by the role of dissipation in emerging liquid properties, as discussed in Sec.~\ref{sec:method}, we bias the equilibrium dynamics~\eqref{eq:dyn_eq} with the sum of the dissipation and the rate of work, scaled by $T$, that would be produced by applying a constant force ${\bf F}_{\rm d}$ to a subset $\Omega$ of particles:
\begin{equation}\label{eq:eps}
	{\cal E} = \frac{1}{\gamma T}\sum_{i\in\Omega} {\bf F}_{\rm d}\cdot\big[ \gamma\dot{\bf r}_i + \nabla_i V \big] ,
\end{equation}
where $V=(1/2)\sum_{i,j}v({\bf r}_i-{\bf r}_j)$. The path probability ${\cal P}\sim\exp\big[-\sum_i\int_0^t{\mathbb A}_i(s){\rm d}s\big]$ corresponding to this biased ensemble is obtained with standard methods~\cite{Martin1973, Dominicis1975}:
\begin{equation}\label{eq:action_exp}
	{\mathbb A}_i = \frac{1}{4\gamma T} \big[\gamma\dot{\bf r}_i + \nabla_iV\big]^2 - \frac{1}{2\gamma} \nabla^2_iV - \frac{\kappa}{\gamma T} \delta_{i\in\Omega} {\bf F}_{\rm d} \cdot\big[ \gamma\dot{\bf r}_i + \nabla_i V \big] ,
\end{equation}
where the two first terms correspond to the unbiased dynamics~\eqref{eq:dyn_eq}, and the third one to the bias in~\eqref{eq:eps}. It can also be written as
\begin{equation}\label{eq:action}
	{\mathbb A}_i = \frac{1}{4\gamma T} \big[\gamma\dot{\bf r}_i - 2\kappa\delta_{i\in\Omega}{\bf F}_{\rm d} + \nabla_iV\big]^2 - \frac{1}{2\gamma} \nabla^2_iV - \frac{\kappa^2}{\gamma T}\delta_{i\in\Omega}{\bf F}_{\rm d}^2 .
\end{equation}
As a result, given that the last term in~\eqref{eq:action} can be absorbed in a normalization factor, we deduce that the trajectories biased by~\eqref{eq:eps} can be generated, at leading order, in a physical dynamics where the external force $2\kappa{\bf F}_{\rm d}$ is applied to every particle in $\Omega$. In particular, it does not feature any long-range interactions which are usually found in auxiliary dynamics~\cite{Jack2015}.


\subsection{Dynamical bias and modified interactions}\label{sec:doob}

To go beyond the case of applying a constant force, we now seek for a dynamical bias which regulates particle interactions in a controlled manner. In particular, we examine cases where the control parameters $\kappa_{ij}$ are specific to particle pairs $\{i,j\}$, so that the biasing factor in path probability now reads $\exp\big[\sum_{i,j}\kappa_{ij}\int_0^t{\cal E}_{ij}(s){\rm d}s\big]$. {Now}, our choice for the biasing function ${\cal E}_{ij}$ is informed by the connection between the rate of work and many-body interactions in driven liquids, as detailed in Sec.~\ref{sec:struc}. Specifically, we observe that the power balance~\eqref{eq:balance} for deterministic drive ($\dot w_{\rm act}=0$) can be written as $\dot w = - \sum_{i\in\Omega,j} \langle{\cal L} v({\bf r}_i-{\bf r}_j)\rangle$ in terms of the evolution operator of the equilibrium dynamics~\eqref{eq:dyn_eq} defined by $\gamma{\cal L} = \sum_i\big[ T \nabla_i - \nabla_i V\big] \cdot\nabla_i $. This motivates us to consider the following bias
\begin{equation}\label{eq:eps_bis}
	{\cal E}_{ij} = \frac{1}{4T} {\cal L} v({\bf r}_i-{\bf r}_j) .
\end{equation}
In the unbiased ensembles of Sec.~\ref{sec:struc}, $\langle {\cal E}_{ij} \rangle$ provides a measure of the rate at which driving forces pump energy into or extract energy from the specific interaction between the $i^{\rm th}$ and $j^{\rm th}$ particles. Here, instead of driving the system with a specific driving force, trajectories are driven by atypical realizations of the noise generated by biased sampling.

To explore how this bias modifies interactions, we first employ a derivation different from the path integral approach in Sec.~\ref{sec:biasexternal}. Based on the procedure in~\cite{Jack2010, Chetrite2013}, the auxiliary physical dynamics, which has the same statistical properties as in the biased ensemble, can be constructed by solving the eigenvalue equation 
\begin{equation}\label{eq:adjointbiased}
	\Big[ {\cal L} + \sum_{i,j}\kappa_{ij}{\cal E}_{ij} \Big]\,{\cal G}(\{{\bf r}_k\},\kappa) = \lambda(\kappa)\,{\cal G}(\{{\bf r}_k\},\kappa) ,
\end{equation}
where the eigenvalue $\lambda$, parametrized by $\kappa_{ij}$, is the scaled cumulant generating function appropriate to ${\cal E}_{ij}$. The auxiliary dynamics is then defined by replacing the interaction potential in~\eqref{eq:dyn_eq} by the following auxiliary potential:
\begin{equation} \label{eq:auxiliary}
	\tilde V = \frac{1}{2}\sum_{i,j} v({\bf r}_i-{\bf r}_j) - 2 T \ln{\cal G} .
\end{equation}
In practice, computing $\cal G$ is a highly non-trivial procedure for many-body systems. The explicit solutions considered so far concern either exclusion processes~\cite{Popkov2010, Popkov2011, Limmer2018a} or particle-based diffusive systems restricted to small noise regimes~\cite{Lecomte2019, Proesmans2019} and non-interacting cases in some specific potentials~\cite{Majumdar2002, Touchette2016, Touchette2018}.

In our case, a simple expression can be obtained for the auxiliary potential $\tilde V_i$ by solving~\eqref{eq:adjointbiased} perturbatively at small bias parameter $\kappa$. Specifically, we expand
\begin{equation}
	\begin{aligned}
		\lambda(\kappa) &= \sum_{ij} \kappa_{ij} \langle{\cal E}_{ij}\rangle + {\cal O}(\kappa^2) ,
		\\
		{\cal G}(\{{\bf r}_k\},\kappa) &= {\cal G}^{(0)} + \sum_{ij}\kappa_{ij}\,{\cal G}^{(1)}_{ij}(\{{\bf r}_k\}) + {\cal O}(\kappa^2) ,
	\end{aligned}
\end{equation}
where ${\cal G}_0$ is the uniform eigenvector associated with the zero eigenvalue. Given that $\langle{\cal E}_{ij}\rangle=0$ in steady state, which follows from the vanishing current condition in the unbiased dynamics ($\langle\dot v\rangle=0$), the leading non-trivial order of~\eqref{eq:adjointbiased} reads
\begin{equation}\label{eq:eigenvalue}
	\sum_{ij} \kappa_{ij} \Big[ {\cal L}\, {\cal G}^{(1)}_{ij} + {\cal G}^{(0)} {\cal E}_{ij} \Big] + {\cal O}(\kappa^2) = 0 .
\end{equation}
Substituting the explicit expressions for the biasing function in~\eqref{eq:eps_bis}, we then deduce that $4T{\cal G}^{(1)}_{ij} = -{\cal G}_0\,v({\bf r}_i-{\bf r}_j)$ is a solution of the eigenvalue problem to order $\kappa$. The auxiliary potential follows as
\begin{equation}\label{eq:centralresult0}
	\tilde V = \frac{1}{2}\sum_{i,j} (1+\kappa_{ij})\, v({\bf r}_i-{\bf r}_j) + {\cal O}(\kappa^2) .
\end{equation}
Therefore, biasing with~\eqref{eq:eps_bis} amounts to changing the strength of particle interaction by a factor $\kappa_{ij}$ specifically for any pair $\{i,j\}$. This is the main result of this section.

While energy flows were sustained by explicit nonequilibrium forces in Sec.~\ref{sec:method}, we now maintain a non-zero average for ${\cal E}_{ij}$ by a biased sampling of trajectories. The corresponding noise realizations can be thought of as an external protocol, which leads to modifying the energy landscape sampled by the biased system as given in~\eqref{eq:centralresult0}. Note that tuning interaction strength between targeted pairs is qualitatively consistent with the effect of external driving. Indeed, phase separation in mixtures of driven and undriven particles, reported both experimentally and numerically, can be rationalized in terms of an effective decrease of specific interactions between these particles~\cite{delJunco2018,Han2016}.

\begin{figure*}
	\centering
	\includegraphics[width=.49\linewidth]{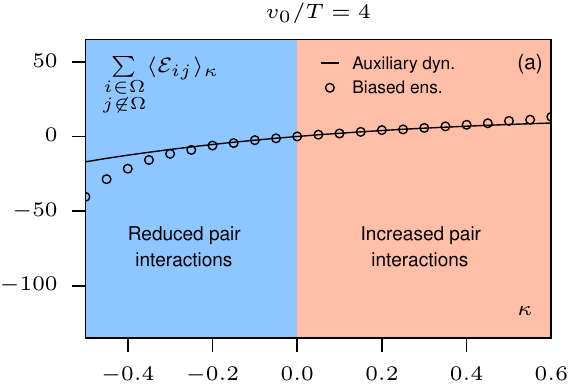}
	\hfill
	\includegraphics[width=.49\linewidth]{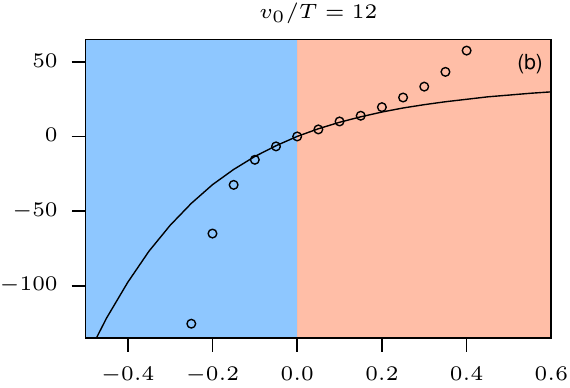}
	\vskip.5cm
	\includegraphics[width=.49\linewidth]{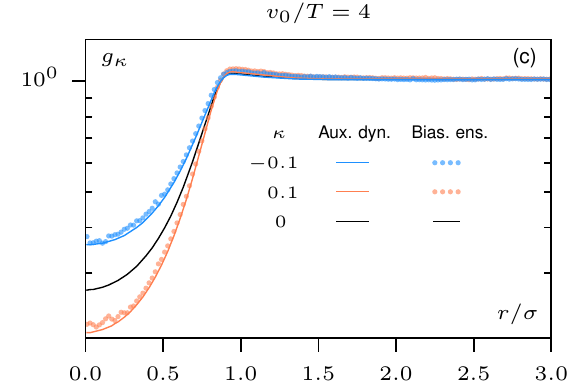}
	\hfill
	\includegraphics[width=.49\linewidth]{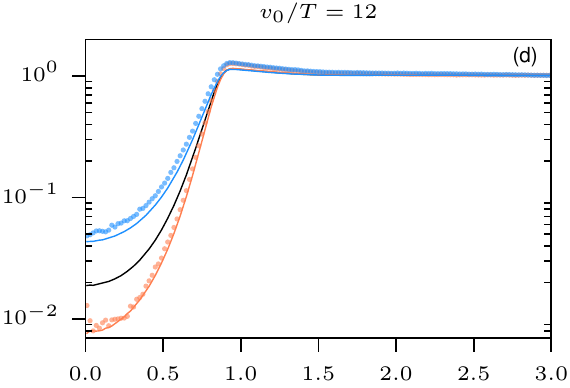}
	\caption{\label{fig:energybias}
		(a-b)~Average biasing observable $\sum_{i\in\Omega,j\not\in\Omega}\langle{\cal E}_{ij}\rangle_\kappa = \sum_{i\in\Omega,j\not\in\Omega}\langle{\cal L} v({\bf r}_i-{\bf r}_j)\rangle_\kappa/T$ as a function of bias parameter $\kappa$, where ${\cal L}$ and $v$ respectively denote the evolution operator and the pair potential of the equilibrium dynamics~\eqref{eq:dyn_eq}. Results from the first-order auxiliary dynamics (solid lines) and from a direct sampling of the biased ensemble (circles) coincide for a finite range of $\kappa$.
		(c-d)~Biased density correlation $g_\kappa$ as a function of interparticle distance $r/\sigma$ obtained from auxiliary dynamics (solid lines) and direct sampling (dotted lines). At leading order, our dynamical bias effectively renormalizes the potential $v$ by a factor $\kappa$ for specific pairs of particles $\{i\in\Omega,j\not\in\Omega\}$, in satisfying agreement with direct sampling. This illustrates the control of liquid structure at small $\kappa$ and weak interactions.
		Simulation details in Appendix~\ref{app:simu}.
	}
\end{figure*}

Moreover, the techniques in Sec.~\ref{sec:biasexternal} allow one to anticipate the trajectories generated at higher-order when now biasing with~\eqref{eq:eps_bis}. To this end, we consider the ensemble where the first-order dynamics, given by the potential~\eqref{eq:centralresult0}, is biased with $\exp\big[\int_0^t \varepsilon(s){\rm d}s\big]$ defined in terms of
\begin{equation}\label{eq:eps_pprime}
	\varepsilon = \frac{1}{4\gamma T}\sum_k\Big[\sum_{i,j}\kappa_{ij} \nabla_kv({\bf r}_i(s)-{\bf r}_j(s))\Big]^2 .
\end{equation}
As detailed in Appendix~\ref{app:far}, this ensemble is equivalent to biasing the original dynamics~\eqref{eq:dyn_eq} with~\eqref{eq:eps_bis}. Thus, the effect of higher-order bias on trajectories amounts to maximizing the squared forces in the integrand of~\eqref{eq:eps_pprime}, which effectively tends to cluster particles for both signs of $\kappa_{ij}$.

Finally, the decomposition between first-order auxiliary potential and higher-order symmetric bias can be extended to a generic class of biases of the form $T{\cal E}_{ij} = {\cal L} A({\bf r}_i-{\bf r}_j)$ for an arbitrary observable $A$: the corresponding first-order auxiliary potential $V+2\sum_{i,j}\kappa_{ij}A({\bf r}_i-{\bf r}_j)$ is now complemented with the higher-order bias~\eqref{eq:eps_pprime} where $A$ replaces $v$. Such a bias is reminiscent of, yet qualitatively different from, the escape rate used to promote dynamical heterogeneity in glassy systems~\cite{Pitard2011, Fullerton2013}. In our case, clustering is favored for both positive and negative bias parameters $\kappa_{ij}$. In particular, this is in contrast with the emergence of a hyperuniform phase, where large scale fluctuations are suppressed, reported when biasing some hydrodynamic theories of diffusive systems~\cite{Jack2015b}.


\begin{figure*}
	\centering
	\begin{flushleft}
		{\large
			\hskip1.35cm
			(a)~Red-blue micelle
			\hskip1.7cm
			(b)~Homogeneous state
			\hskip1.75cm
			(c)~Mixed clustering
		}
	\end{flushleft}
	\vskip-.1cm
	\includegraphics[width=.9\linewidth, trim=1.9cm 2.75cm 1.7cm 2.1cm, clip=true]{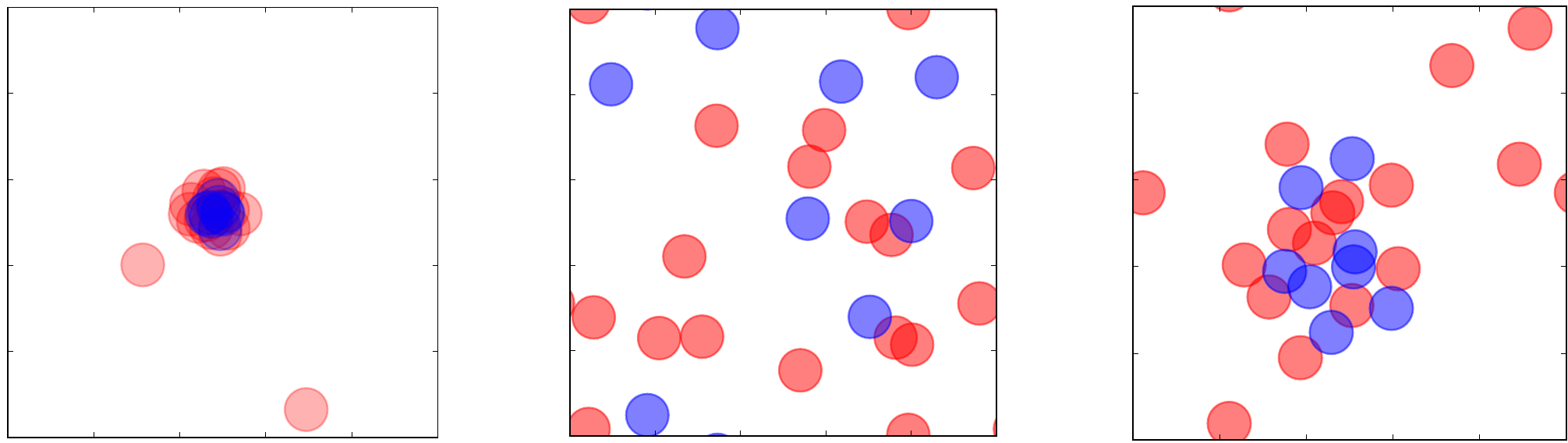}
	\begin{flushleft}
		\vskip0cm
		{\large
			\hskip2.5cm
			$\kappa=-3$
			\hskip4.5cm
			$\kappa=0$
			\hskip4.7cm
			$\kappa=3$
		}
	\end{flushleft}
	\caption{\label{fig:outofperturbation}
	Configurations obtained from a direct sampling of the biased ensemble where the pair interactions between red and blue particles are selectively modified. In the unbiased dynamics ($\kappa=0$), interactions are purely repulsive with a soft core which has a similar strength for all particles, either red or blue, so that the system is homogeneous. The dynamical bias promotes clustering for both signs of $\kappa$, yet it changes interaction selectively for either sign. The repulsion is increased between red and blue particles for $\kappa=3$, and their interactions become effectively attractive for $\kappa=-3$.  As a result, the clusters which emerge spontaneously have different structures: either a random composition of mixed reds and blues ($\kappa=3$) or a micelle-like structure with a blue core ($\kappa=-3$). This illustrates how biasing specific pairs leads to supervised spatial organization.
		Simulation details in Appendix~\ref{app:simu} and movies in~\cite{movie}.
	}
\end{figure*}

\subsection{Numerical sampling of biased structures}\label{sec:bias_num}

To illustrate the potential of our bias to control liquid properties, we focus in what follows on the specific case $\kappa_{ij}=\kappa\delta_{i\in\Omega}\delta_{j\not\in\Omega}$ where all pairs between a subset $\Omega$ and other particles are biased with the same strength $\kappa$. Here, the set $\Omega$ could for instance refer to some tracer particles immersed in the liquid, to connect with the settings in Sec.~\ref{sec:method}.

To confirm numerically the validity of our approach, we first probe the range of the first-order auxiliary dynamics where interactions {are predicted to be} simply renormalized. We compare measurements of $\sum_{i\in\Omega, j\not\in\Omega}\langle{\cal E}_{ij}\rangle_\kappa$, where $\langle\cdot\rangle_\kappa$ denotes an average in the biased ensemble, obtained from simulations {with} {the renormalized potentials in}~\eqref{eq:centralresult0} and from a direct sampling of the biased ensemble. The latter is implemented with a cloning algorithm which regularly selects and multiplies rare realizations for efficient sampling~\cite{Giadina2006, tailleur2007probing, Hurtado2009, Nemoto2016, Ray2018, Klymko2018, Brewer2018}. For convenience, interactions are now given by the soft-core potential $v({\bf r}) = v_0\exp\big[-1 / (1 - (|{\bf r}|/\sigma)^2)\big]\Theta(\sigma-|{\bf r}|)$. For weak interactions ($v_0=4T$), we observe a very satisfying agreement between the two measurements for a finite range of $\kappa$, as reported in Fig.~\ref{fig:energybias}(a), which supports the validity of our perturbation up to interaction change between $-20\%$ and $+40\%$. The range of validity decreases as $v_0/T$ increases, as shown in Fig.~\ref{fig:energybias}(b), and we expect a similar trend when also increasing the number of biased pairs.

To explore further the features of this biased ensemble, we now compare the density correlations of biased pairs $g_\kappa({\bf r})\sim \sum_{i\in\Omega,j\not\in\Omega}\langle \delta({\bf r}-{\bf r}_i+{\bf r}_j)\rangle_\kappa$ obtained from both direct sampling and first-order auxiliary dynamics. For $\kappa=\pm0.1$, we observe that the structural modification induced by the bias becomes more dramatic as $v_0/T$ increases. The agreement between the cloning and auxiliary dynamics is good for the whole curve when $v_0/T=4$, whereas a deviation appears beyond $r\simeq\sigma$ when $v_0/T=12$, as shown in Figs.~\ref{fig:energybias}(c-d). In both cases, the region of particle overlap $r<\sigma$ is well reproduced. These results corroborate the ability of the first-order auxiliary dynamics to capture interaction changes as a simple renormalization of potential strength. In contrast, the tendency for particles to cluster, manifest numerically in the increased peak value at $r\simeq\sigma$, is a higher-order effect missed by this auxiliary dynamics when $v_0/T=12$. Yet, note that the peak value is comparable for $\kappa=\pm0.1$, in agreement with~\eqref{eq:eps_pprime} being symmetric in $\kappa$. Altogether, these results demonstrate that our bias modulates the liquid structure in a controlled manner for small bias and weak interactions as predicted by~\eqref{eq:centralresult0}.

Finally, we probe numerically the effect of large bias ($|\kappa|>1$) using direct sampling, to explore configurations significantly distinct from the one of the equilibrium dynamics~\eqref{eq:dyn_eq}. The particles spontaneously tend to cluster for both positive and negative $\kappa$, as shown in Fig.~\ref{fig:outofperturbation} and movies in~\cite{movie}. This confirms the propensity of trajectories to maximize interaction forces at high bias, as captured by~\eqref{eq:eps_pprime}. Importantly, the shape of clusters differs depending on the sign of $\kappa$: a micelle-like structure featuring the particles in $\Omega$ at the core (blue) surrounded by others (red) appears for $\kappa=-3$, whereas clusters have a random composition for $\kappa=3$. Again, this agrees with the renormalized interactions being either increased ($\kappa>0$) or decreased ($\kappa<0$). In practice, the interaction strength changes sign when $\kappa<-1$ according to~\eqref{eq:centralresult0}, so that the red-blue pairs are effectively attractive for $\kappa=-3$. To optimize the overall energy, the most favorable configuration then consists in maximizing (minimizing) overlap of particles in $\Omega$ (not in $\Omega$), which in turn stabilizes a cluster of blues surrounded by reds. In general, two types of configurations should generically be stabilized for a given interaction potential $v$, depending on the sign of the bias. Overall, this establishes a reliable proof of principle for the design of tailored self-assembled structures with our specific choice of biased ensembles.


\begin{figure*}
	\centering
	\begin{flushleft}
		{\large
			\hskip4.9cm
			Isotropic state
			\hskip6.48cm
			Polar state
		}
	\end{flushleft}
	\vskip-.6cm
	\raisebox{2.1cm}{\rotatebox{90}{\large $D_{\rm r}>D_{\rm r}^*$}}
	\includegraphics[width=.318\linewidth]{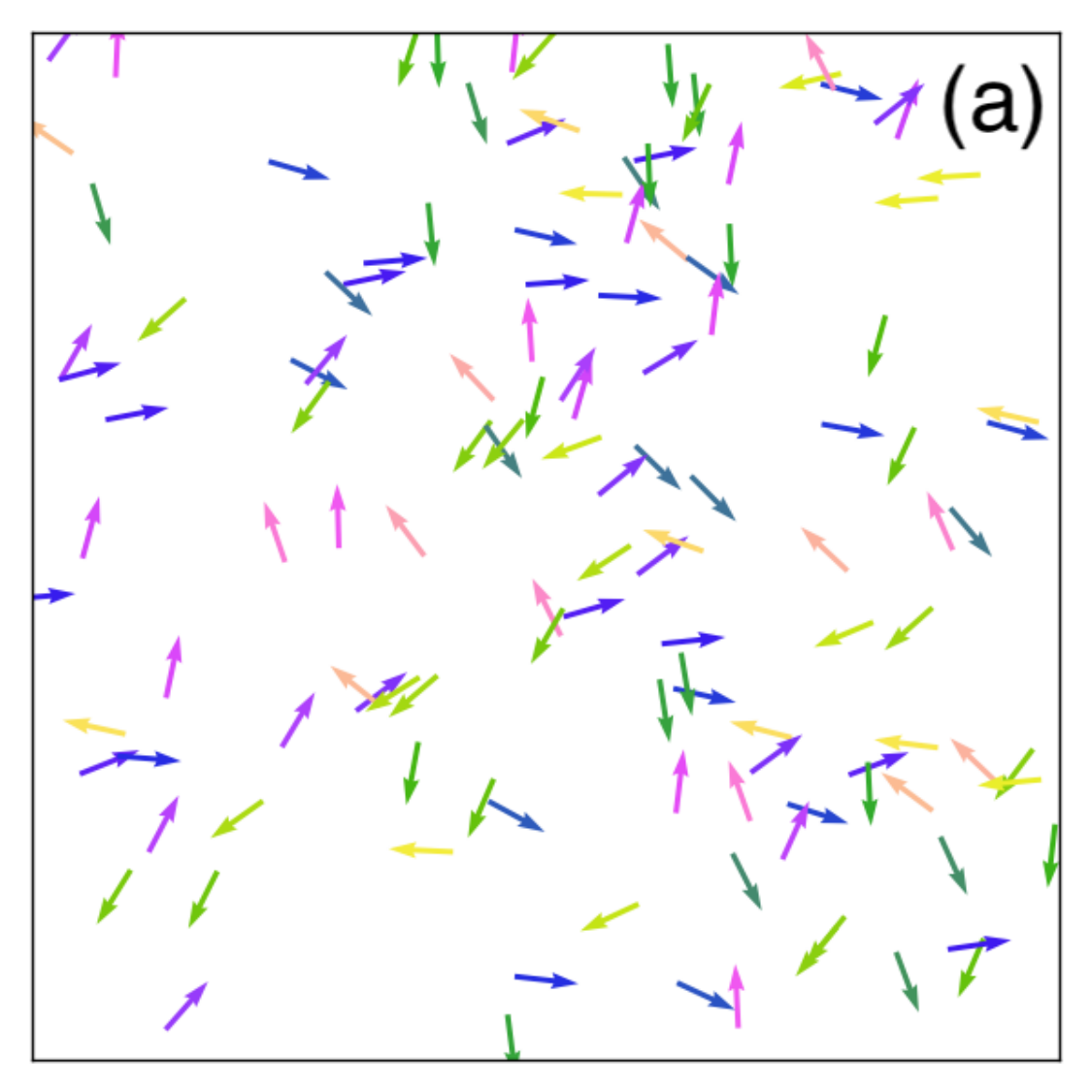}
	\includegraphics[width=.318\linewidth]{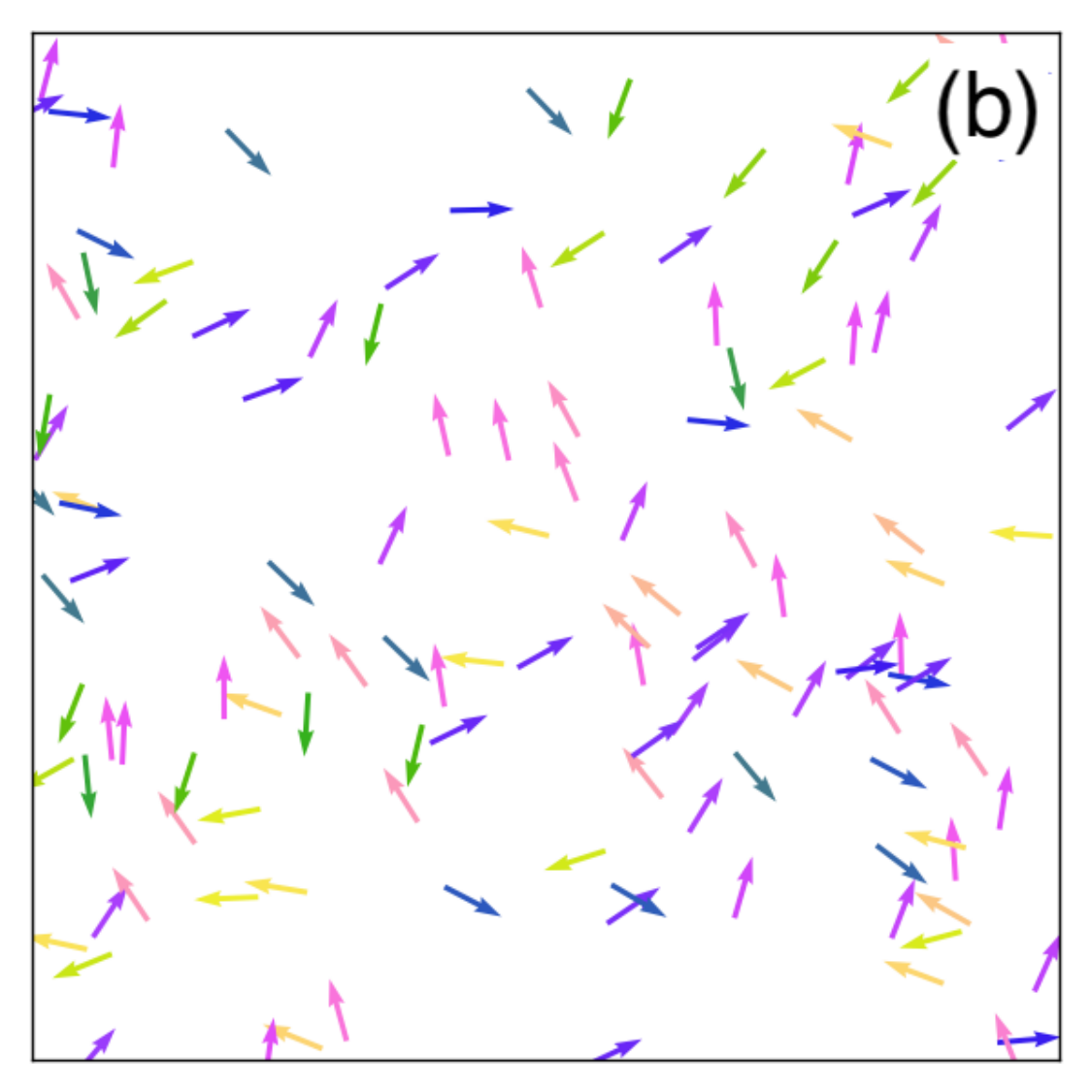}
	\raisebox{-.1cm}{\rule{.2mm}{6.2cm}}
	\includegraphics[width=.318\linewidth]{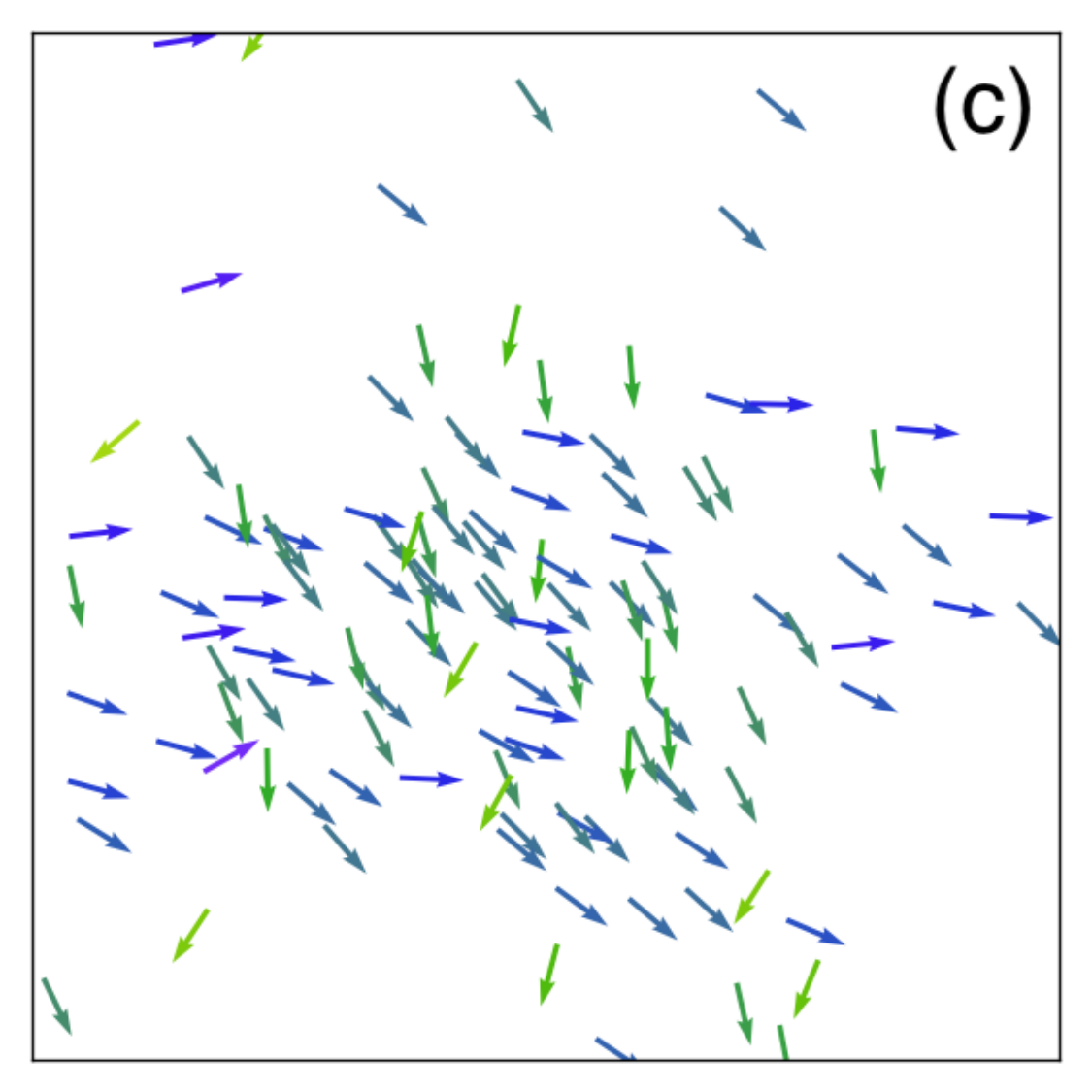}
	\vskip-.25cm
	\hskip.51cm
	\rule{5.81cm}{.2mm}
	\vskip-.05cm
	\raisebox{2.2cm}{\rotatebox{90}{\large $D_{\rm r}<D_{\rm r}^*$}}
	\includegraphics[width=.318\linewidth]{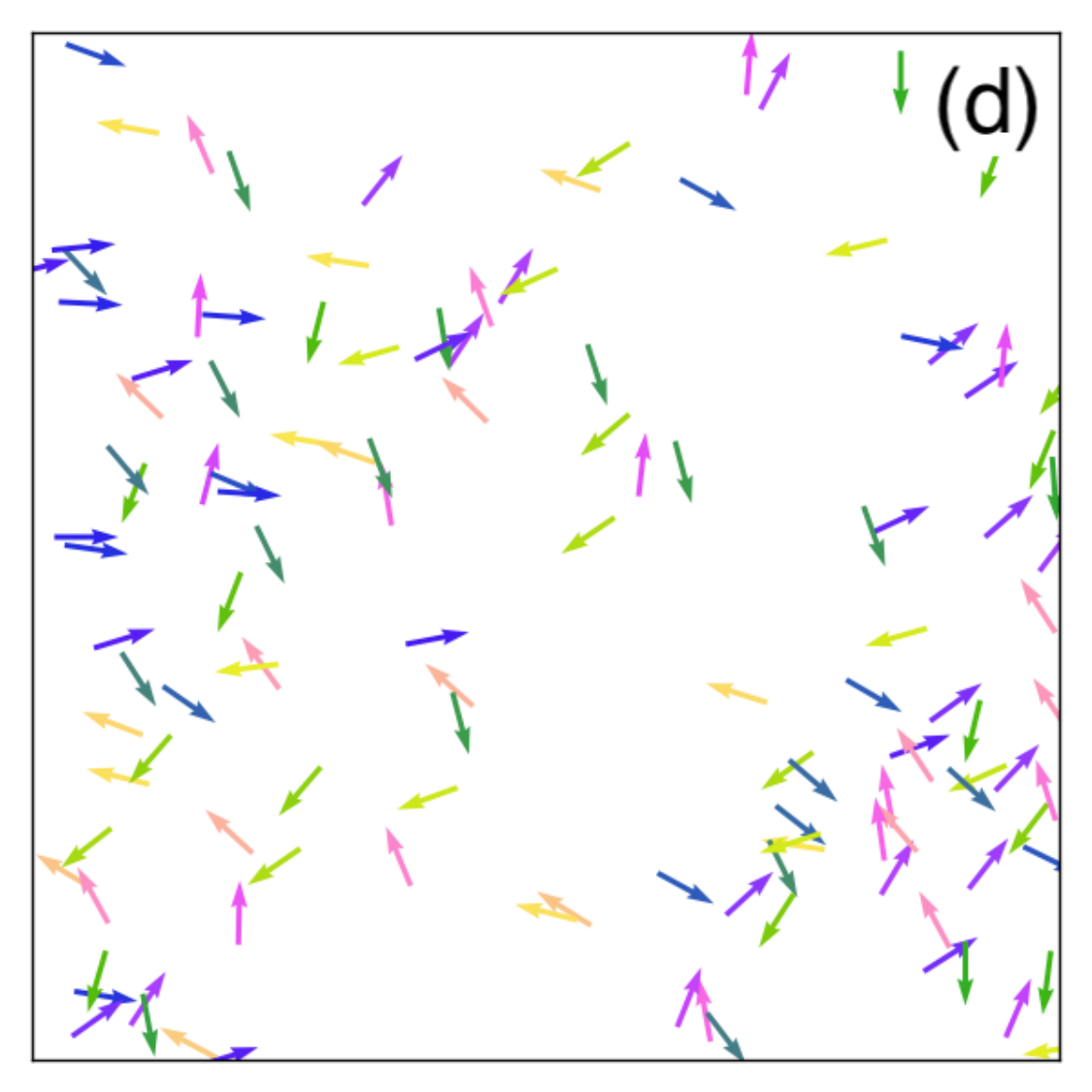}
	\raisebox{-.5cm}{\rule{.2mm}{6.3cm}}
	\includegraphics[width=.318\linewidth]{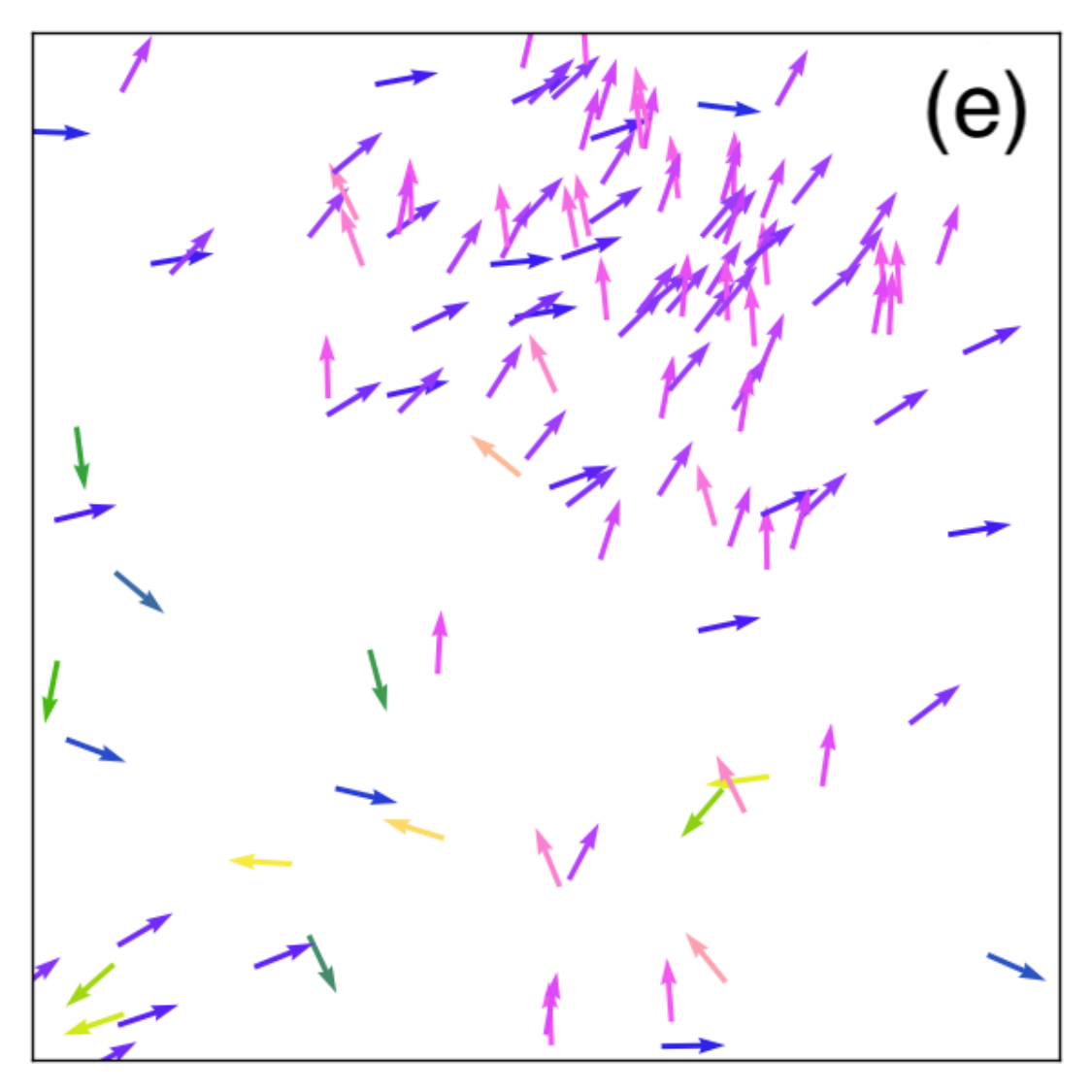}
	\includegraphics[width=.318\linewidth]{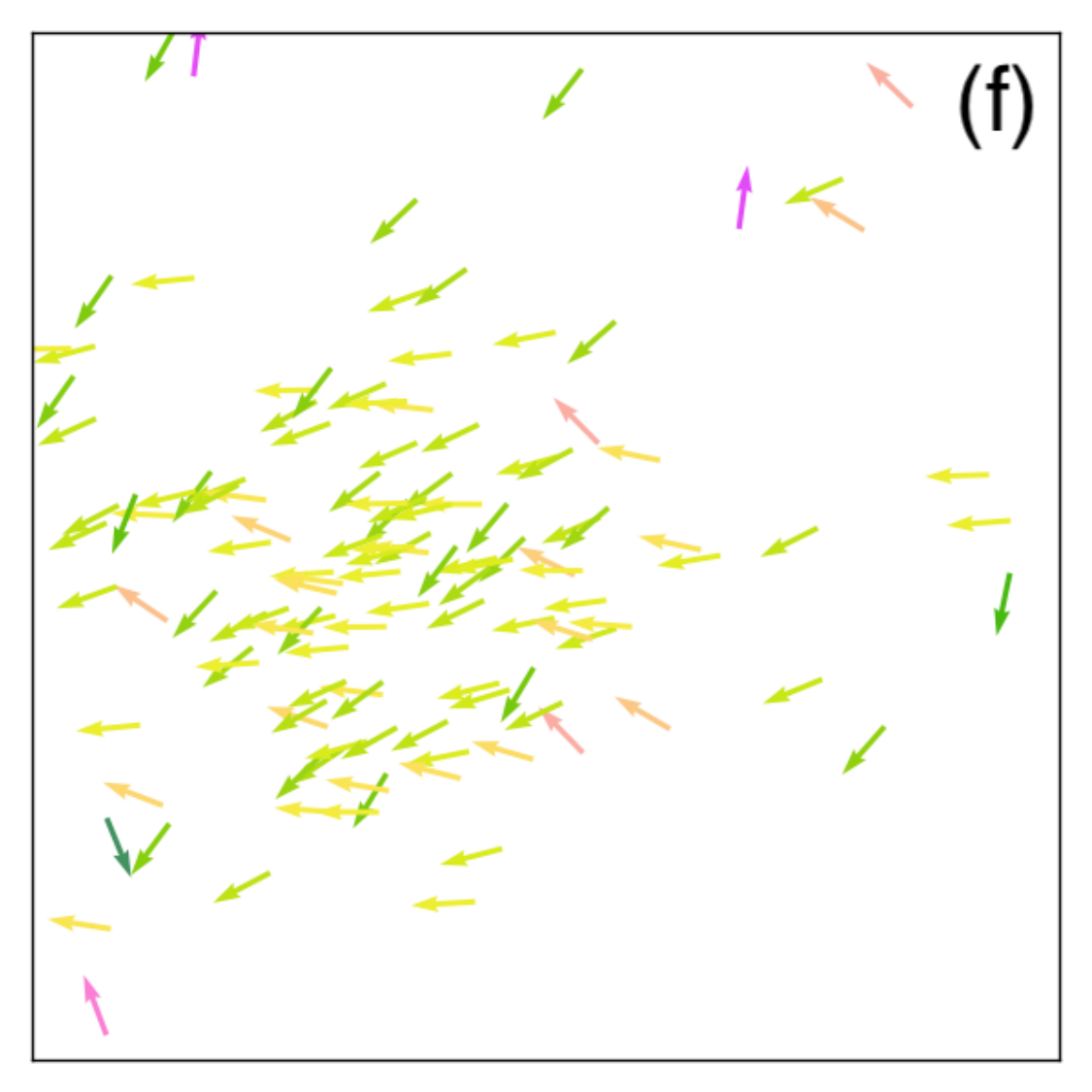}
	\begin{flushleft}
		\vskip-.6cm
		{\large
			\hskip2.55cm
			$\kappa=-0.2$
			\hskip4.48cm
			$\kappa=0$
			\hskip4.55cm
			$\kappa=0.2$
		}
	\end{flushleft}
	\caption{\label{fig:vicsek}
		Configurations obtained from a direct sampling of the biased ensemble for aligning self-propelled particles. The color code refers to the orientation of particles. In the unbiased dynamics ($\kappa=0$), we observe isotropic and polar states respectively at large noise ($D_{\rm r}>D_{\rm r}^*$) and small noise ($D_{\rm r}<D_{\rm r}^*$). Here, the critical noise is $D_{\rm r}^*=8$ and we take the noise values $D_{\rm r}=\{7,9\}$ for respectively the polar and isotropic regimes. The dynamical bias leads to renormalize interactions in a controlled manner, which effectively changes the transition threshold as $D_{\rm r}^*\to D_{\rm r}^*(1+\kappa)$ at leading order. As a result, one can stabilize either isotropic or polar states, respectively for $\kappa<0$ and $\kappa>0$, thus illustrating the ability to trigger or inhibit collective effects in nonequilibrium systems.
		Simulation details in Appendix~\ref{app:simu} and movies in~\cite{movie}.
}
\end{figure*}

\subsection{Bias-induced collective motion}\label{sec:vicsek}

As a final illustration of how collective effects can be controlled by dynamical bias, we consider a model of self-propelled particles where interactions are now only mediated via the angular dynamics~\cite{Farrell2012}:
\begin{equation}\label{eq:vicsek}
	\dot{\bf r}_i= V_0 {\bf u}(\theta_i),
	\quad
	\dot\theta_i = \mu_{\rm r} \sum_j {\cal T}(\theta_j-\theta_i, {\bf r}_i-{\bf r}_j) + \eta_i(t) ,
\end{equation}
where $V_0$ denotes the self-propulsion velocity, ${\bf u}(\theta) = (\cos\theta, \sin\theta)$ is the unit vector, and $\mu_{\rm r}$ is the rotational mobility. The term $\eta_i$ is a zero-mean Gaussian white noise with correlations $\langle\eta_i(t)\eta_j(0)\rangle = 2 D_{\rm r} \delta_{ij}\delta(t)$ given in terms of the rotational diffusion coefficient $D_{\rm r}$. To promote alignment between neighboring particles, we choose the pair-wise torque as ${\cal T}(\theta,{\bf r})=\Theta(\sigma-|{\bf r}|)\sin\theta/(\pi\sigma^2)$. This dynamics was originally introduced as a generalization of the Vicsek model to continuous time~\cite{Vicsek1995}. Thus, it exhibits a transition between a isotropic state for small density $\rho_0$ and large noise $D_{\rm r}$, and a polar state for large density $\rho_0$ and small noise $D_{\rm r}$. In practice, the linear stability analysis of the corresponding hydrodynamic equations predicts the transition to occur when $2 D_{\rm r}=\mu_{\rm r}\rho_0$~\cite{Farrell2012}. In what follows, our aim is to show that such a transition can also be mediated by tuning interactions with a dynamical bias.

To this end, we take the biasing factor in path probability as $\exp\big[\kappa \int_0^t{\cal E}_\theta(s){\rm d}s]$, where the biasing observable ${\cal E}_\theta$ reads
\begin{equation}
	\begin{aligned}\label{eq:eps_theta}
		{\cal E}_\theta = - \frac{1}{2} \sum_{i,j}& \bigg[ \frac{\partial}{\partial\theta_i} + \frac{\mu_{\rm r}}{D_{\rm r}} \sum_k {\cal T}(\theta_i-\theta_k, {\bf r}_i-{\bf r}_k) \bigg]
		\\
		&\times{\cal T}(\theta_i-\theta_j, {\bf r}_i-{\bf r}_j) .
	\end{aligned}
\end{equation}
The average value $\langle{\cal E}_\theta\rangle$ is proportional to $\frac{\rm d}{{\rm d}t}\sum_{i,j}\langle\cos(\theta_i-\theta_j)\Theta(\sigma-|{\bf r}_i-{\bf r}_j|)\rangle$, which vanishes in steady state. Then, following the procedure detailed in Sec.~\ref{sec:doob}, we deduce that biasing the dynamics~\eqref{eq:vicsek} with ${\cal E}_\theta$ amounts to considering renormalized interactions of the form 
\begin{equation}\label{eq:torque}
	\tilde{\cal T} = (1+\kappa){\cal T} + {\cal O}(\kappa^2) .
\end{equation}
Thus, by promoting a non-zero average for ${\cal E}_\theta$, aligning interactions can be tuned in a controlled manner at first order in $\kappa$. Higher order bias leads to maximize the squared torque, as presented in Appendix~\ref{app:far}.

We test this prediction numerically using a direct sampling of biased trajectories. We consider values of $\{D_{\rm r},\mu_{\rm r},\rho_0\}$ above and below the threshold $D_{\rm r}^*=\mu_{\rm r}\rho_0/2$, where the system exhibits either isotropic or polar states in the unbiased dynamics ($\kappa=0$), as shown in Fig.~\ref{fig:vicsek}. Specifically, when the original system is isotropic ($D_{\rm r}>D_{\rm r}^*$), we observe a transition to polar for $\kappa>0$, and, conversely, when it is polar ($D_{\rm r}<D_{\rm r}^*$) a transition to isotropic for $\kappa<0$. This confirms our result~\eqref{eq:torque} where the bias amounts to changing the angular mobility as $\tilde\mu_{\rm r}=(1+\kappa)\mu_{\rm r} + {\cal O}(\kappa^2)$ for weak $\kappa$, so that the linear instability is either triggered or suppressed by solely tuning $\kappa$, all other parameters being held the same. Thus, these results demonstrate how biasing the dynamics with an appropriate observable leads to control the emergence of spontaneous organization, with potential interest for other nonequilibrium dynamics.


\section{Conclusion}

Developing techniques to characterize and control the behavior of systems operating far from equilibrium remains a central and outstanding problem. Despite the apparently complex interplay between internal dissipation and emerging properties, we have demonstrated that tracer diffusion and density correlations can simply be connected to dissipation in driven liquids. We have also constructed a mapping between deterministic and active drives for a specific active matter model, thus showing how our approach can potentially be extended to a broad class of systems. Importantly, our results open promising perspectives to evaluate dissipation simply from the structure of the system. Inspired by recent works~\cite{Nardini2017, Spinney2018}, one could also introduce a map of dissipation, directly related to the statistics of interaction forces, to resolve spatially where energy is released in the thermostat. Though the corresponding integrated map would not cover the total dissipation, it would already provide insightful information about locations of low and high dissipation with respect to a constant background set by the squared driving amplitude.

In practice, monitoring dissipation with a well-defined parameter remains an open challenge for many-body systems. To this end, biased ensembles enable one to specify the statistics of dissipation by introducing an additional control parameter, analogously to the change from micro-canonical to canonical ensemble in equilibrium thermodynamics~\cite{Chetrite2013, Jack2010}. This is done by selecting rare noise realizations which drive the system away from typical behaviour, without introducing any driving force. Pioneering works were focused on favoring dynamical heterogeneities, without affecting the structure, of kinetically constrained models~\cite{garrahan2007, Hedges2009, Pitard2011, Speck2012, Bodineau2012a}. Yet, more recent studies have shown the potential to also modify density correlations in diffusive systems~\cite{Jack2014, Cagnetta2017, Nemoto2019}.

Using these large-deviation techniques, we have put forward a particular set of biased ensembles which allows one to regulate the liquid structure in a controlled manner. The explicit form of the bias is motivated by the relations between dissipation and structure that we have derived for driven liquids. At leading order, any bias in this class simply leads to introducing additional interactions in the dynamics. Furthermore, higher-order bias systematically constrains the trajectories to favor the formation of clusters. Based on minimal case studies, we have sampled the biased configurations, using state-of-the-art numerics~\cite{Giadina2006, tailleur2007probing, Hurtado2009, Nemoto2016, Ray2018, Klymko2018, Brewer2018}, to illustrate the ability to stabilize specific structures and collective effects in a controlled manner.

Since dynamical bias consists in favoring rare noise fluctuations, the corresponding dynamics effectively provides useful insights on how to promote atypical configurations with an external drive. In practice, the driving protocol should simply mimic the biased noise realizations. This line of thought has already been exploited for efficient sampling of the biased ensemble~\cite{Nemoto2016, Jack2017, Jack2018, Ferre2018}, where control forces make rare events become typical. Moreover, since our analytic framework encompasses the case of a specific bias for each pair of particle, it could potentially be regarded as a fruitful route to promote the spontaneous self-assembly of complex structures at the cost of energy dissipation. For instance, inspired by recent works~\cite{Murugan2015, Murugan2017b}, one might consider our approach to design energetic landscapes, in terms of the pair-specific bias parameters, which selectively stabilize some target molecules.

Overall, these results illustrate how specifying the amount of energy dissipated by nonequilibrium forces allows one to constrain the dynamics and structure of driven liquids. This paves the way towards controlling the emerging properties of such systems by tuning dissipation accurately. It remains to investigate whether similar results can be obtained in more complex systems which could, for instance, potentially include anisotropic building blocks, such as driven chiral objects or active liquid crystals~\cite{Joshi2017, VanZuiden2016, Nguyen2014b}.


\acknowledgements{The authors acknowledge insightful discussions with Michael E. Cates, David Martin, Robert L. Jack, and Vincenzo Vitelli. This work was granted access to the HPC resources of CINES/TGCC under the allocation 2018-A0042A10457 made by GENCI and of MesoPSL financed by the Region Ile de France and the project Equip@Meso (reference ANR-10-EQPX-29-01) of the program Investissements d'Avenir supervised by the Agence Nationale pour la Recherche. SV and LT were supported by the University of Chicago Materials Research Science and Engineering Center, which is funded by National Science Foundation under award number DMR-1420709. SV acknowledges support from the Sloan Foundation and startup funds from the University of Chicago. SV and LT acknowledge support from the National Science Foundation under award number DMR-1848306. \'EF benefits from an Oppenheimer Research Fellowship from the University of Cambridge, and a Junior Research Fellowship from St Catherine's College.}


\appendix

\section{Numerical simulations}\label{app:simu}

In Sec.~\ref{sec:map}, a custom code of molecular dynamics, based on finite time difference, is used to perform the simulations in a two-dimensional box $10^2\sigma\times 10^2\sigma$ with periodic boundary conditions. The time step is $\delta t = 10^{-4}$ and the initial condition is homogeneous. Parameter values: $\rho_0=0.7$, $T=0$, $\gamma=1$, $f=3\times 10^{-2}$, $\tau=10^3$, $v_0=5$, $n=10^2$, $\sigma=1$.

In Secs.~\ref{sec:diff} and~\ref{sec:struc}, numerical simulations of the dynamics~\eqref{eq:dyn} are performed using the LAMMPS simulation package in a two-dimensional box $10^2\sigma\times 10^2\sigma$, where $\sigma$ is the particle diameter, with periodic boundary conditions at average density $\rho_0=0.45$. 
{Our custom code actually implements overdamped Langevin equations of motion with finite time difference. It simply utilizes the efficient force computation routines that are built as a part of the Molecular dynamics package}. The system is first relaxed for $10^3$ conjugate gradient descent steps, and later equilibrated during $50\tau$. We evaluated average values over $10$ independent trajectories with duration $150\tau$. The density pair correlations were constructed using $10$ independent trajectories each with duration $50\tau$. We performed error analysis from the independent simulations and obtained negligible errors for all the data in Figs.~\ref{fig:fig1}-\ref{fig:fig2}-\ref{fig:fig2_ac}. The time step is $5\times 10^{-4}$ and the bin size for computing the pair correlations is $0.01 \sigma$. We performed simulations at other values of the time step $\{10^{-4},10^{-5}\}$ and of the bin size $\{5\times10^{-3} \sigma, 2\times10^{-2} \sigma\}$ to confirm that our calculations of the $\alpha$ coefficients are well converged. Parameter values: $T=1$, $\gamma=10^2$, $v_0=1$, $\sigma=1$.

In Sec.~\ref{sec:bias_num}, a custom code of molecular dynamics, based on finite time difference, is used to perform the simulations in a two-dimensional box $10\sigma\times 10\sigma$ with periodic boundary conditions. We bias the pair potential between $8$ blue particles and $16$ red particles. To sample the biased ensemble, we use the cloning algorithm described in Appendix A of~\cite{Nemoto2016}. The time interval for cloning is $\Delta t = 10 \delta t$ and the number of clones is $1600$. The time step is $\delta t = 10^{-4}$, the initial relaxation time is $10^4\Delta t$, and the total simulation time is $10^6 \Delta t$. Parameter values: $T=1$, $\gamma=1$, $v_0=4$ (Fig.~\ref{fig:outofperturbation}), $\sigma=1$.

In Sec.~\ref{sec:vicsek}, a custom code of molecular dynamics based on finite time difference is used to perform the simulations. $N=128$ particles are simulated in a two dimensional box of size $4\sigma\times 4\sigma$ with periodic boundary conditions. To sample the biased ensemble, we use the cloning algorithm described in Appendix A of~\cite{Nemoto2016}. The time interval for cloning is $\Delta t = 10 \delta t$ and the number of clones is $200$. The time step is $\delta t = 10^{-3}$, and the total simulation time is $500 \Delta t$. Parameter values: $V_0=2$, $\mu_{\rm r}=2$, $\rho_0=8$, $\sigma=1$.


\section{Dissipation and diffusion}\label{app:diff}

This appendix is devoted to the derivation of the dissipation rate $\cal J$ and the diffusion coefficient $D$ of a driven tracer, as defined in Sec.~\ref{sec:method}. We employ a perturbative treatment at weak interactions, originally introduced for a particle driven at constant force in~\cite{Demery2011, Demery2014}. To this aim, the tracer-bath interaction potential $v$ is scaled by a small dimensionless parameter $h\ll1$ in what follows. Besides, we focus on the regime of dilute tracers, so that interactions among them, either direct or mediated by the bath, can be safely neglected.

The dynamic action associated with the tracer dynamics~(\ref{eq:rho}-\ref{eq:EvolutionTracer}) follows from standard path integral methods~\cite{Martin1973, Dominicis1975}. It can be separated into contributions from the free tracer motion and from interactions, respectively denoted by ${\cal A}_0$ and ${\cal A}_{\rm int}$:
\begin{equation}\label{eq:action_app}
	\begin{aligned}
		{\cal A}_0 &= \int \bar{\bf r}_0 \cdot \big[ {\rm i}(\dot{\bf r}_0 - {\bf F}_{\rm d}/\gamma) + D_0 \bar{\bf r}_0 \big] {\rm d}t ,
		\\
		{\cal A}_{\rm int} &= \frac{h^2}{\gamma} \int \frac{{\rm d}{\bf q}}{(2\pi)^d} |{\bf q}|^2 |v({\bf q})|^2 \int_{-\infty}^\infty {\rm d}s\int_{-\infty}^s{\rm d}u
		\\
		&\quad\times {\rm e}^{-D_{\rm G}|{\bf q}|^2K({\bf q})(s-u)+{\rm i}{\bf q}\cdot[{\bf r}_0(s)-{\bf r}_0(u)]}
		\\
		&\quad\times \bar{\bf r}_0(s) \cdot \bigg[ \frac{\bar{\bf r}_0(u)}{\gamma K({\bf q})} - \frac{{\bf q}}{\gamma_{\rm G}} \bigg] ,
	\end{aligned}
\end{equation}
where $D_0=T/\gamma$ is the tracer diffusion coefficient in the absence of interactions ($v=0$), and $\bar{\bf r}_0$ is the process conjugated with the tracer position ${\bf r}_0$. For weak interactions $h\ll1$, any average value can be then expanded in terms of $h$ as $\langle\cdot\rangle=\langle\cdot\rangle_0 - h^2 \langle{\cal A}_{\rm int}\cdot\rangle_0 + {\cal O}(h^4)$, where $\langle\cdot\rangle_0$ is the average taken with respect to ${\cal A}_0$ only. As a result, determining the first correction from interactions in any observable amounts to computing the corresponding average $\langle{\cal A}_{\rm int}\cdot\rangle_0$.

Considering the dissipation rate per particle ${\cal J}/N=\langle\dot{\bf r}_0\rangle\cdot{\bf F}_{\rm d}$, the leading order is $\langle\dot{\bf r}_0\rangle_0 \cdot {\bf F}_{\rm d} = |{\bf F}_{\rm d}|^2/\gamma = f^2/\gamma$, and the first correction reads $ - h^2 \langle{\cal A}_{\rm int}\dot{\bf r}_0\rangle_0 \cdot {\bf F}_{\rm d} $. Given the explicit form of ${\cal A}_{\rm int}$ in~\eqref{eq:action_app}, the correlations of interest are
\begin{equation}
	\begin{aligned}
		\Big\langle\dot{\bf r}_0(t) &\big[{\bf q}\cdot\bar{\bf r}_0(s)\big] {\rm e}^{{\rm i}{\bf q}\cdot[{\bf r}_0(s)-{\bf r}_0(u)]}\Big\rangle_0
		\\
		&= {\rm i}{\bf q}\delta(t-s){\rm e}^{ - D_0|{\bf q}|^2(t-u) + \frac{{\rm i}{\bf q}}{\gamma}\cdot\int_u^t {\bf F}_{\rm d}(w) {\rm d}w} ,
		\\
		\Big\langle\dot{\bf r}_0(t) &\big[\bar{\bf r}_0(u)\cdot\bar{\bf r}_0(s)\big] {\rm e}^{{\rm i}{\bf q}\cdot[{\bf r}_0(s)-{\bf r}_0(u)]}\Big\rangle_0
		\\
		&= -{\rm i}{\bf q}\delta(t-s){\rm e}^{ - D_0|{\bf q}|^2(t-u) + \frac{{\rm i}{\bf q}}{\gamma}\cdot\int_u^t {\bf F}_{\rm d}(w) {\rm d}w} ,
	\end{aligned}
\end{equation}
where we have used that the tracer statistics is Gaussian in the absence of interactions, following~\cite{Demery2011, Demery2014}. From this result, we get
\begin{equation}
	\begin{aligned}
		&{\cal J} - f^2/\gamma
		\\
		&\,= \frac{Nh^2}{d\gamma^2} \int \frac{{\rm d}{\bf q}}{(2\pi)^d} {\rm i}{\bf q}\cdot{\bf F}_{\rm d}(t) |{\bf q}|^2 |v({\bf q})|^2 \frac{D_0+D_{\rm G}K({\bf q})}{D_0K({\bf q})}
		\\
		&\,\quad\times \int_{-\infty}^t {\rm d}u {\rm e}^{-|{\bf q}|^2 [D_0+D_{\rm G} K({\bf q})](t-u) + \frac{{\rm i}{\bf q}}{\gamma}\cdot\int_u^t{\bf F}_{\rm d}(w){\rm d}w}
		\\
		&\,\quad + {\cal O}(h^4) ,
	\end{aligned}
\end{equation}
where we have used $\gamma_{\rm G}=\gamma/\rho_0$ and $D_{\rm G}=\rho_0D_0$. Expanding at small $f$, we deduce
\begin{equation}\label{eq:diss}
	\begin{aligned}
		&{\cal J} - f^2/\gamma
		\\
		&\,= - \frac{Nh^2}{d\gamma^3} \int \frac{{\rm d}{\bf q}}{(2\pi)^d} |{\bf q}|^4 |v({\bf q})|^2 \frac{D_0+D_{\rm G}K({\bf q})}{D_0K({\bf q})}
		\\
		&\,\quad\times \int_{-\infty}^t {\rm d}u {\rm e}^{-|{\bf q}|^2 [D_0+D_{\rm G} K({\bf q})](t-u)} \int_u^t{\rm d}w{\bf F}_{\rm d}(t)\cdot{\bf F}_{\rm d}(w)
		\\
		&\,\quad + {\cal O}(h^4,f^4) .
	\end{aligned}
\end{equation}
Substituting the explicit expression of the deterministic drive~\eqref{eq:theta} in~\eqref{eq:diss}, and then integrating over $u$ and $w$, we obtain
\begin{equation}
	\begin{aligned}
		&{\cal J} - f^2/\gamma
		\\
		&\,= - \frac{N(hf)^2}{d\gamma^3} \int \frac{{\rm d}{\bf q}}{(2\pi)^d} \frac{|{\bf q}|^4 |v({\bf q})|^2}{|{\bf q}|^4\big[D_0+D_{\rm G}K({\bf q})\big]^2 + \omega^2}
		\\
		&\,\quad\times \frac{D_0+D_{\rm G}K({\bf q})}{D_0K({\bf q})} + {\cal O}(h^4,f^4) .
	\end{aligned}
\end{equation}
For the case of active drive with correlations~\eqref{eq:theta_ac}, we exploit the equivalence with a disordered drive detailed in Sec.~\ref{sec:map}. Substituting the explicit drive~\eqref{eq:theta_dis} in~\eqref{eq:diss}, and then averaging over disorder in the limit of many oscillators ($n\gg1$), we get
\begin{equation}
	\begin{aligned}
		&{\cal J} - f^2/\gamma
		\\
		\,&= - \frac{N(hf)^2}{d\gamma^3} \int \frac{{\rm d}{\bf q}{\rm d}\omega'}{(2\pi)^{d+1}} \frac{|{\bf q}|^4 |v({\bf q})|^2 \phi(\omega')}{|{\bf q}|^4\big[D_0+D_{\rm G}K({\bf q})\big]^2 + (\omega')^2}
		\\
		&\quad\times \frac{D_0+D_{\rm G}K({\bf q})}{D_0K({\bf q})} + {\cal O}(h^4,f^4) ,
	\end{aligned}
\end{equation}
where $\phi(\omega') = 2\tau/\big[1+(\omega'\tau)^2\big]$, yielding
\begin{equation}
	\begin{aligned}
		&{\cal J} - f^2/\gamma
		\\
		\,&= - \frac{N\tau(hf)^2}{d\gamma^3} \int \frac{{\rm d}{\bf q}}{(2\pi)^d} \frac{|{\bf q}|^2 |v({\bf q})|^2}{D_0 K({\bf q})}
		\\
		&\quad\times \frac{1}{\tau|{\bf q}|^2\big[D_0+D_{\rm G}K({\bf q})\big] + 1} + {\cal O}(h^4,f^4) .
	\end{aligned}
\end{equation}
The asymptotic results for the rate of work $\dot w = f^2/\gamma-{\cal J}$, presented in Sec.~\ref{sec:diff} for both deterministic and active drives, follow directly.

We now turn to deriving the diffusion coefficient $D$. It is defined in terms of the mean-squared displacement (MSD) $\langle\Delta{\bf r}_0^2(t)\rangle=\big\langle\big[\langle{\bf r}_0(t)\rangle - {\bf r}_0(t)\big]^2\big\rangle$ as $D=\underset{t\to\infty}{\lim}\langle\Delta{\bf r}_0^2(t)\rangle / 2 d t$. At leading order, the MSD reads $\langle\Delta{\bf r}_0^2(t)\rangle_0 = 2dD_0t$. To obtain the first order, we need to compute the following correlations
\begin{equation}
	\begin{aligned}
		\Big\langle\Delta{\bf r}^2_0(t)& \big[{\bf q}\cdot\bar{\bf r}_0(s)\big] {\rm e}^{{\rm i}{\bf q}\cdot[{\bf r}_0(s)-{\bf r}_0(u)]}\Big\rangle_0
		\\
		&= - 4 (D_0/\gamma)|{\bf q}|^2(s-u) \Theta(t-s)
		\\
		&\quad\times{\rm e}^{ - D_0|{\bf q}|^2(t-u) + \frac{{\rm i}{\bf q}}{\gamma}\cdot\int_u^t {\bf F}_{\rm d}(w) {\rm d}w},
		\\
		\Big\langle\Delta{\bf r}^2_0(t)& \big[\bar{\bf r}_0(u)\cdot\bar{\bf r}_0(s)\big] {\rm e}^{{\rm i}{\bf q}\cdot[{\bf r}_0(s)-{\bf r}_0(u)]}\Big\rangle_0
		\\
		&= (2/\gamma^2)\Theta(t-s) \big[2D_0|{\bf q}|^2(s-u)-1\big]
		\\
		&\quad\times{\rm e}^{ - D_0|{\bf q}|^2(t-u) + \frac{{\rm i}{\bf q}}{\gamma}\cdot\int_u^t {\bf F}_{\rm d}(w) {\rm d}w} ,
	\end{aligned}
\end{equation}
where we have used again that ${\cal A}_0$ is Gaussian in terms of $\bar{\bf r}_0$, yielding
\begin{equation}
	\begin{aligned}
		&\big\langle\Delta{\bf r}^2_0(t)\big\rangle - 2 d D_0 t
		\\
		&\,= \frac{2h^2}{\gamma^2} \int \frac{{\rm d}{\bf q}}{(2\pi)^d} \frac{|{\bf q}|^2 |v({\bf q})|^2}{K({\bf q})}
		\\
		&\,\quad\times \int_{-\infty}^t{\rm d}s \int_{-\infty}^s{\rm d}u \big\{ 2|{\bf q}|^2\big[D_0+D_{\rm G}K({\bf q})\big](s-u) - 1\big\}
		\\
		&\,\quad\times {\rm e}^{ -|{\bf q}|^2[D_0+D_{\rm G}K({\bf q})](s-u) + \frac{{\rm i}{\bf q}}{\gamma}\cdot\int_u^s{\bf F}_{\rm d}(w) {\rm d}w } + {\cal O}(h^4) .
	\end{aligned}
\end{equation}
Expanding at small $f$, we get
\begin{equation}
	\begin{aligned}
		&\big\langle\Delta{\bf r}^2_0(t)\big\rangle - 2 d D_{\rm eq} t
		\\
		&\,= - \frac{2h^2}{\gamma^4} \int \frac{{\rm d}{\bf q}}{(2\pi)^d} \frac{|{\bf q}|^4 |v({\bf q})|^2}{K({\bf q})}
		\\
		&\,\quad\times \int_{-\infty}^t{\rm d}s \int_{-\infty}^s{\rm d}u \big\{ 2|{\bf q}|^2\big[D_0+D_{\rm G}K({\bf q})\big](s-u) - 1\big\}
		\\
		&\,\quad\times {\rm e}^{ -|{\bf q}|^2[D_0+D_{\rm G}K({\bf q})](s-u)} \int_u^s {\rm d}w_1{\rm d}w_2 {\bf F}_{\rm d}(w_1)\cdot{\bf F}_{\rm d}(w_2)
		\\
		&\,\quad + {\cal O}(h^4,f^4) ,
	\end{aligned}
\end{equation}
where $D_{\rm eq}$ refers to the diffusion coefficient in the absence of driving force ($f=0$). For the deterministic drive~\eqref{eq:theta}, the explicit time integrations give
\begin{equation}
	\begin{aligned}
		D - D_{\rm eq} &= \frac{(hf)^2}{d\gamma^4} \int \frac{{\rm d}{\bf q}}{(2\pi)^d} \frac{|{\bf q}|^2|v({\bf q})|^2}{K({\bf q})\big[D_0+D_{\rm G}K({\bf q})\big]}
		\\
		&\quad\times \frac{5|{\bf q}|^4\big[D_0+D_{\rm G}K({\bf q})\big]^2 + \omega^2}{ \big\{ |{\bf q}|^4\big[D_0+D_{\rm G}K({\bf q})\big]^2 + \omega^2 \big\}^2}
		\\
		&\quad + {\cal O}(h^4,f^4) .
	\end{aligned}
\end{equation}
Using the mapping in Sec.~\ref{sec:map} for the case of active drive with correlations~\eqref{eq:theta_ac}, we deduce
\begin{equation}
	\begin{aligned}
		D - D_{\rm eq} &= \frac{(hf)^2}{d\gamma^4} \int \frac{{\rm d}{\bf q}{\rm d}\omega'}{(2\pi)^{d+1}} \frac{|{\bf q}|^2|v({\bf q})|^2\phi(\omega')}{K({\bf q})\big[D_0+D_{\rm G}K({\bf q})\big]}
		\\
		&\quad\times \frac{5|{\bf q}|^4\big[D_0+D_{\rm G}K({\bf q})\big]^2 + (\omega')^2}{ \big\{ |{\bf q}|^4\big[D_0+D_{\rm G}K({\bf q})\big]^2 + (\omega')^2 \big\}^2}
		\\
		&\quad + {\cal O}(h^4,f^4) ,
	\end{aligned}
\end{equation}
where again $\phi(\omega') = 2\tau/\big[1+(\omega'\tau)^2\big]$, yielding
\begin{equation}
	\begin{aligned}
		D - D_{\rm eq} &= \frac{\tau(hf)^2}{d\gamma^4} \int \frac{{\rm d}{\bf q}}{(2\pi)^d} \frac{|v({\bf q})|^2}{K({\bf q})\big[D_0+D_{\rm G}K({\bf q})\big]^2}
		\\
		&\quad\times \frac{5\tau|{\bf q}|^2\big[D_0+D_{\rm G}K({\bf q})\big] + 3}{ \big\{ \tau|{\bf q}|^2\big[D_0+D_{\rm G}K({\bf q})\big] + 1 \big\}^2}
		\\
		&\quad + {\cal O}(h^4,f^4) .
	\end{aligned}
\end{equation}
Finally, we obtain the expressions in the asymptotic regimes, as reported in Sec.~\ref{sec:diff} for both deterministic and active drives.


\section{Equivalence of biased ensembles}\label{app:far}

In this Appendix, we demonstrate the equivalence between specific dynamical biased ensembles. First, we consider ensembles related to the equilibrium dynamics~\eqref{eq:dyn_eq}. Ensemble (a) corresponds to biasing with the factor $\exp\big[\sum_{i,j} \kappa_{ij}\int_0^t{\cal E}_{ij}(s){\rm d}s\big]$ in the path probability, where
\begin{equation}
	{\cal E}_{ij} = \frac{1}{\gamma T} \sum_k\big[T\nabla_k - \nabla_kV\big]\cdot\nabla_k A({\bf r}_i-{\bf r}_j) .
\end{equation}
Ensemble (b) is associated with the first-order auxiliary dynamics, whose potential reads $V+2\sum_{i,j}\kappa_{ij}A({\bf r}_i-{\bf r}_j)$, biased with $\exp\big[\int_0^t\varepsilon'(s){\rm d}s\big]$ where
\begin{equation}
	\varepsilon' = \frac{1}{\gamma T}\sum_k\Big[\sum_{i,j}\kappa_{ij} \nabla_kA({\bf r}_i(s)-{\bf r}_j(s))\Big]^2 .
\end{equation}
Obtaining the equivalence between (a) and (b) amounts to showing that their path probabilities are similar. The corresponding dynamic actions, denoted by ${\cal A}^{( \sigma)}(t) = \sum_k\int_0^t{\mathbb A}_k^{(\sigma)}(s){\rm d}s$ for $\sigma=\{{\rm a,b}\}$, are given by
\begin{equation}\label{eq:Aa}
	\begin{aligned}
		{\mathbb A}_k^{(\rm a)} &= \frac{1}{4\gamma T} \big[\gamma\dot{\bf r}_k + \nabla_kV\big]^2 - \frac{1}{2\gamma} \nabla^2_kV
		\\
		&\quad - \frac{1}{\gamma T} \sum_{i,j}\kappa_{ij}\big[T \nabla_k - \nabla_k V\big] \cdot\nabla_k A({\bf r}_i-{\bf r}_j) ,
	\end{aligned}
\end{equation}
and
\begin{equation}\label{eq:Ab}
	\begin{aligned}
		{\mathbb A}_k^{(\rm b)}	&= \frac{1}{4\gamma T} \Big[\gamma\dot{\bf r}_k + \nabla_k V + 2\sum_{i,j}\kappa_{ij}\nabla_kA({\bf r}_i-{\bf r}_j)\Big]^2
		\\
		&\quad - \frac{1}{2\gamma} \nabla^2_k \Big[V + 2\sum_{i,j}\kappa_{ij} A({\bf r}_i-{\bf r}_j)\Big]
		\\
		&\quad - \frac{1}{\gamma T}\Big[\sum_{i,j}\kappa_{ij}\nabla_kA({\bf r}_i-{\bf r}_j)\Big]^2 .
	\end{aligned}
\end{equation}
Expanding ${\mathbb A}_k^{(\rm b)}$ in~\eqref{eq:Ab} and comparing with ${\mathbb A}_k^{(\rm b)}$ in~\eqref{eq:Aa}, it appears that ${\cal A}^{(\rm a)}$ and ${\cal A}^{(\rm b)}$ are indeed equal up to a boundary term proportional to $\sum_{i,j}\kappa_{ij}\big[A({\bf r}_i(t)-{\bf r}_j(t)) - A({\bf r}_i(0)-{\bf r}_j(0))\big]$ which can be neglected at large $t$: this establishes the equivalence between ensembles (a) and (b).

We now turn to demonstrate the equivalence between two ensembles related to the Vicsek-like dynamics~\eqref{eq:vicsek}. Ensemble (c) is biased with the factor $\exp\big[\kappa\int_0^t{\cal E}_\theta(s){\rm d}s\big]$, where ${\cal E}_\theta$ is defined in~\eqref{eq:eps_theta}. Ensemble (d) corresponds to the first-order auxiliary dynamics, with torque given by $(1+\kappa){\cal T}$, biased with $\exp\big[\int_0^t\varepsilon_\theta(s){\rm d}s\big]$ where
\begin{equation}
	 \varepsilon_\theta = \frac{(\kappa\mu_{\rm r})^2}{4D_{\rm r}} \sum_{i,j,k} {\cal T}(\theta_i-\theta_k,{\bf r}_i-{\bf r}_k) {\cal T}(\theta_i-\theta_j,{\bf r}_i-{\bf r}_j) .
\end{equation}
The dynamic actions for each ensemble, denoted by ${\cal B}^{( \sigma)}(t) = \sum_i\int_0^t{\mathbb B}_i^{(\sigma)}(s){\rm d}s$ for $\sigma=\{{\rm c,d}\}$, are given by
\begin{widetext}
\begin{equation}\label{eq:Bc}
	\begin{aligned}
		{\mathbb B}_i^{(\rm c)} &= \frac{1}{4D_{\rm r}} \Big[ \dot\theta_i - \mu_{\rm r}\sum_j{\cal T}(\theta_i-\theta_j,{\bf r}_i-{\bf r}_j) \Big]^2 + \frac{\mu_{\rm r}}{2} \frac{\partial}{\partial\theta_i}\sum_j{\cal T}(\theta_i-\theta_j,{\bf r}_i-{\bf r}_j)
		\\
		&\quad + \frac{\kappa\mu_{\rm r}}{2} \sum_j \bigg[ \frac{\partial}{\partial\theta_i} + \frac{\mu_{\rm r}}{D_{\rm r}}\sum_k{\cal T}(\theta_i-\theta_k,{\bf r}_i-{\bf r}_k) \bigg] {\cal T}(\theta_i-\theta_j,{\bf r}_i-{\bf r}_j) ,
	\end{aligned}
\end{equation}
and
\begin{equation}\label{eq:Bd}
	\begin{aligned}
		{\mathbb B}_i^{(\rm d)} &= \frac{1}{4D_{\rm r}} \Big[ \dot\theta_i - (1+\kappa)\mu_{\rm r}\sum_j{\cal T}(\theta_i-\theta_j,{\bf r}_i-{\bf r}_j) \Big]^2 + \frac{(1+\kappa)\mu_{\rm r}}{2} \frac{\partial}{\partial\theta_i}\sum_j{\cal T}(\theta_i-\theta_j,{\bf r}_i-{\bf r}_j)
		\\
		&\quad - \frac{(\kappa\mu_{\rm r}	)^2}{4D_{\rm r}} \sum_{j,k} {\cal T}(\theta_i-\theta_k,{\bf r}_i-{\bf r}_k) {\cal T}(\theta_i-\theta_j,{\bf r}_i-{\bf r}_j) .
	\end{aligned}
\end{equation}
Expanding ${\mathbb B}_i^{(\rm d)}$ in~\eqref{eq:Bc} and comparing with ${\mathbb B}_i^{(\rm c)}$ in~\eqref{eq:Bd}, it appears that ${\cal B}^{(\rm c)}$ and ${\cal B}^{(\rm d)}$ only differ by a term proportional to $\sum_{i,j} \dot\theta_i {\cal T}(\theta_i-\theta_j,{\bf r}_i-{\bf r}_j)$ which can safely be neglected at large $t$, thus proving the equivalence between ensembles (c) and (d).
\end{widetext}


\bibliography{Driven_references.bib}

\begin{thebibliography}{119}%
\makeatletter
\providecommand \@ifxundefined [1]{%
 \@ifx{#1\undefined}
}%
\providecommand \@ifnum [1]{%
 \ifnum #1\expandafter \@firstoftwo
 \else \expandafter \@secondoftwo
 \fi
}%
\providecommand \@ifx [1]{%
 \ifx #1\expandafter \@firstoftwo
 \else \expandafter \@secondoftwo
 \fi
}%
\providecommand \natexlab [1]{#1}%
\providecommand \enquote  [1]{``#1''}%
\providecommand \bibnamefont  [1]{#1}%
\providecommand \bibfnamefont [1]{#1}%
\providecommand \citenamefont [1]{#1}%
\providecommand \href@noop [0]{\@secondoftwo}%
\providecommand \href [0]{\begingroup \@sanitize@url \@href}%
\providecommand \@href[1]{\@@startlink{#1}\@@href}%
\providecommand \@@href[1]{\endgroup#1\@@endlink}%
\providecommand \@sanitize@url [0]{\catcode `\\12\catcode `\$12\catcode
  `\&12\catcode `\#12\catcode `\^12\catcode `\_12\catcode `\%12\relax}%
\providecommand \@@startlink[1]{}%
\providecommand \@@endlink[0]{}%
\providecommand \url  [0]{\begingroup\@sanitize@url \@url }%
\providecommand \@url [1]{\endgroup\@href {#1}{\urlprefix }}%
\providecommand \urlprefix  [0]{URL }%
\providecommand \Eprint [0]{\href }%
\providecommand \doibase [0]{http://dx.doi.org/}%
\providecommand \selectlanguage [0]{\@gobble}%
\providecommand \bibinfo  [0]{\@secondoftwo}%
\providecommand \bibfield  [0]{\@secondoftwo}%
\providecommand \translation [1]{[#1]}%
\providecommand \BibitemOpen [0]{}%
\providecommand \bibitemStop [0]{}%
\providecommand \bibitemNoStop [0]{.\EOS\space}%
\providecommand \EOS [0]{\spacefactor3000\relax}%
\providecommand \BibitemShut  [1]{\csname bibitem#1\endcsname}%
\let\auto@bib@innerbib\@empty
\bibitem [{\citenamefont {Toyabe}\ \emph {et~al.}(2010)\citenamefont {Toyabe},
  \citenamefont {Okamoto}, \citenamefont {Watanabe-Nakayama}, \citenamefont
  {Taketani}, \citenamefont {Kudo},\ and\ \citenamefont
  {Muneyuki}}]{Toyabe2010}%
  \BibitemOpen
  \bibfield  {author} {\bibinfo {author} {\bibfnamefont {Shoichi}\ \bibnamefont
  {Toyabe}}, \bibinfo {author} {\bibfnamefont {Tetsuaki}\ \bibnamefont
  {Okamoto}}, \bibinfo {author} {\bibfnamefont {Takahiro}\ \bibnamefont
  {Watanabe-Nakayama}}, \bibinfo {author} {\bibfnamefont {Hiroshi}\
  \bibnamefont {Taketani}}, \bibinfo {author} {\bibfnamefont {Seishi}\
  \bibnamefont {Kudo}}, \ and\ \bibinfo {author} {\bibfnamefont {Eiro}\
  \bibnamefont {Muneyuki}},\ }\bibfield  {title} {\enquote {\bibinfo {title}
  {Nonequilibrium energetics of a single ${\mathbf{f}}_{1}$-atpase molecule},}\
  }\href {\doibase 10.1103/PhysRevLett.104.198103} {\bibfield  {journal}
  {\bibinfo  {journal} {Phys. Rev. Lett.}\ }\textbf {\bibinfo {volume} {104}},\
  \bibinfo {pages} {198103} (\bibinfo {year} {2010})}\BibitemShut {NoStop}%
\bibitem [{\citenamefont {Fodor}\ \emph
  {et~al.}(2016{\natexlab{a}})\citenamefont {Fodor}, \citenamefont {Ahmed},
  \citenamefont {Almonacid}, \citenamefont {Bussonnier}, \citenamefont {Gov},
  \citenamefont {Verlhac}, \citenamefont {Betz}, \citenamefont {Visco},\ and\
  \citenamefont {van Wijland}}]{Ahmed2016}%
  \BibitemOpen
  \bibfield  {author} {\bibinfo {author} {\bibfnamefont {{\'E}.}~\bibnamefont
  {Fodor}}, \bibinfo {author} {\bibfnamefont {W.~W.}\ \bibnamefont {Ahmed}},
  \bibinfo {author} {\bibfnamefont {M.}~\bibnamefont {Almonacid}}, \bibinfo
  {author} {\bibfnamefont {M.}~\bibnamefont {Bussonnier}}, \bibinfo {author}
  {\bibfnamefont {N.~S.}\ \bibnamefont {Gov}}, \bibinfo {author} {\bibfnamefont
  {M.-H.}\ \bibnamefont {Verlhac}}, \bibinfo {author} {\bibfnamefont
  {T.}~\bibnamefont {Betz}}, \bibinfo {author} {\bibfnamefont {P.}~\bibnamefont
  {Visco}}, \ and\ \bibinfo {author} {\bibfnamefont {F.}~\bibnamefont {van
  Wijland}},\ }\bibfield  {title} {\enquote {\bibinfo {title} {Nonequilibrium
  dissipation in living oocytes},}\ }\href {\doibase
  10.1209/0295-5075/116/30008} {\bibfield  {journal} {\bibinfo  {journal} {EPL
  (Europhys. Lett.)}\ }\textbf {\bibinfo {volume} {116}},\ \bibinfo {pages}
  {30008} (\bibinfo {year} {2016}{\natexlab{a}})}\BibitemShut {NoStop}%
\bibitem [{\citenamefont {Battle}\ \emph {et~al.}(2016)\citenamefont {Battle},
  \citenamefont {Broedersz}, \citenamefont {Fakhri}, \citenamefont {Geyer},
  \citenamefont {Howard}, \citenamefont {Schmidt},\ and\ \citenamefont
  {MacKintosh}}]{Battle604}%
  \BibitemOpen
  \bibfield  {author} {\bibinfo {author} {\bibfnamefont {Christopher}\
  \bibnamefont {Battle}}, \bibinfo {author} {\bibfnamefont {Chase}\
  \bibnamefont {Broedersz}}, \bibinfo {author} {\bibfnamefont {Mikta}\
  \bibnamefont {Fakhri}}, \bibinfo {author} {\bibfnamefont {Veikko}\
  \bibnamefont {Geyer}}, \bibinfo {author} {\bibfnamefont {Jonaton}\
  \bibnamefont {Howard}}, \bibinfo {author} {\bibfnamefont {Christoph}\
  \bibnamefont {Schmidt}}, \ and\ \bibinfo {author} {\bibfnamefont {Fred}\
  \bibnamefont {MacKintosh}},\ }\bibfield  {title} {\enquote {\bibinfo {title}
  {{Broken detailed balance at mesoscopic scales in active biological
  systems}},}\ }\href {\doibase 10.1126/science.aac8167} {\bibfield  {journal}
  {\bibinfo  {journal} {Science}\ }\textbf {\bibinfo {volume} {351}},\ \bibinfo
  {pages} {604--607} (\bibinfo {year} {2016})}\BibitemShut {NoStop}%
\bibitem [{\citenamefont {Gnesotto}\ \emph {et~al.}(2018)\citenamefont
  {Gnesotto}, \citenamefont {Mura}, \citenamefont {Gladrow},\ and\
  \citenamefont {Broedersz}}]{Mura2018}%
  \BibitemOpen
  \bibfield  {author} {\bibinfo {author} {\bibfnamefont {F~S}\ \bibnamefont
  {Gnesotto}}, \bibinfo {author} {\bibfnamefont {F}~\bibnamefont {Mura}},
  \bibinfo {author} {\bibfnamefont {J}~\bibnamefont {Gladrow}}, \ and\ \bibinfo
  {author} {\bibfnamefont {C~P}\ \bibnamefont {Broedersz}},\ }\bibfield
  {title} {\enquote {\bibinfo {title} {Broken detailed balance and
  non-equilibrium dynamics in living systems: a review},}\ }\href {\doibase
  10.1088/1361-6633/aab3ed} {\bibfield  {journal} {\bibinfo  {journal} {Rep.
  Prog. Phys.}\ }\textbf {\bibinfo {volume} {81}},\ \bibinfo {pages} {066601}
  (\bibinfo {year} {2018})}\BibitemShut {NoStop}%
\bibitem [{\citenamefont {Lele}\ \emph {et~al.}(2013)\citenamefont {Lele},
  \citenamefont {Hosu},\ and\ \citenamefont {Berg}}]{Lele2013}%
  \BibitemOpen
  \bibfield  {author} {\bibinfo {author} {\bibfnamefont {Pushkar~P.}\
  \bibnamefont {Lele}}, \bibinfo {author} {\bibfnamefont {Basarab~G.}\
  \bibnamefont {Hosu}}, \ and\ \bibinfo {author} {\bibfnamefont {Howard~C.}\
  \bibnamefont {Berg}},\ }\bibfield  {title} {\enquote {\bibinfo {title}
  {Dynamics of mechanosensing in the bacterial flagellar motor},}\ }\href
  {\doibase 10.1073/pnas.1305885110} {\bibfield  {journal} {\bibinfo  {journal}
  {Proc. Natl. Acad. Sci. USA}\ }\textbf {\bibinfo {volume} {110}},\ \bibinfo
  {pages} {11839--11844} (\bibinfo {year} {2013})}\BibitemShut {NoStop}%
\bibitem [{\citenamefont {Lan}\ \emph {et~al.}(2012)\citenamefont {Lan},
  \citenamefont {Sartori}, \citenamefont {Neumann}, \citenamefont {Sourjik},\
  and\ \citenamefont {Tu}}]{Lan2012}%
  \BibitemOpen
  \bibfield  {author} {\bibinfo {author} {\bibfnamefont {Ganhui}\ \bibnamefont
  {Lan}}, \bibinfo {author} {\bibfnamefont {Pablo}\ \bibnamefont {Sartori}},
  \bibinfo {author} {\bibfnamefont {Silke}\ \bibnamefont {Neumann}}, \bibinfo
  {author} {\bibfnamefont {Victor}\ \bibnamefont {Sourjik}}, \ and\ \bibinfo
  {author} {\bibfnamefont {Yuhai}\ \bibnamefont {Tu}},\ }\bibfield  {title}
  {\enquote {\bibinfo {title} {{The energy--speed--accuracy trade-off in
  sensory adaptation}},}\ }\href {\doibase 10.1038/nphys2276} {\bibfield
  {journal} {\bibinfo  {journal} {Nat. Phys.}\ }\textbf {\bibinfo {volume}
  {8}},\ \bibinfo {pages} {422--428} (\bibinfo {year} {2012})}\BibitemShut
  {NoStop}%
\bibitem [{\citenamefont {Wang}\ \emph {et~al.}(2017)\citenamefont {Wang},
  \citenamefont {Shi}, \citenamefont {He}, \citenamefont {Wang}, \citenamefont
  {Zhang},\ and\ \citenamefont {Yuan}}]{Wang2017}%
  \BibitemOpen
  \bibfield  {author} {\bibinfo {author} {\bibfnamefont {Fangbin}\ \bibnamefont
  {Wang}}, \bibinfo {author} {\bibfnamefont {Hui}\ \bibnamefont {Shi}},
  \bibinfo {author} {\bibfnamefont {Rui}\ \bibnamefont {He}}, \bibinfo {author}
  {\bibfnamefont {Renjie}\ \bibnamefont {Wang}}, \bibinfo {author}
  {\bibfnamefont {Rongjing}\ \bibnamefont {Zhang}}, \ and\ \bibinfo {author}
  {\bibfnamefont {Junhua}\ \bibnamefont {Yuan}},\ }\bibfield  {title} {\enquote
  {\bibinfo {title} {Non-equilibrium effect in the allosteric regulation of the
  bacterial flagellar switch},}\ }\href {\doibase 10.1038/nphys4081} {\bibfield
   {journal} {\bibinfo  {journal} {Nat. Phys.}\ }\textbf {\bibinfo {volume}
  {13}},\ \bibinfo {pages} {710} (\bibinfo {year} {2017})}\BibitemShut
  {NoStop}%
\bibitem [{\citenamefont {Soares~e Silva}\ \emph {et~al.}(2011)\citenamefont
  {Soares~e Silva}, \citenamefont {Depken}, \citenamefont {Stuhrmann},
  \citenamefont {Korsten}, \citenamefont {MacKintosh},\ and\ \citenamefont
  {Koenderink}}]{Silva2011}%
  \BibitemOpen
  \bibfield  {author} {\bibinfo {author} {\bibfnamefont {Marina}\ \bibnamefont
  {Soares~e Silva}}, \bibinfo {author} {\bibfnamefont {Martin}\ \bibnamefont
  {Depken}}, \bibinfo {author} {\bibfnamefont {Bj{\"o}rn}\ \bibnamefont
  {Stuhrmann}}, \bibinfo {author} {\bibfnamefont {Marijn}\ \bibnamefont
  {Korsten}}, \bibinfo {author} {\bibfnamefont {Fred~C.}\ \bibnamefont
  {MacKintosh}}, \ and\ \bibinfo {author} {\bibfnamefont {Gijsje~H.}\
  \bibnamefont {Koenderink}},\ }\bibfield  {title} {\enquote {\bibinfo {title}
  {Active multistage coarsening of actin networks driven by myosin motors},}\
  }\href {\doibase 10.1073/pnas.1016616108} {\bibfield  {journal} {\bibinfo
  {journal} {Proc. Natl. Acad. Sci. USA}\ }\textbf {\bibinfo {volume} {108}},\
  \bibinfo {pages} {9408--9413} (\bibinfo {year} {2011})}\BibitemShut {NoStop}%
\bibitem [{\citenamefont {Sanchez}\ \emph {et~al.}(2012)\citenamefont
  {Sanchez}, \citenamefont {Chen}, \citenamefont {DeCamp}, \citenamefont
  {Heymann},\ and\ \citenamefont {Dogic}}]{Sanchez2012}%
  \BibitemOpen
  \bibfield  {author} {\bibinfo {author} {\bibfnamefont {Tim}\ \bibnamefont
  {Sanchez}}, \bibinfo {author} {\bibfnamefont {Daniel~TN}\ \bibnamefont
  {Chen}}, \bibinfo {author} {\bibfnamefont {Stephen~J}\ \bibnamefont
  {DeCamp}}, \bibinfo {author} {\bibfnamefont {Michael}\ \bibnamefont
  {Heymann}}, \ and\ \bibinfo {author} {\bibfnamefont {Zvonimir}\ \bibnamefont
  {Dogic}},\ }\bibfield  {title} {\enquote {\bibinfo {title} {Spontaneous
  motion in hierarchically assembled active matter},}\ }\href {\doibase
  10.1038/nature11591} {\bibfield  {journal} {\bibinfo  {journal} {Nature}\
  }\textbf {\bibinfo {volume} {491}},\ \bibinfo {pages} {431} (\bibinfo {year}
  {2012})}\BibitemShut {NoStop}%
\bibitem [{\citenamefont {Blanchoin}\ \emph {et~al.}(2014)\citenamefont
  {Blanchoin}, \citenamefont {Boujemaa-Paterski}, \citenamefont {Sykes},\ and\
  \citenamefont {Plastino}}]{Blanchoin2014}%
  \BibitemOpen
  \bibfield  {author} {\bibinfo {author} {\bibfnamefont {Laurent}\ \bibnamefont
  {Blanchoin}}, \bibinfo {author} {\bibfnamefont {Rajaa}\ \bibnamefont
  {Boujemaa-Paterski}}, \bibinfo {author} {\bibfnamefont {C{\'e}cile}\
  \bibnamefont {Sykes}}, \ and\ \bibinfo {author} {\bibfnamefont {Julie}\
  \bibnamefont {Plastino}},\ }\bibfield  {title} {\enquote {\bibinfo {title}
  {Actin dynamics, architecture, and mechanics in cell motility},}\ }\href
  {\doibase 10.1152/physrev.00018.2013} {\bibfield  {journal} {\bibinfo
  {journal} {Physiological reviews}\ }\textbf {\bibinfo {volume} {94}},\
  \bibinfo {pages} {235--263} (\bibinfo {year} {2014})}\BibitemShut {NoStop}%
\bibitem [{\citenamefont {Murrell}\ \emph {et~al.}(2015)\citenamefont
  {Murrell}, \citenamefont {Oakes}, \citenamefont {Lenz},\ and\ \citenamefont
  {Gardel}}]{Murrell2015}%
  \BibitemOpen
  \bibfield  {author} {\bibinfo {author} {\bibfnamefont {Michael}\ \bibnamefont
  {Murrell}}, \bibinfo {author} {\bibfnamefont {Patrick~W}\ \bibnamefont
  {Oakes}}, \bibinfo {author} {\bibfnamefont {Martin}\ \bibnamefont {Lenz}}, \
  and\ \bibinfo {author} {\bibfnamefont {Margaret~L}\ \bibnamefont {Gardel}},\
  }\bibfield  {title} {\enquote {\bibinfo {title} {Forcing cells into shape:
  the mechanics of actomyosin contractility},}\ }\href {\doibase
  10.1038/nrm4012} {\bibfield  {journal} {\bibinfo  {journal} {Nature Reviews
  Molecular Cell Biology}\ }\textbf {\bibinfo {volume} {16}},\ \bibinfo {pages}
  {486--498} (\bibinfo {year} {2015})}\BibitemShut {NoStop}%
\bibitem [{\citenamefont {DeCamp}\ \emph {et~al.}(2015)\citenamefont {DeCamp},
  \citenamefont {Redner}, \citenamefont {Baskaran}, \citenamefont {Hagan},\
  and\ \citenamefont {Dogic}}]{Decamp2015}%
  \BibitemOpen
  \bibfield  {author} {\bibinfo {author} {\bibfnamefont {Stephen~J}\
  \bibnamefont {DeCamp}}, \bibinfo {author} {\bibfnamefont {Gabriel~S}\
  \bibnamefont {Redner}}, \bibinfo {author} {\bibfnamefont {Aparna}\
  \bibnamefont {Baskaran}}, \bibinfo {author} {\bibfnamefont {Michael~F}\
  \bibnamefont {Hagan}}, \ and\ \bibinfo {author} {\bibfnamefont {Zvonimir}\
  \bibnamefont {Dogic}},\ }\bibfield  {title} {\enquote {\bibinfo {title}
  {Orientational order of motile defects in active nematics},}\ }\href
  {\doibase 10.1038/nmat4387} {\bibfield  {journal} {\bibinfo  {journal} {Nat.
  Mat.}\ }\textbf {\bibinfo {volume} {14}},\ \bibinfo {pages} {1110} (\bibinfo
  {year} {2015})}\BibitemShut {NoStop}%
\bibitem [{\citenamefont {Cates}\ and\ \citenamefont
  {Tailleur}(2015)}]{Cates2015}%
  \BibitemOpen
  \bibfield  {author} {\bibinfo {author} {\bibfnamefont {Michael~E.}\
  \bibnamefont {Cates}}\ and\ \bibinfo {author} {\bibfnamefont {Julien}\
  \bibnamefont {Tailleur}},\ }\bibfield  {title} {\enquote {\bibinfo {title}
  {{Motility-Induced Phase Separation}},}\ }\href {\doibase
  10.1146/annurev-conmatphys-031214-014710} {\bibfield  {journal} {\bibinfo
  {journal} {Annu. Rev. CMP}\ }\textbf {\bibinfo {volume} {6}},\ \bibinfo
  {pages} {219--244} (\bibinfo {year} {2015})}\BibitemShut {NoStop}%
\bibitem [{\citenamefont {Solon}\ \emph {et~al.}(2015)\citenamefont {Solon},
  \citenamefont {Fily}, \citenamefont {Baskaran}, \citenamefont {Cates},
  \citenamefont {Kafri}, \citenamefont {Kardar},\ and\ \citenamefont
  {Tailleur}}]{Solon2015a}%
  \BibitemOpen
  \bibfield  {author} {\bibinfo {author} {\bibfnamefont {A.~P.}\ \bibnamefont
  {Solon}}, \bibinfo {author} {\bibfnamefont {Y.}~\bibnamefont {Fily}},
  \bibinfo {author} {\bibfnamefont {A.}~\bibnamefont {Baskaran}}, \bibinfo
  {author} {\bibfnamefont {M.~E.}\ \bibnamefont {Cates}}, \bibinfo {author}
  {\bibfnamefont {Y.}~\bibnamefont {Kafri}}, \bibinfo {author} {\bibfnamefont
  {M.}~\bibnamefont {Kardar}}, \ and\ \bibinfo {author} {\bibfnamefont
  {J.}~\bibnamefont {Tailleur}},\ }\bibfield  {title} {\enquote {\bibinfo
  {title} {{Pressure is not a state function for generic active fluids}},}\
  }\href {\doibase 10.1038/nphys3377} {\bibfield  {journal} {\bibinfo
  {journal} {Nat. Phys.}\ }\textbf {\bibinfo {volume} {11}},\ \bibinfo {pages}
  {673--678} (\bibinfo {year} {2015})}\BibitemShut {NoStop}%
\bibitem [{\citenamefont {Nguyen}\ and\ \citenamefont
  {Vaikuntanathan}(2016)}]{Nguyen2016}%
  \BibitemOpen
  \bibfield  {author} {\bibinfo {author} {\bibfnamefont {Michael}\ \bibnamefont
  {Nguyen}}\ and\ \bibinfo {author} {\bibfnamefont {Suriyanarayanan}\
  \bibnamefont {Vaikuntanathan}},\ }\bibfield  {title} {\enquote {\bibinfo
  {title} {{Design principles for nonequilibrium self-assembly.}}}\ }\href
  {\doibase 10.1073/pnas.1609983113} {\bibfield  {journal} {\bibinfo  {journal}
  {Proc. Natl. Acad. Sci. USA}\ }\textbf {\bibinfo {volume} {113}},\ \bibinfo
  {pages} {14231--14236} (\bibinfo {year} {2016})}\BibitemShut {NoStop}%
\bibitem [{\citenamefont {Fodor}\ \emph
  {et~al.}(2016{\natexlab{b}})\citenamefont {Fodor}, \citenamefont {Nardini},
  \citenamefont {Cates}, \citenamefont {Tailleur}, \citenamefont {Visco},\ and\
  \citenamefont {van Wijland}}]{Fodor2016}%
  \BibitemOpen
  \bibfield  {author} {\bibinfo {author} {\bibfnamefont {\'Etienne}\
  \bibnamefont {Fodor}}, \bibinfo {author} {\bibfnamefont {Cesare}\
  \bibnamefont {Nardini}}, \bibinfo {author} {\bibfnamefont {Michael~E.}\
  \bibnamefont {Cates}}, \bibinfo {author} {\bibfnamefont {Julien}\
  \bibnamefont {Tailleur}}, \bibinfo {author} {\bibfnamefont {Paolo}\
  \bibnamefont {Visco}}, \ and\ \bibinfo {author} {\bibfnamefont
  {Fr\'ed\'eric}\ \bibnamefont {van Wijland}},\ }\bibfield  {title} {\enquote
  {\bibinfo {title} {How far from equilibrium is active matter?}}\ }\href
  {\doibase 10.1103/PhysRevLett.117.038103} {\bibfield  {journal} {\bibinfo
  {journal} {Phys. Rev. Lett.}\ }\textbf {\bibinfo {volume} {117}},\ \bibinfo
  {pages} {038103} (\bibinfo {year} {2016}{\natexlab{b}})}\BibitemShut
  {NoStop}%
\bibitem [{\citenamefont {Murugan}\ and\ \citenamefont
  {Vaikuntanathan}(2017)}]{Murugan2017}%
  \BibitemOpen
  \bibfield  {author} {\bibinfo {author} {\bibfnamefont {Arvind}\ \bibnamefont
  {Murugan}}\ and\ \bibinfo {author} {\bibfnamefont {Suriyanarayanan}\
  \bibnamefont {Vaikuntanathan}},\ }\bibfield  {title} {\enquote {\bibinfo
  {title} {{Topologically protected modes in non-equilibrium stochastic
  systems}},}\ }\href {\doibase 10.1038/ncomms13881} {\bibfield  {journal}
  {\bibinfo  {journal} {Nat. Commun.}\ }\textbf {\bibinfo {volume} {8}},\
  \bibinfo {pages} {13881} (\bibinfo {year} {2017})}\BibitemShut {NoStop}%
\bibitem [{\citenamefont {Nardini}\ \emph {et~al.}(2017)\citenamefont
  {Nardini}, \citenamefont {Fodor}, \citenamefont {Tjhung}, \citenamefont {van
  Wijland}, \citenamefont {Tailleur},\ and\ \citenamefont
  {Cates}}]{Nardini2017}%
  \BibitemOpen
  \bibfield  {author} {\bibinfo {author} {\bibfnamefont {Cesare}\ \bibnamefont
  {Nardini}}, \bibinfo {author} {\bibfnamefont {{\'E}tienne}\ \bibnamefont
  {Fodor}}, \bibinfo {author} {\bibfnamefont {Elsen}\ \bibnamefont {Tjhung}},
  \bibinfo {author} {\bibfnamefont {Fr\'ed\'eric}\ \bibnamefont {van Wijland}},
  \bibinfo {author} {\bibfnamefont {Julien}\ \bibnamefont {Tailleur}}, \ and\
  \bibinfo {author} {\bibfnamefont {Michael~E.}\ \bibnamefont {Cates}},\
  }\bibfield  {title} {\enquote {\bibinfo {title} {Entropy production in field
  theories without time-reversal symmetry: Quantifying the non-equilibrium
  character of active matter},}\ }\href {\doibase 10.1103/PhysRevX.7.021007}
  {\bibfield  {journal} {\bibinfo  {journal} {Phys. Rev. X}\ }\textbf {\bibinfo
  {volume} {7}},\ \bibinfo {pages} {021007} (\bibinfo {year}
  {2017})}\BibitemShut {NoStop}%
\bibitem [{\citenamefont {{Nguyen}}\ and\ \citenamefont
  {{Vaikuntanathan}}(2018)}]{Nguyen2018}%
  \BibitemOpen
  \bibfield  {author} {\bibinfo {author} {\bibfnamefont {M.}~\bibnamefont
  {{Nguyen}}}\ and\ \bibinfo {author} {\bibfnamefont {S.}~\bibnamefont
  {{Vaikuntanathan}}},\ }\bibfield  {title} {\enquote {\bibinfo {title}
  {{Dissipation induced transitions in elastic strings}},}\ }\href@noop {}
  {\bibfield  {journal} {\bibinfo  {journal} {ArXiv e-prints}\ } (\bibinfo
  {year} {2018})},\ \Eprint {http://arxiv.org/abs/1803.04368}
  {arXiv:1803.04368} \BibitemShut {NoStop}%
\bibitem [{\citenamefont {Marchetti}\ \emph {et~al.}(2013)\citenamefont
  {Marchetti}, \citenamefont {Joanny}, \citenamefont {Ramaswamy}, \citenamefont
  {Liverpool}, \citenamefont {Prost}, \citenamefont {Rao},\ and\ \citenamefont
  {Simha}}]{Marchetti2013}%
  \BibitemOpen
  \bibfield  {author} {\bibinfo {author} {\bibfnamefont {M.~C.}\ \bibnamefont
  {Marchetti}}, \bibinfo {author} {\bibfnamefont {J.~F.}\ \bibnamefont
  {Joanny}}, \bibinfo {author} {\bibfnamefont {S.}~\bibnamefont {Ramaswamy}},
  \bibinfo {author} {\bibfnamefont {T.~B.}\ \bibnamefont {Liverpool}}, \bibinfo
  {author} {\bibfnamefont {J.}~\bibnamefont {Prost}}, \bibinfo {author}
  {\bibfnamefont {Madan}\ \bibnamefont {Rao}}, \ and\ \bibinfo {author}
  {\bibfnamefont {R.~Aditi}\ \bibnamefont {Simha}},\ }\bibfield  {title}
  {\enquote {\bibinfo {title} {Hydrodynamics of soft active matter},}\ }\href
  {\doibase 10.1103/RevModPhys.85.1143} {\bibfield  {journal} {\bibinfo
  {journal} {Rev. Mod. Phys.}\ }\textbf {\bibinfo {volume} {85}},\ \bibinfo
  {pages} {1143--1189} (\bibinfo {year} {2013})}\BibitemShut {NoStop}%
\bibitem [{\citenamefont {Han}\ \emph {et~al.}(2017)\citenamefont {Han},
  \citenamefont {Yan}, \citenamefont {Granick},\ and\ \citenamefont
  {Luijten}}]{Han2016}%
  \BibitemOpen
  \bibfield  {author} {\bibinfo {author} {\bibfnamefont {Ming}\ \bibnamefont
  {Han}}, \bibinfo {author} {\bibfnamefont {Jing}\ \bibnamefont {Yan}},
  \bibinfo {author} {\bibfnamefont {Steve}\ \bibnamefont {Granick}}, \ and\
  \bibinfo {author} {\bibfnamefont {Erik}\ \bibnamefont {Luijten}},\ }\bibfield
   {title} {\enquote {\bibinfo {title} {Effective temperature concept evaluated
  in an active colloid mixture},}\ }\href {\doibase 10.1073/pnas.1706702114}
  {\bibfield  {journal} {\bibinfo  {journal} {Proc. Natl. Acad. Sci. USA}\
  }\textbf {\bibinfo {volume} {114}},\ \bibinfo {pages} {7513--7518} (\bibinfo
  {year} {2017})}\BibitemShut {NoStop}%
\bibitem [{\citenamefont {Bechinger}\ \emph {et~al.}(2016)\citenamefont
  {Bechinger}, \citenamefont {Di~Leonardo}, \citenamefont {L\"owen},
  \citenamefont {Reichhardt}, \citenamefont {Volpe},\ and\ \citenamefont
  {Volpe}}]{Bechinger2016}%
  \BibitemOpen
  \bibfield  {author} {\bibinfo {author} {\bibfnamefont {Clemens}\ \bibnamefont
  {Bechinger}}, \bibinfo {author} {\bibfnamefont {Roberto}\ \bibnamefont
  {Di~Leonardo}}, \bibinfo {author} {\bibfnamefont {Hartmut}\ \bibnamefont
  {L\"owen}}, \bibinfo {author} {\bibfnamefont {Charles}\ \bibnamefont
  {Reichhardt}}, \bibinfo {author} {\bibfnamefont {Giorgio}\ \bibnamefont
  {Volpe}}, \ and\ \bibinfo {author} {\bibfnamefont {Giovanni}\ \bibnamefont
  {Volpe}},\ }\bibfield  {title} {\enquote {\bibinfo {title} {Active particles
  in complex and crowded environments},}\ }\href {\doibase
  10.1103/RevModPhys.88.045006} {\bibfield  {journal} {\bibinfo  {journal}
  {Rev. Mod. Phys.}\ }\textbf {\bibinfo {volume} {88}},\ \bibinfo {pages}
  {045006} (\bibinfo {year} {2016})}\BibitemShut {NoStop}%
\bibitem [{\citenamefont {del Junco}\ \emph {et~al.}(2018)\citenamefont {del
  Junco}, \citenamefont {Tociu},\ and\ \citenamefont
  {Vaikuntanathan}}]{delJunco2018}%
  \BibitemOpen
  \bibfield  {author} {\bibinfo {author} {\bibfnamefont {Clara}\ \bibnamefont
  {del Junco}}, \bibinfo {author} {\bibfnamefont {Laura}\ \bibnamefont
  {Tociu}}, \ and\ \bibinfo {author} {\bibfnamefont {Suriyanarayanan}\
  \bibnamefont {Vaikuntanathan}},\ }\bibfield  {title} {\enquote {\bibinfo
  {title} {Energy dissipation and fluctuations in a driven liquid},}\ }\href
  {\doibase 10.1073/pnas.1713573115} {\bibfield  {journal} {\bibinfo  {journal}
  {Proc. Natl. Acad. Sci. USA}\ }\textbf {\bibinfo {volume} {115}},\ \bibinfo
  {pages} {3569--3574} (\bibinfo {year} {2018})}\BibitemShut {NoStop}%
\bibitem [{\citenamefont {Fodor}\ and\ \citenamefont
  {Marchetti}(2018)}]{Marchetti2018}%
  \BibitemOpen
  \bibfield  {author} {\bibinfo {author} {\bibfnamefont {{\'E}tienne}\
  \bibnamefont {Fodor}}\ and\ \bibinfo {author} {\bibfnamefont {M.~C}\
  \bibnamefont {Marchetti}},\ }\bibfield  {title} {\enquote {\bibinfo {title}
  {The statistical physics of active matter: From self-catalytic colloids to
  living cells},}\ }\href {\doibase 10.1016/j.physa.2017.12.137} {\bibfield
  {journal} {\bibinfo  {journal} {Physica A}\ }\textbf {\bibinfo {volume}
  {504}},\ \bibinfo {pages} {106--120} (\bibinfo {year} {2018})}\BibitemShut
  {NoStop}%
\bibitem [{\citenamefont {Vicsek}\ \emph {et~al.}(1995)\citenamefont {Vicsek},
  \citenamefont {Czir\'ok}, \citenamefont {Ben-Jacob}, \citenamefont {Cohen},\
  and\ \citenamefont {Shochet}}]{Vicsek1995}%
  \BibitemOpen
  \bibfield  {author} {\bibinfo {author} {\bibfnamefont {Tam\'as}\ \bibnamefont
  {Vicsek}}, \bibinfo {author} {\bibfnamefont {Andr\'as}\ \bibnamefont
  {Czir\'ok}}, \bibinfo {author} {\bibfnamefont {Eshel}\ \bibnamefont
  {Ben-Jacob}}, \bibinfo {author} {\bibfnamefont {Inon}\ \bibnamefont {Cohen}},
  \ and\ \bibinfo {author} {\bibfnamefont {Ofer}\ \bibnamefont {Shochet}},\
  }\bibfield  {title} {\enquote {\bibinfo {title} {Novel type of phase
  transition in a system of self-driven particles},}\ }\href {\doibase
  10.1103/PhysRevLett.75.1226} {\bibfield  {journal} {\bibinfo  {journal}
  {Phys. Rev. Lett.}\ }\textbf {\bibinfo {volume} {75}},\ \bibinfo {pages}
  {1226--1229} (\bibinfo {year} {1995})}\BibitemShut {NoStop}%
\bibitem [{\citenamefont {Tailleur}\ and\ \citenamefont
  {Cates}(2008)}]{Tailleur2008}%
  \BibitemOpen
  \bibfield  {author} {\bibinfo {author} {\bibfnamefont {J.}~\bibnamefont
  {Tailleur}}\ and\ \bibinfo {author} {\bibfnamefont {M.~E.}\ \bibnamefont
  {Cates}},\ }\bibfield  {title} {\enquote {\bibinfo {title} {Statistical
  mechanics of interacting run-and-tumble bacteria},}\ }\href {\doibase
  10.1103/PhysRevLett.100.218103} {\bibfield  {journal} {\bibinfo  {journal}
  {Phys. Rev. Lett.}\ }\textbf {\bibinfo {volume} {100}},\ \bibinfo {pages}
  {218103} (\bibinfo {year} {2008})}\BibitemShut {NoStop}%
\bibitem [{\citenamefont {van Zuiden}\ \emph {et~al.}(2016)\citenamefont {van
  Zuiden}, \citenamefont {Paulose}, \citenamefont {Irvine}, \citenamefont
  {Bartolo},\ and\ \citenamefont {Vitelli}}]{VanZuiden2016}%
  \BibitemOpen
  \bibfield  {author} {\bibinfo {author} {\bibfnamefont {Benjamin~C}\
  \bibnamefont {van Zuiden}}, \bibinfo {author} {\bibfnamefont {Jayson}\
  \bibnamefont {Paulose}}, \bibinfo {author} {\bibfnamefont {William T~M}\
  \bibnamefont {Irvine}}, \bibinfo {author} {\bibfnamefont {Denis}\
  \bibnamefont {Bartolo}}, \ and\ \bibinfo {author} {\bibfnamefont {Vincenzo}\
  \bibnamefont {Vitelli}},\ }\bibfield  {title} {\enquote {\bibinfo {title}
  {{Spatiotemporal order and emergent edge currents in active spinner
  materials.}}}\ }\href {\doibase 10.1073/pnas.1609572113} {\bibfield
  {journal} {\bibinfo  {journal} {Proc. Natl. Acad. Sci. USA}\ }\textbf
  {\bibinfo {volume} {113}},\ \bibinfo {pages} {12919--12924} (\bibinfo {year}
  {2016})}\BibitemShut {NoStop}%
\bibitem [{\citenamefont {Takatori}\ and\ \citenamefont
  {Brady}(2015)}]{Takatori2015}%
  \BibitemOpen
  \bibfield  {author} {\bibinfo {author} {\bibfnamefont {S.~C.}\ \bibnamefont
  {Takatori}}\ and\ \bibinfo {author} {\bibfnamefont {J.~F.}\ \bibnamefont
  {Brady}},\ }\bibfield  {title} {\enquote {\bibinfo {title} {Towards a
  thermodynamics of active matter},}\ }\href {\doibase
  10.1103/PhysRevE.91.032117} {\bibfield  {journal} {\bibinfo  {journal} {Phys.
  Rev. E}\ }\textbf {\bibinfo {volume} {91}},\ \bibinfo {pages} {032117}
  (\bibinfo {year} {2015})}\BibitemShut {NoStop}%
\bibitem [{\citenamefont {Solon}\ \emph
  {et~al.}(2018{\natexlab{a}})\citenamefont {Solon}, \citenamefont
  {Stenhammar}, \citenamefont {Cates}, \citenamefont {Kafri},\ and\
  \citenamefont {Tailleur}}]{Solon2018}%
  \BibitemOpen
  \bibfield  {author} {\bibinfo {author} {\bibfnamefont {Alexandre~P}\
  \bibnamefont {Solon}}, \bibinfo {author} {\bibfnamefont {Joakim}\
  \bibnamefont {Stenhammar}}, \bibinfo {author} {\bibfnamefont {Michael~E}\
  \bibnamefont {Cates}}, \bibinfo {author} {\bibfnamefont {Yariv}\ \bibnamefont
  {Kafri}}, \ and\ \bibinfo {author} {\bibfnamefont {Julien}\ \bibnamefont
  {Tailleur}},\ }\bibfield  {title} {\enquote {\bibinfo {title} {Generalized
  thermodynamics of motility-induced phase separation: phase equilibria,
  laplace pressure, and change of ensembles},}\ }\href {\doibase
  10.1088/1367-2630/aaccdd} {\bibfield  {journal} {\bibinfo  {journal} {New J.
  Phys.}\ }\textbf {\bibinfo {volume} {20}},\ \bibinfo {pages} {075001}
  (\bibinfo {year} {2018}{\natexlab{a}})}\BibitemShut {NoStop}%
\bibitem [{\citenamefont {Solon}\ \emph
  {et~al.}(2018{\natexlab{b}})\citenamefont {Solon}, \citenamefont
  {Stenhammar}, \citenamefont {Cates}, \citenamefont {Kafri},\ and\
  \citenamefont {Tailleur}}]{Solon2018b}%
  \BibitemOpen
  \bibfield  {author} {\bibinfo {author} {\bibfnamefont {Alexandre~P.}\
  \bibnamefont {Solon}}, \bibinfo {author} {\bibfnamefont {Joakim}\
  \bibnamefont {Stenhammar}}, \bibinfo {author} {\bibfnamefont {Michael~E.}\
  \bibnamefont {Cates}}, \bibinfo {author} {\bibfnamefont {Yariv}\ \bibnamefont
  {Kafri}}, \ and\ \bibinfo {author} {\bibfnamefont {Julien}\ \bibnamefont
  {Tailleur}},\ }\bibfield  {title} {\enquote {\bibinfo {title} {Generalized
  thermodynamics of phase equilibria in scalar active matter},}\ }\href
  {\doibase 10.1103/PhysRevE.97.020602} {\bibfield  {journal} {\bibinfo
  {journal} {Phys. Rev. E}\ }\textbf {\bibinfo {volume} {97}},\ \bibinfo
  {pages} {020602} (\bibinfo {year} {2018}{\natexlab{b}})}\BibitemShut
  {NoStop}%
\bibitem [{\citenamefont {Farage}\ \emph {et~al.}(2015)\citenamefont {Farage},
  \citenamefont {Krinninger},\ and\ \citenamefont {Brader}}]{Farage2015}%
  \BibitemOpen
  \bibfield  {author} {\bibinfo {author} {\bibfnamefont {T.~F.~F.}\
  \bibnamefont {Farage}}, \bibinfo {author} {\bibfnamefont {P.}~\bibnamefont
  {Krinninger}}, \ and\ \bibinfo {author} {\bibfnamefont {J.~M.}\ \bibnamefont
  {Brader}},\ }\bibfield  {title} {\enquote {\bibinfo {title} {Effective
  interactions in active brownian suspensions},}\ }\href {\doibase
  10.1103/PhysRevE.91.042310} {\bibfield  {journal} {\bibinfo  {journal} {Phys.
  Rev. E}\ }\textbf {\bibinfo {volume} {91}},\ \bibinfo {pages} {042310}
  (\bibinfo {year} {2015})}\BibitemShut {NoStop}%
\bibitem [{\citenamefont {Wittmann}\ \emph
  {et~al.}(2017{\natexlab{a}})\citenamefont {Wittmann}, \citenamefont
  {Marconi}, \citenamefont {Maggi},\ and\ \citenamefont {Brader}}]{Brader2017}%
  \BibitemOpen
  \bibfield  {author} {\bibinfo {author} {\bibfnamefont {Ren\'e}\ \bibnamefont
  {Wittmann}}, \bibinfo {author} {\bibfnamefont {U~Marini~Bettolo}\
  \bibnamefont {Marconi}}, \bibinfo {author} {\bibfnamefont {C}~\bibnamefont
  {Maggi}}, \ and\ \bibinfo {author} {\bibfnamefont {J~M}\ \bibnamefont
  {Brader}},\ }\bibfield  {title} {\enquote {\bibinfo {title} {Effective
  equilibrium states in the colored-noise model for active matter ii. a unified
  framework for phase equilibria, structure and mechanical properties},}\
  }\href {\doibase 10.1088/1742-5468/aa8c37} {\bibfield  {journal} {\bibinfo
  {journal} {J. Stat. Mech.}\ }\textbf {\bibinfo {volume} {2017}},\ \bibinfo
  {pages} {113208} (\bibinfo {year} {2017}{\natexlab{a}})}\BibitemShut
  {NoStop}%
\bibitem [{\citenamefont {McLennan}(1959)}]{McLennan1959}%
  \BibitemOpen
  \bibfield  {author} {\bibinfo {author} {\bibfnamefont {James~A.}\
  \bibnamefont {McLennan}},\ }\bibfield  {title} {\enquote {\bibinfo {title}
  {Statistical mechanics of the steady state},}\ }\href {\doibase
  10.1103/PhysRev.115.1405} {\bibfield  {journal} {\bibinfo  {journal} {Phys.
  Rev.}\ }\textbf {\bibinfo {volume} {115}},\ \bibinfo {pages} {1405--1409}
  (\bibinfo {year} {1959})}\BibitemShut {NoStop}%
\bibitem [{\citenamefont {Komatsu}\ and\ \citenamefont
  {Nakagawa}(2008)}]{Komatsu2008}%
  \BibitemOpen
  \bibfield  {author} {\bibinfo {author} {\bibfnamefont {Teruhisa~S.}\
  \bibnamefont {Komatsu}}\ and\ \bibinfo {author} {\bibfnamefont {Naoko}\
  \bibnamefont {Nakagawa}},\ }\bibfield  {title} {\enquote {\bibinfo {title}
  {Expression for the stationary distribution in nonequilibrium steady
  states},}\ }\href {\doibase 10.1103/PhysRevLett.100.030601} {\bibfield
  {journal} {\bibinfo  {journal} {Phys. Rev. Lett.}\ }\textbf {\bibinfo
  {volume} {100}},\ \bibinfo {pages} {030601} (\bibinfo {year}
  {2008})}\BibitemShut {NoStop}%
\bibitem [{\citenamefont {Maes}\ \emph {et~al.}(2009)\citenamefont {Maes},
  \citenamefont {Neto\v{c}n\'y},\ and\ \citenamefont
  {Shergelashvili}}]{Maes2009}%
  \BibitemOpen
  \bibfield  {author} {\bibinfo {author} {\bibfnamefont {Christian}\
  \bibnamefont {Maes}}, \bibinfo {author} {\bibfnamefont {Karel}\ \bibnamefont
  {Neto\v{c}n\'y}}, \ and\ \bibinfo {author} {\bibfnamefont {Bidzina~M.}\
  \bibnamefont {Shergelashvili}},\ }\bibfield  {title} {\enquote {\bibinfo
  {title} {Nonequilibrium relation between potential and stationary
  distribution for driven diffusion},}\ }\href {\doibase
  10.1103/PhysRevE.80.011121} {\bibfield  {journal} {\bibinfo  {journal} {Phys.
  Rev. E}\ }\textbf {\bibinfo {volume} {80}},\ \bibinfo {pages} {011121}
  (\bibinfo {year} {2009})}\BibitemShut {NoStop}%
\bibitem [{\citenamefont {Maes}\ and\ \citenamefont
  {Netočný}(2010)}]{Maes2010}%
  \BibitemOpen
  \bibfield  {author} {\bibinfo {author} {\bibfnamefont {Christian}\
  \bibnamefont {Maes}}\ and\ \bibinfo {author} {\bibfnamefont {Karel}\
  \bibnamefont {Netočný}},\ }\bibfield  {title} {\enquote {\bibinfo {title}
  {Rigorous meaning of mclennan ensembles},}\ }\href {\doibase
  10.1063/1.3274819} {\bibfield  {journal} {\bibinfo  {journal} {J. Math.
  Phys.}\ }\textbf {\bibinfo {volume} {51}},\ \bibinfo {pages} {015219}
  (\bibinfo {year} {2010})}\BibitemShut {NoStop}%
\bibitem [{\citenamefont {Touchette}(2009)}]{Touchette2009}%
  \BibitemOpen
  \bibfield  {author} {\bibinfo {author} {\bibfnamefont {Hugo}\ \bibnamefont
  {Touchette}},\ }\bibfield  {title} {\enquote {\bibinfo {title} {The large
  deviation approach to statistical mechanics},}\ }\href {\doibase
  10.1016/j.physrep.2009.05.002} {\bibfield  {journal} {\bibinfo  {journal}
  {Phys. Rep.}\ }\textbf {\bibinfo {volume} {478}},\ \bibinfo {pages} {1--69}
  (\bibinfo {year} {2009})}\BibitemShut {NoStop}%
\bibitem [{\citenamefont {Jack}\ and\ \citenamefont
  {Sollich}(2010)}]{Jack2010}%
  \BibitemOpen
  \bibfield  {author} {\bibinfo {author} {\bibfnamefont {Robert~L.}\
  \bibnamefont {Jack}}\ and\ \bibinfo {author} {\bibfnamefont {Peter}\
  \bibnamefont {Sollich}},\ }\bibfield  {title} {\enquote {\bibinfo {title}
  {Large deviations and ensembles of trajectories in stochastic models},}\
  }\href {\doibase 10.1143/PTPS.184.304} {\bibfield  {journal} {\bibinfo
  {journal} {Prog. Theor. Phys. Supp.}\ }\textbf {\bibinfo {volume} {184}},\
  \bibinfo {pages} {304--317} (\bibinfo {year} {2010})}\BibitemShut {NoStop}%
\bibitem [{\citenamefont {Garrahan}\ \emph {et~al.}(2007)\citenamefont
  {Garrahan}, \citenamefont {Jack}, \citenamefont {Lecomte}, \citenamefont
  {Pitard}, \citenamefont {van Duijvendijk},\ and\ \citenamefont {van
  Wijland}}]{garrahan2007}%
  \BibitemOpen
  \bibfield  {author} {\bibinfo {author} {\bibfnamefont {J.~P.}\ \bibnamefont
  {Garrahan}}, \bibinfo {author} {\bibfnamefont {R.~L.}\ \bibnamefont {Jack}},
  \bibinfo {author} {\bibfnamefont {V.}~\bibnamefont {Lecomte}}, \bibinfo
  {author} {\bibfnamefont {E.}~\bibnamefont {Pitard}}, \bibinfo {author}
  {\bibfnamefont {K.}~\bibnamefont {van Duijvendijk}}, \ and\ \bibinfo {author}
  {\bibfnamefont {F.}~\bibnamefont {van Wijland}},\ }\bibfield  {title}
  {\enquote {\bibinfo {title} {Dynamical first-order phase transition in
  kinetically constrained models of glasses},}\ }\href {\doibase
  10.1103/PhysRevLett.98.195702} {\bibfield  {journal} {\bibinfo  {journal}
  {Phys. Rev. Lett.}\ }\textbf {\bibinfo {volume} {98}},\ \bibinfo {pages}
  {195702} (\bibinfo {year} {2007})}\BibitemShut {NoStop}%
\bibitem [{\citenamefont {Hedges}\ \emph {et~al.}(2009)\citenamefont {Hedges},
  \citenamefont {Jack}, \citenamefont {Garrahan},\ and\ \citenamefont
  {Chandler}}]{Hedges2009}%
  \BibitemOpen
  \bibfield  {author} {\bibinfo {author} {\bibfnamefont {Lester~O.}\
  \bibnamefont {Hedges}}, \bibinfo {author} {\bibfnamefont {Robert~L.}\
  \bibnamefont {Jack}}, \bibinfo {author} {\bibfnamefont {Juan~P.}\
  \bibnamefont {Garrahan}}, \ and\ \bibinfo {author} {\bibfnamefont {David}\
  \bibnamefont {Chandler}},\ }\bibfield  {title} {\enquote {\bibinfo {title}
  {Dynamic order-disorder in atomistic models of structural glass formers},}\
  }\href {\doibase 10.1126/science.1166665} {\bibfield  {journal} {\bibinfo
  {journal} {Science}\ }\textbf {\bibinfo {volume} {323}},\ \bibinfo {pages}
  {1309--1313} (\bibinfo {year} {2009})}\BibitemShut {NoStop}%
\bibitem [{\citenamefont {Pitard}\ \emph {et~al.}(2011)\citenamefont {Pitard},
  \citenamefont {Lecomte},\ and\ \citenamefont {van Wijland}}]{Pitard2011}%
  \BibitemOpen
  \bibfield  {author} {\bibinfo {author} {\bibfnamefont {E.}~\bibnamefont
  {Pitard}}, \bibinfo {author} {\bibfnamefont {V.}~\bibnamefont {Lecomte}}, \
  and\ \bibinfo {author} {\bibfnamefont {F.}~\bibnamefont {van Wijland}},\
  }\bibfield  {title} {\enquote {\bibinfo {title} {Dynamic transition in an
  atomic glass former: A molecular-dynamics evidence},}\ }\href
  {http://stacks.iop.org/0295-5075/96/i=5/a=56002} {\bibfield  {journal}
  {\bibinfo  {journal} {EPL (Europhys. Lett.)}\ }\textbf {\bibinfo {volume}
  {96}},\ \bibinfo {pages} {56002} (\bibinfo {year} {2011})}\BibitemShut
  {NoStop}%
\bibitem [{\citenamefont {Speck}\ \emph {et~al.}(2012)\citenamefont {Speck},
  \citenamefont {Malins},\ and\ \citenamefont {Royall}}]{Speck2012}%
  \BibitemOpen
  \bibfield  {author} {\bibinfo {author} {\bibfnamefont {Thomas}\ \bibnamefont
  {Speck}}, \bibinfo {author} {\bibfnamefont {Alex}\ \bibnamefont {Malins}}, \
  and\ \bibinfo {author} {\bibfnamefont {C.~Patrick}\ \bibnamefont {Royall}},\
  }\bibfield  {title} {\enquote {\bibinfo {title} {First-order phase transition
  in a model glass former: Coupling of local structure and dynamics},}\ }\href
  {\doibase 10.1103/PhysRevLett.109.195703} {\bibfield  {journal} {\bibinfo
  {journal} {Phys. Rev. Lett.}\ }\textbf {\bibinfo {volume} {109}},\ \bibinfo
  {pages} {195703} (\bibinfo {year} {2012})}\BibitemShut {NoStop}%
\bibitem [{\citenamefont {Bodineau}\ \emph {et~al.}(2012)\citenamefont
  {Bodineau}, \citenamefont {Lecomte},\ and\ \citenamefont
  {Toninelli}}]{Bodineau2012a}%
  \BibitemOpen
  \bibfield  {author} {\bibinfo {author} {\bibfnamefont {Thierry}\ \bibnamefont
  {Bodineau}}, \bibinfo {author} {\bibfnamefont {Vivien}\ \bibnamefont
  {Lecomte}}, \ and\ \bibinfo {author} {\bibfnamefont {Cristina}\ \bibnamefont
  {Toninelli}},\ }\bibfield  {title} {\enquote {\bibinfo {title} {{Finite Size
  Scaling of the Dynamical Free-Energy in a Kinetically Constrained Model}},}\
  }\href {\doibase 10.1007/s10955-012-0458-1} {\bibfield  {journal} {\bibinfo
  {journal} {J. Stat. Phys.}\ }\textbf {\bibinfo {volume} {147}},\ \bibinfo
  {pages} {1--17} (\bibinfo {year} {2012})}\BibitemShut {NoStop}%
\bibitem [{\citenamefont {Limmer}\ and\ \citenamefont
  {Chandler}(2014)}]{Limmer2014}%
  \BibitemOpen
  \bibfield  {author} {\bibinfo {author} {\bibfnamefont {David~T.}\
  \bibnamefont {Limmer}}\ and\ \bibinfo {author} {\bibfnamefont {David}\
  \bibnamefont {Chandler}},\ }\bibfield  {title} {\enquote {\bibinfo {title}
  {Theory of amorphous ices},}\ }\href {\doibase 10.1073/pnas.1407277111}
  {\bibfield  {journal} {\bibinfo  {journal} {Proc. Natl. Acad Sci. USA}\
  }\textbf {\bibinfo {volume} {111}},\ \bibinfo {pages} {9413--9418} (\bibinfo
  {year} {2014})}\BibitemShut {NoStop}%
\bibitem [{\citenamefont {Nemoto}\ \emph {et~al.}(2017)\citenamefont {Nemoto},
  \citenamefont {Jack},\ and\ \citenamefont {Lecomte}}]{Jack2017}%
  \BibitemOpen
  \bibfield  {author} {\bibinfo {author} {\bibfnamefont {Takahiro}\
  \bibnamefont {Nemoto}}, \bibinfo {author} {\bibfnamefont {Robert~L.}\
  \bibnamefont {Jack}}, \ and\ \bibinfo {author} {\bibfnamefont {Vivien}\
  \bibnamefont {Lecomte}},\ }\bibfield  {title} {\enquote {\bibinfo {title}
  {Finite-size scaling of a first-order dynamical phase transition: Adaptive
  population dynamics and an effective model},}\ }\href {\doibase
  10.1103/PhysRevLett.118.115702} {\bibfield  {journal} {\bibinfo  {journal}
  {Phys. Rev. Lett.}\ }\textbf {\bibinfo {volume} {118}},\ \bibinfo {pages}
  {115702} (\bibinfo {year} {2017})}\BibitemShut {NoStop}%
\bibitem [{\citenamefont {Tailleur}\ and\ \citenamefont
  {Kurchan}(2007)}]{tailleur2007probing}%
  \BibitemOpen
  \bibfield  {author} {\bibinfo {author} {\bibfnamefont {Julien}\ \bibnamefont
  {Tailleur}}\ and\ \bibinfo {author} {\bibfnamefont {Jorge}\ \bibnamefont
  {Kurchan}},\ }\bibfield  {title} {\enquote {\bibinfo {title} {Probing rare
  physical trajectories with lyapunov weighted dynamics},}\ }\href {\doibase
  10.1038/nphys515} {\bibfield  {journal} {\bibinfo  {journal} {Nat. Phys.}\
  }\textbf {\bibinfo {volume} {3}},\ \bibinfo {pages} {203} (\bibinfo {year}
  {2007})}\BibitemShut {NoStop}%
\bibitem [{\citenamefont {Laffargue}\ \emph {et~al.}(2013)\citenamefont
  {Laffargue}, \citenamefont {Thu-Lam}, \citenamefont {Kurchan},\ and\
  \citenamefont {Tailleur}}]{laffargue2013}%
  \BibitemOpen
  \bibfield  {author} {\bibinfo {author} {\bibfnamefont {T.}~\bibnamefont
  {Laffargue}}, \bibinfo {author} {\bibfnamefont {K.-D.~N.}\ \bibnamefont
  {Thu-Lam}}, \bibinfo {author} {\bibfnamefont {J.}~\bibnamefont {Kurchan}}, \
  and\ \bibinfo {author} {\bibfnamefont {J.}~\bibnamefont {Tailleur}},\
  }\bibfield  {title} {\enquote {\bibinfo {title} {Large deviations of lyapunov
  exponents},}\ }\href {http://stacks.iop.org/1751-8121/46/i=25/a=254002}
  {\bibfield  {journal} {\bibinfo  {journal} {J. Phys. A}\ }\textbf {\bibinfo
  {volume} {46}},\ \bibinfo {pages} {254002} (\bibinfo {year}
  {2013})}\BibitemShut {NoStop}%
\bibitem [{\citenamefont {Cagnetta}\ \emph {et~al.}(2017)\citenamefont
  {Cagnetta}, \citenamefont {Corberi}, \citenamefont {Gonnella},\ and\
  \citenamefont {Suma}}]{Cagnetta2017}%
  \BibitemOpen
  \bibfield  {author} {\bibinfo {author} {\bibfnamefont {F.}~\bibnamefont
  {Cagnetta}}, \bibinfo {author} {\bibfnamefont {F.}~\bibnamefont {Corberi}},
  \bibinfo {author} {\bibfnamefont {G.}~\bibnamefont {Gonnella}}, \ and\
  \bibinfo {author} {\bibfnamefont {A.}~\bibnamefont {Suma}},\ }\bibfield
  {title} {\enquote {\bibinfo {title} {Large fluctuations and dynamic phase
  transition in a system of self-propelled particles},}\ }\href {\doibase
  10.1103/PhysRevLett.119.158002} {\bibfield  {journal} {\bibinfo  {journal}
  {Phys. Rev. Lett.}\ }\textbf {\bibinfo {volume} {119}},\ \bibinfo {pages}
  {158002} (\bibinfo {year} {2017})}\BibitemShut {NoStop}%
\bibitem [{\citenamefont {Nemoto}\ \emph {et~al.}(2019)\citenamefont {Nemoto},
  \citenamefont {Fodor}, \citenamefont {Cates}, \citenamefont {Jack},\ and\
  \citenamefont {Tailleur}}]{Nemoto2019}%
  \BibitemOpen
  \bibfield  {author} {\bibinfo {author} {\bibfnamefont {Takahiro}\
  \bibnamefont {Nemoto}}, \bibinfo {author} {\bibfnamefont {\'Etienne}\
  \bibnamefont {Fodor}}, \bibinfo {author} {\bibfnamefont {Michael~E.}\
  \bibnamefont {Cates}}, \bibinfo {author} {\bibfnamefont {Robert~L.}\
  \bibnamefont {Jack}}, \ and\ \bibinfo {author} {\bibfnamefont {Julien}\
  \bibnamefont {Tailleur}},\ }\bibfield  {title} {\enquote {\bibinfo {title}
  {Optimizing active work: Dynamical phase transitions, collective motion, and
  jamming},}\ }\href {\doibase 10.1103/PhysRevE.99.022605} {\bibfield
  {journal} {\bibinfo  {journal} {Phys. Rev. E}\ }\textbf {\bibinfo {volume}
  {99}},\ \bibinfo {pages} {022605} (\bibinfo {year} {2019})}\BibitemShut
  {NoStop}%
\bibitem [{\citenamefont {Chetrite}\ and\ \citenamefont
  {Touchette}(2013)}]{Chetrite2013}%
  \BibitemOpen
  \bibfield  {author} {\bibinfo {author} {\bibfnamefont {Rapha\"el}\
  \bibnamefont {Chetrite}}\ and\ \bibinfo {author} {\bibfnamefont {Hugo}\
  \bibnamefont {Touchette}},\ }\bibfield  {title} {\enquote {\bibinfo {title}
  {Nonequilibrium microcanonical and canonical ensembles and their
  equivalence},}\ }\href {\doibase 10.1103/PhysRevLett.111.120601} {\bibfield
  {journal} {\bibinfo  {journal} {Phys. Rev. Lett.}\ }\textbf {\bibinfo
  {volume} {111}},\ \bibinfo {pages} {120601} (\bibinfo {year}
  {2013})}\BibitemShut {NoStop}%
\bibitem [{\citenamefont {Dean}(1996)}]{Dean1996}%
  \BibitemOpen
  \bibfield  {author} {\bibinfo {author} {\bibfnamefont {David~S}\ \bibnamefont
  {Dean}},\ }\bibfield  {title} {\enquote {\bibinfo {title} {{Langevin equation
  for the density of a system of interacting Langevin processes}},}\ }\href
  {\doibase 10.1088/0305-4470/29/24/001} {\bibfield  {journal} {\bibinfo
  {journal} {J. Phys. A.: Math. Gen.}\ }\textbf {\bibinfo {volume} {29}},\
  \bibinfo {pages} {L613--L617} (\bibinfo {year} {1996})}\BibitemShut {NoStop}%
\bibitem [{\citenamefont {D\'emery}\ and\ \citenamefont
  {Dean}(2011)}]{Demery2011}%
  \BibitemOpen
  \bibfield  {author} {\bibinfo {author} {\bibfnamefont {Vincent}\ \bibnamefont
  {D\'emery}}\ and\ \bibinfo {author} {\bibfnamefont {David~S.}\ \bibnamefont
  {Dean}},\ }\bibfield  {title} {\enquote {\bibinfo {title} {Perturbative
  path-integral study of active- and passive-tracer diffusion in fluctuating
  fields},}\ }\href {\doibase 10.1103/PhysRevE.84.011148} {\bibfield  {journal}
  {\bibinfo  {journal} {Phys. Rev. E}\ }\textbf {\bibinfo {volume} {84}},\
  \bibinfo {pages} {011148} (\bibinfo {year} {2011})}\BibitemShut {NoStop}%
\bibitem [{\citenamefont {D\'emery}\ \emph {et~al.}(2014)\citenamefont
  {D\'emery}, \citenamefont {B\'enichou},\ and\ \citenamefont
  {Jacquin}}]{Demery2014}%
  \BibitemOpen
  \bibfield  {author} {\bibinfo {author} {\bibfnamefont {Vincent}\ \bibnamefont
  {D\'emery}}, \bibinfo {author} {\bibfnamefont {Olivier}\ \bibnamefont
  {B\'enichou}}, \ and\ \bibinfo {author} {\bibfnamefont {Hugo}\ \bibnamefont
  {Jacquin}},\ }\bibfield  {title} {\enquote {\bibinfo {title} {Generalized
  langevin equations for a driven tracer in dense soft colloids: construction
  and applications},}\ }\href {\doibase 10.1088/1367-2630/16/5/053032}
  {\bibfield  {journal} {\bibinfo  {journal} {New J. Phys.}\ }\textbf {\bibinfo
  {volume} {16}},\ \bibinfo {pages} {053032} (\bibinfo {year}
  {2014})}\BibitemShut {NoStop}%
\bibitem [{\citenamefont {Harada}\ and\ \citenamefont
  {Sasa}(2005)}]{Harada2005}%
  \BibitemOpen
  \bibfield  {author} {\bibinfo {author} {\bibfnamefont {Takahiro}\
  \bibnamefont {Harada}}\ and\ \bibinfo {author} {\bibfnamefont {Shin-ichi}\
  \bibnamefont {Sasa}},\ }\bibfield  {title} {\enquote {\bibinfo {title}
  {Equality connecting energy dissipation with a violation of the
  fluctuation-response relation},}\ }\href {\doibase
  10.1103/PhysRevLett.95.130602} {\bibfield  {journal} {\bibinfo  {journal}
  {Phys. Rev. Lett.}\ }\textbf {\bibinfo {volume} {95}},\ \bibinfo {pages}
  {130602} (\bibinfo {year} {2005})}\BibitemShut {NoStop}%
\bibitem [{\citenamefont {Mizuno}\ \emph {et~al.}(2007)\citenamefont {Mizuno},
  \citenamefont {Tardin}, \citenamefont {Schmidt},\ and\ \citenamefont
  {MacKintosh}}]{Mizuno2007}%
  \BibitemOpen
  \bibfield  {author} {\bibinfo {author} {\bibfnamefont {Daisuke}\ \bibnamefont
  {Mizuno}}, \bibinfo {author} {\bibfnamefont {Catherine}\ \bibnamefont
  {Tardin}}, \bibinfo {author} {\bibfnamefont {C.~F.}\ \bibnamefont {Schmidt}},
  \ and\ \bibinfo {author} {\bibfnamefont {F.~C.}\ \bibnamefont {MacKintosh}},\
  }\bibfield  {title} {\enquote {\bibinfo {title} {Nonequilibrium mechanics of
  active cytoskeletal networks},}\ }\href {\doibase 10.1126/science.1134404}
  {\bibfield  {journal} {\bibinfo  {journal} {Science}\ }\textbf {\bibinfo
  {volume} {315}},\ \bibinfo {pages} {370--373} (\bibinfo {year}
  {2007})}\BibitemShut {NoStop}%
\bibitem [{\citenamefont {Fodor}\ \emph {et~al.}(2015)\citenamefont {Fodor},
  \citenamefont {Guo}, \citenamefont {Gov}, \citenamefont {Visco},
  \citenamefont {Weitz},\ and\ \citenamefont {van Wijland}}]{Visco2015}%
  \BibitemOpen
  \bibfield  {author} {\bibinfo {author} {\bibfnamefont {{\'E}.}~\bibnamefont
  {Fodor}}, \bibinfo {author} {\bibfnamefont {M.}~\bibnamefont {Guo}}, \bibinfo
  {author} {\bibfnamefont {N.~S.}\ \bibnamefont {Gov}}, \bibinfo {author}
  {\bibfnamefont {P.}~\bibnamefont {Visco}}, \bibinfo {author} {\bibfnamefont
  {D.~A.}\ \bibnamefont {Weitz}}, \ and\ \bibinfo {author} {\bibfnamefont
  {F.}~\bibnamefont {van Wijland}},\ }\bibfield  {title} {\enquote {\bibinfo
  {title} {Activity-driven fluctuations in living cells},}\ }\href {\doibase
  10.1209/0295-5075/110/48005} {\bibfield  {journal} {\bibinfo  {journal} {EPL
  (Europhys. Lett.)}\ }\textbf {\bibinfo {volume} {110}},\ \bibinfo {pages}
  {48005} (\bibinfo {year} {2015})}\BibitemShut {NoStop}%
\bibitem [{\citenamefont {Turlier}\ \emph {et~al.}(2016)\citenamefont
  {Turlier}, \citenamefont {Fedosov}, \citenamefont {Audoly}, \citenamefont
  {Auth}, \citenamefont {Gov}, \citenamefont {Sykes}, \citenamefont {Joanny},
  \citenamefont {Gompper},\ and\ \citenamefont {Betz}}]{Turlier2016}%
  \BibitemOpen
  \bibfield  {author} {\bibinfo {author} {\bibfnamefont {H.}~\bibnamefont
  {Turlier}}, \bibinfo {author} {\bibfnamefont {D.~A.}\ \bibnamefont
  {Fedosov}}, \bibinfo {author} {\bibfnamefont {B.}~\bibnamefont {Audoly}},
  \bibinfo {author} {\bibfnamefont {T.}~\bibnamefont {Auth}}, \bibinfo {author}
  {\bibfnamefont {N.~S.}\ \bibnamefont {Gov}}, \bibinfo {author} {\bibfnamefont
  {C.}~\bibnamefont {Sykes}}, \bibinfo {author} {\bibfnamefont {J.-F.}\
  \bibnamefont {Joanny}}, \bibinfo {author} {\bibfnamefont {G.}~\bibnamefont
  {Gompper}}, \ and\ \bibinfo {author} {\bibfnamefont {T.}~\bibnamefont
  {Betz}},\ }\bibfield  {title} {\enquote {\bibinfo {title} {Equilibrium
  physics breakdown reveals the active nature of red blood cell flickering},}\
  }\href {\doibase 10.1038/nphys3621} {\bibfield  {journal} {\bibinfo
  {journal} {Nat. Phys.}\ }\textbf {\bibinfo {volume} {12}},\ \bibinfo {pages}
  {513} (\bibinfo {year} {2016})}\BibitemShut {NoStop}%
\bibitem [{\citenamefont {Ahmed}\ \emph {et~al.}(2018)\citenamefont {Ahmed},
  \citenamefont {Fodor}, \citenamefont {Almonacid}, \citenamefont {Bussonnier},
  \citenamefont {Verlhac}, \citenamefont {Gov}, \citenamefont {Visco},
  \citenamefont {van Wijland},\ and\ \citenamefont {Betz}}]{Ahmed2018}%
  \BibitemOpen
  \bibfield  {author} {\bibinfo {author} {\bibfnamefont {W.~W.}\ \bibnamefont
  {Ahmed}}, \bibinfo {author} {\bibfnamefont {{\'E}.}~\bibnamefont {Fodor}},
  \bibinfo {author} {\bibfnamefont {M.}~\bibnamefont {Almonacid}}, \bibinfo
  {author} {\bibfnamefont {M.}~\bibnamefont {Bussonnier}}, \bibinfo {author}
  {\bibfnamefont {M.-H.}\ \bibnamefont {Verlhac}}, \bibinfo {author}
  {\bibfnamefont {N.~S.}\ \bibnamefont {Gov}}, \bibinfo {author} {\bibfnamefont
  {P.}~\bibnamefont {Visco}}, \bibinfo {author} {\bibfnamefont
  {F.}~\bibnamefont {van Wijland}}, \ and\ \bibinfo {author} {\bibfnamefont
  {T.}~\bibnamefont {Betz}},\ }\bibfield  {title} {\enquote {\bibinfo {title}
  {Active mechanics reveal molecular-scale force kinetics in living oocytes},}\
  }\href {\doibase 10.1016/j.bpj.2018.02.009} {\bibfield  {journal} {\bibinfo
  {journal} {Biophys. J.}\ }\textbf {\bibinfo {volume} {114}},\ \bibinfo
  {pages} {1664} (\bibinfo {year} {2018})}\BibitemShut {NoStop}%
\bibitem [{\citenamefont {Gingrich}\ \emph {et~al.}(2017)\citenamefont
  {Gingrich}, \citenamefont {Rotskoff},\ and\ \citenamefont
  {Horowitz}}]{Gingrich2017}%
  \BibitemOpen
  \bibfield  {author} {\bibinfo {author} {\bibfnamefont {Todd~R}\ \bibnamefont
  {Gingrich}}, \bibinfo {author} {\bibfnamefont {Grant~M}\ \bibnamefont
  {Rotskoff}}, \ and\ \bibinfo {author} {\bibfnamefont {Jordan~M}\ \bibnamefont
  {Horowitz}},\ }\bibfield  {title} {\enquote {\bibinfo {title} {Inferring
  dissipation from current fluctuations},}\ }\href {\doibase
  10.1088/1751-8121/aa672f} {\bibfield  {journal} {\bibinfo  {journal} {J.
  Phys. A: Math. Theor.}\ }\textbf {\bibinfo {volume} {50}},\ \bibinfo {pages}
  {184004} (\bibinfo {year} {2017})}\BibitemShut {NoStop}%
\bibitem [{\citenamefont {{Rold{\'a}n}}\ \emph {et~al.}(2018)\citenamefont
  {{Rold{\'a}n}}, \citenamefont {{Barral}}, \citenamefont {{Martin}},
  \citenamefont {{Parrondo}},\ and\ \citenamefont
  {{J{\"u}licher}}}]{Roldan2018}%
  \BibitemOpen
  \bibfield  {author} {\bibinfo {author} {\bibfnamefont {{\'E}.}~\bibnamefont
  {{Rold{\'a}n}}}, \bibinfo {author} {\bibfnamefont {J.}~\bibnamefont
  {{Barral}}}, \bibinfo {author} {\bibfnamefont {P.}~\bibnamefont {{Martin}}},
  \bibinfo {author} {\bibfnamefont {J.~M.~R.}\ \bibnamefont {{Parrondo}}}, \
  and\ \bibinfo {author} {\bibfnamefont {F.}~\bibnamefont {{J{\"u}licher}}},\
  }\bibfield  {title} {\enquote {\bibinfo {title} {{Arrow of Time in Active
  Fluctuations}},}\ }\href@noop {} {\bibfield  {journal} {\bibinfo  {journal}
  {ArXiv e-prints}\ } (\bibinfo {year} {2018})},\ \Eprint
  {http://arxiv.org/abs/1803.04743} {arXiv:1803.04743} \BibitemShut {NoStop}%
\bibitem [{\citenamefont {{Mart{\'{\i}}nez}}\ \emph {et~al.}(2018)\citenamefont
  {{Mart{\'{\i}}nez}}, \citenamefont {{Bisker}}, \citenamefont {{Horowitz}},\
  and\ \citenamefont {{Parrondo}}}]{Parrondo2018}%
  \BibitemOpen
  \bibfield  {author} {\bibinfo {author} {\bibfnamefont {I.~A.}\ \bibnamefont
  {{Mart{\'{\i}}nez}}}, \bibinfo {author} {\bibfnamefont {G.}~\bibnamefont
  {{Bisker}}}, \bibinfo {author} {\bibfnamefont {J.~M.}\ \bibnamefont
  {{Horowitz}}}, \ and\ \bibinfo {author} {\bibfnamefont {J.~M.~R.}\
  \bibnamefont {{Parrondo}}},\ }\bibfield  {title} {\enquote {\bibinfo {title}
  {{Inferring broken detailed balance in the absence of observable
  currents}},}\ }\href@noop {} {\bibfield  {journal} {\bibinfo  {journal}
  {ArXiv e-prints}\ } (\bibinfo {year} {2018})},\ \Eprint
  {http://arxiv.org/abs/1809.02084} {arXiv:1809.02084} \BibitemShut {NoStop}%
\bibitem [{\citenamefont {{Li}}\ \emph {et~al.}(2019)\citenamefont {{Li}},
  \citenamefont {{Horowitz}}, \citenamefont {{Gingrich}},\ and\ \citenamefont
  {{Fakhri}}}]{Li2018}%
  \BibitemOpen
  \bibfield  {author} {\bibinfo {author} {\bibfnamefont {J.}~\bibnamefont
  {{Li}}}, \bibinfo {author} {\bibfnamefont {J.~M.}\ \bibnamefont
  {{Horowitz}}}, \bibinfo {author} {\bibfnamefont {T.~R.}\ \bibnamefont
  {{Gingrich}}}, \ and\ \bibinfo {author} {\bibfnamefont {N.}~\bibnamefont
  {{Fakhri}}},\ }\bibfield  {title} {\enquote {\bibinfo {title} {{Quantifying
  dissipation using fluctuating currents}},}\ }\href {\doibase
  10.1038/s41467-019-09631-x} {\bibfield  {journal} {\bibinfo  {journal} {Nat.
  Commun.}\ }\textbf {\bibinfo {volume} {10}},\ \bibinfo {pages} {1666}
  (\bibinfo {year} {2019})}\BibitemShut {NoStop}%
\bibitem [{\citenamefont {Giardin\`a}\ \emph {et~al.}(2006)\citenamefont
  {Giardin\`a}, \citenamefont {Kurchan},\ and\ \citenamefont
  {Peliti}}]{Giadina2006}%
  \BibitemOpen
  \bibfield  {author} {\bibinfo {author} {\bibfnamefont {Cristian}\
  \bibnamefont {Giardin\`a}}, \bibinfo {author} {\bibfnamefont {Jorge}\
  \bibnamefont {Kurchan}}, \ and\ \bibinfo {author} {\bibfnamefont {Luca}\
  \bibnamefont {Peliti}},\ }\bibfield  {title} {\enquote {\bibinfo {title}
  {Direct evaluation of large-deviation functions},}\ }\href {\doibase
  10.1103/PhysRevLett.96.120603} {\bibfield  {journal} {\bibinfo  {journal}
  {Phys. Rev. Lett.}\ }\textbf {\bibinfo {volume} {96}},\ \bibinfo {pages}
  {120603} (\bibinfo {year} {2006})}\BibitemShut {NoStop}%
\bibitem [{\citenamefont {Hurtado}\ and\ \citenamefont
  {Garrido}(2009)}]{Hurtado2009}%
  \BibitemOpen
  \bibfield  {author} {\bibinfo {author} {\bibfnamefont {Pablo~I.}\
  \bibnamefont {Hurtado}}\ and\ \bibinfo {author} {\bibfnamefont {Pedro~L.}\
  \bibnamefont {Garrido}},\ }\bibfield  {title} {\enquote {\bibinfo {title}
  {Test of the additivity principle for current fluctuations in a model of heat
  conduction},}\ }\href {\doibase 10.1103/PhysRevLett.102.250601} {\bibfield
  {journal} {\bibinfo  {journal} {Phys. Rev. Lett.}\ }\textbf {\bibinfo
  {volume} {102}},\ \bibinfo {pages} {250601} (\bibinfo {year}
  {2009})}\BibitemShut {NoStop}%
\bibitem [{\citenamefont {Nemoto}\ \emph {et~al.}(2016)\citenamefont {Nemoto},
  \citenamefont {Bouchet}, \citenamefont {Jack},\ and\ \citenamefont
  {Lecomte}}]{Nemoto2016}%
  \BibitemOpen
  \bibfield  {author} {\bibinfo {author} {\bibfnamefont {Takahiro}\
  \bibnamefont {Nemoto}}, \bibinfo {author} {\bibfnamefont {Freddy}\
  \bibnamefont {Bouchet}}, \bibinfo {author} {\bibfnamefont {Robert~L.}\
  \bibnamefont {Jack}}, \ and\ \bibinfo {author} {\bibfnamefont {Vivien}\
  \bibnamefont {Lecomte}},\ }\bibfield  {title} {\enquote {\bibinfo {title}
  {Population-dynamics method with a multicanonical feedback control},}\ }\href
  {\doibase 10.1103/PhysRevE.93.062123} {\bibfield  {journal} {\bibinfo
  {journal} {Phys. Rev. E}\ }\textbf {\bibinfo {volume} {93}},\ \bibinfo
  {pages} {062123} (\bibinfo {year} {2016})}\BibitemShut {NoStop}%
\bibitem [{\citenamefont {Ray}\ \emph {et~al.}(2018{\natexlab{a}})\citenamefont
  {Ray}, \citenamefont {Chan},\ and\ \citenamefont {Limmer}}]{Ray2018}%
  \BibitemOpen
  \bibfield  {author} {\bibinfo {author} {\bibfnamefont {Ushnish}\ \bibnamefont
  {Ray}}, \bibinfo {author} {\bibfnamefont {Garnet Kin-Lic}\ \bibnamefont
  {Chan}}, \ and\ \bibinfo {author} {\bibfnamefont {David~T.}\ \bibnamefont
  {Limmer}},\ }\bibfield  {title} {\enquote {\bibinfo {title} {Exact
  fluctuations of nonequilibrium steady states from approximate auxiliary
  dynamics},}\ }\href {\doibase 10.1103/PhysRevLett.120.210602} {\bibfield
  {journal} {\bibinfo  {journal} {Phys. Rev. Lett.}\ }\textbf {\bibinfo
  {volume} {120}},\ \bibinfo {pages} {210602} (\bibinfo {year}
  {2018}{\natexlab{a}})}\BibitemShut {NoStop}%
\bibitem [{\citenamefont {Klymko}\ \emph {et~al.}(2018)\citenamefont {Klymko},
  \citenamefont {Geissler}, \citenamefont {Garrahan},\ and\ \citenamefont
  {Whitelam}}]{Klymko2018}%
  \BibitemOpen
  \bibfield  {author} {\bibinfo {author} {\bibfnamefont {Katherine}\
  \bibnamefont {Klymko}}, \bibinfo {author} {\bibfnamefont {Phillip~L.}\
  \bibnamefont {Geissler}}, \bibinfo {author} {\bibfnamefont {Juan~P.}\
  \bibnamefont {Garrahan}}, \ and\ \bibinfo {author} {\bibfnamefont {Stephen}\
  \bibnamefont {Whitelam}},\ }\bibfield  {title} {\enquote {\bibinfo {title}
  {Rare behavior of growth processes via umbrella sampling of trajectories},}\
  }\href {\doibase 10.1103/PhysRevE.97.032123} {\bibfield  {journal} {\bibinfo
  {journal} {Phys. Rev. E}\ }\textbf {\bibinfo {volume} {97}},\ \bibinfo
  {pages} {032123} (\bibinfo {year} {2018})}\BibitemShut {NoStop}%
\bibitem [{\citenamefont {Brewer}\ \emph
  {et~al.}(2018{\natexlab{a}})\citenamefont {Brewer}, \citenamefont {Clark},
  \citenamefont {Bradford},\ and\ \citenamefont {Jack}}]{Brewer2018}%
  \BibitemOpen
  \bibfield  {author} {\bibinfo {author} {\bibfnamefont {Tobias}\ \bibnamefont
  {Brewer}}, \bibinfo {author} {\bibfnamefont {Stephen~R}\ \bibnamefont
  {Clark}}, \bibinfo {author} {\bibfnamefont {Russell}\ \bibnamefont
  {Bradford}}, \ and\ \bibinfo {author} {\bibfnamefont {Robert~L}\ \bibnamefont
  {Jack}},\ }\bibfield  {title} {\enquote {\bibinfo {title} {Efficient
  characterisation of large deviations using population dynamics},}\ }\href
  {http://stacks.iop.org/1742-5468/2018/i=5/a=053204} {\bibfield  {journal}
  {\bibinfo  {journal} {J. Stat. Mech.}\ }\textbf {\bibinfo {volume} {2018}},\
  \bibinfo {pages} {053204} (\bibinfo {year} {2018}{\natexlab{a}})}\BibitemShut
  {NoStop}%
\bibitem [{\citenamefont {Farrell}\ \emph {et~al.}(2012)\citenamefont
  {Farrell}, \citenamefont {Marchetti}, \citenamefont {Marenduzzo},\ and\
  \citenamefont {Tailleur}}]{Farrell2012}%
  \BibitemOpen
  \bibfield  {author} {\bibinfo {author} {\bibfnamefont {F.~D.~C.}\
  \bibnamefont {Farrell}}, \bibinfo {author} {\bibfnamefont {M.~C.}\
  \bibnamefont {Marchetti}}, \bibinfo {author} {\bibfnamefont {D.}~\bibnamefont
  {Marenduzzo}}, \ and\ \bibinfo {author} {\bibfnamefont {J.}~\bibnamefont
  {Tailleur}},\ }\bibfield  {title} {\enquote {\bibinfo {title} {Pattern
  formation in self-propelled particles with density-dependent motility},}\
  }\href {\doibase 10.1103/PhysRevLett.108.248101} {\bibfield  {journal}
  {\bibinfo  {journal} {Phys. Rev. Lett.}\ }\textbf {\bibinfo {volume} {108}},\
  \bibinfo {pages} {248101} (\bibinfo {year} {2012})}\BibitemShut {NoStop}%
\bibitem [{\citenamefont {Fily}\ and\ \citenamefont
  {Marchetti}(2012)}]{Fily2012}%
  \BibitemOpen
  \bibfield  {author} {\bibinfo {author} {\bibfnamefont {Yaouen}\ \bibnamefont
  {Fily}}\ and\ \bibinfo {author} {\bibfnamefont {M.~Cristina}\ \bibnamefont
  {Marchetti}},\ }\bibfield  {title} {\enquote {\bibinfo {title} {Athermal
  phase separation of self-propelled particles with no alignment},}\ }\href
  {\doibase 10.1103/PhysRevLett.108.235702} {\bibfield  {journal} {\bibinfo
  {journal} {Phys. Rev. Lett.}\ }\textbf {\bibinfo {volume} {108}},\ \bibinfo
  {pages} {235702} (\bibinfo {year} {2012})}\BibitemShut {NoStop}%
\bibitem [{\citenamefont {Redner}\ \emph {et~al.}(2013)\citenamefont {Redner},
  \citenamefont {Hagan},\ and\ \citenamefont {Baskaran}}]{Redner2013}%
  \BibitemOpen
  \bibfield  {author} {\bibinfo {author} {\bibfnamefont {Gabriel~S.}\
  \bibnamefont {Redner}}, \bibinfo {author} {\bibfnamefont {Michael~F.}\
  \bibnamefont {Hagan}}, \ and\ \bibinfo {author} {\bibfnamefont {Aparna}\
  \bibnamefont {Baskaran}},\ }\bibfield  {title} {\enquote {\bibinfo {title}
  {Structure and dynamics of a phase-separating active colloidal fluid},}\
  }\href {\doibase 10.1103/PhysRevLett.110.055701} {\bibfield  {journal}
  {\bibinfo  {journal} {Phys. Rev. Lett.}\ }\textbf {\bibinfo {volume} {110}},\
  \bibinfo {pages} {055701} (\bibinfo {year} {2013})}\BibitemShut {NoStop}%
\bibitem [{\citenamefont {{Maggi}}\ \emph {et~al.}(2015)\citenamefont
  {{Maggi}}, \citenamefont {{Marini Bettolo Marconi}}, \citenamefont {{Gnan}},\
  and\ \citenamefont {{Di Leonardo}}}]{Maggi2015}%
  \BibitemOpen
  \bibfield  {author} {\bibinfo {author} {\bibfnamefont {C.}~\bibnamefont
  {{Maggi}}}, \bibinfo {author} {\bibfnamefont {U.}~\bibnamefont {{Marini
  Bettolo Marconi}}}, \bibinfo {author} {\bibfnamefont {N.}~\bibnamefont
  {{Gnan}}}, \ and\ \bibinfo {author} {\bibfnamefont {R.}~\bibnamefont {{Di
  Leonardo}}},\ }\bibfield  {title} {\enquote {\bibinfo {title}
  {Multidimensional stationary probability distribution for interacting active
  particles},}\ }\href {\doibase 10.1038/srep10742} {\bibfield  {journal}
  {\bibinfo  {journal} {Sci. Rep.}\ }\textbf {\bibinfo {volume} {5}},\ \bibinfo
  {pages} {10742} (\bibinfo {year} {2015})}\BibitemShut {NoStop}%
\bibitem [{mov()}]{movie}%
  \BibitemOpen
  \href@noop {} {}\bibinfo {note} {See Supplemental Material at [URL will be
  inserted by publisher] for movies corresponding to
  Figs.~\ref{fig0},~\ref{fig:outofperturbation}
  and~\ref{fig:vicsek}.}\BibitemShut {Stop}%
\bibitem [{\citenamefont {Chandler}(1993)}]{Chandler1993}%
  \BibitemOpen
  \bibfield  {author} {\bibinfo {author} {\bibfnamefont {David}\ \bibnamefont
  {Chandler}},\ }\bibfield  {title} {\enquote {\bibinfo {title} {{Gaussian
  field model of fluids with an application to polymeric fluids}},}\ }\href
  {\doibase 10.1103/PhysRevE.48.2898} {\bibfield  {journal} {\bibinfo
  {journal} {Phys. Rev. E}\ }\textbf {\bibinfo {volume} {48}},\ \bibinfo
  {pages} {2898--2905} (\bibinfo {year} {1993})}\BibitemShut {NoStop}%
\bibitem [{\citenamefont {Kr\"uger}\ and\ \citenamefont
  {Dean}(2017)}]{Kruger2017}%
  \BibitemOpen
  \bibfield  {author} {\bibinfo {author} {\bibfnamefont {Matthias}\
  \bibnamefont {Kr\"uger}}\ and\ \bibinfo {author} {\bibfnamefont {David~S.}\
  \bibnamefont {Dean}},\ }\bibfield  {title} {\enquote {\bibinfo {title} {A
  gaussian theory for fluctuations in simple liquids},}\ }\href {\doibase
  10.1063/1.4979659} {\bibfield  {journal} {\bibinfo  {journal} {J. Chem.
  Phys.}\ }\textbf {\bibinfo {volume} {146}},\ \bibinfo {pages} {134507}
  (\bibinfo {year} {2017})}\BibitemShut {NoStop}%
\bibitem [{\citenamefont {D\'emery}(2015)}]{Demery2015}%
  \BibitemOpen
  \bibfield  {author} {\bibinfo {author} {\bibfnamefont {Vincent}\ \bibnamefont
  {D\'emery}},\ }\bibfield  {title} {\enquote {\bibinfo {title} {Mean-field
  microrheology of a very soft colloidal suspension: Inertia induces shear
  thickening},}\ }\href {\doibase 10.1103/PhysRevE.91.062301} {\bibfield
  {journal} {\bibinfo  {journal} {Phys. Rev. E}\ }\textbf {\bibinfo {volume}
  {91}},\ \bibinfo {pages} {062301} (\bibinfo {year} {2015})}\BibitemShut
  {NoStop}%
\bibitem [{\citenamefont {Martin}\ \emph {et~al.}(2018)\citenamefont {Martin},
  \citenamefont {Nardini}, \citenamefont {Cates},\ and\ \citenamefont
  {Fodor}}]{Martin2018}%
  \BibitemOpen
  \bibfield  {author} {\bibinfo {author} {\bibfnamefont {D.}~\bibnamefont
  {Martin}}, \bibinfo {author} {\bibfnamefont {C.}~\bibnamefont {Nardini}},
  \bibinfo {author} {\bibfnamefont {M.~E.}\ \bibnamefont {Cates}}, \ and\
  \bibinfo {author} {\bibfnamefont {{\'E}.}~\bibnamefont {Fodor}},\ }\bibfield
  {title} {\enquote {\bibinfo {title} {Extracting maximum power from active
  colloidal heat engines},}\ }\href {\doibase 10.1209/0295-5075/121/60005}
  {\bibfield  {journal} {\bibinfo  {journal} {EPL (Europhys. Lett.)}\ }\textbf
  {\bibinfo {volume} {121}},\ \bibinfo {pages} {60005} (\bibinfo {year}
  {2018})}\BibitemShut {NoStop}%
\bibitem [{\citenamefont {D{\'{e}}mery}\ and\ \citenamefont
  {Fodor}(2019)}]{Demery2019}%
  \BibitemOpen
  \bibfield  {author} {\bibinfo {author} {\bibfnamefont {Vincent}\ \bibnamefont
  {D{\'{e}}mery}}\ and\ \bibinfo {author} {\bibfnamefont {{\'{E}}tienne}\
  \bibnamefont {Fodor}},\ }\bibfield  {title} {\enquote {\bibinfo {title}
  {Driven probe under harmonic confinement in a colloidal bath},}\ }\href
  {\doibase 10.1088/1742-5468/ab02e9} {\bibfield  {journal} {\bibinfo
  {journal} {J. Stat. Mech.}\ }\textbf {\bibinfo {volume} {2019}},\ \bibinfo
  {pages} {033202} (\bibinfo {year} {2019})}\BibitemShut {NoStop}%
\bibitem [{\citenamefont {Sekimoto}(1998)}]{Sekimoto1998}%
  \BibitemOpen
  \bibfield  {author} {\bibinfo {author} {\bibfnamefont {Ken}\ \bibnamefont
  {Sekimoto}},\ }\bibfield  {title} {\enquote {\bibinfo {title} {Langevin
  equation and thermodynamics},}\ }\href {\doibase 10.1143/PTPS.130.17}
  {\bibfield  {journal} {\bibinfo  {journal} {Prog. Theor. Phys. Supp.}\
  }\textbf {\bibinfo {volume} {130}},\ \bibinfo {pages} {17--27} (\bibinfo
  {year} {1998})}\BibitemShut {NoStop}%
\bibitem [{\citenamefont {Seifert}(2012)}]{Seifert2012}%
  \BibitemOpen
  \bibfield  {author} {\bibinfo {author} {\bibfnamefont {Udo}\ \bibnamefont
  {Seifert}},\ }\bibfield  {title} {\enquote {\bibinfo {title} {{Stochastic
  thermodynamics, fluctuation theorems and molecular machines}},}\ }\href
  {\doibase 10.1088/0034-4885/75/12/126001} {\bibfield  {journal} {\bibinfo
  {journal} {Rep. Prog. Phys.}\ }\textbf {\bibinfo {volume} {75}},\ \bibinfo
  {pages} {126001} (\bibinfo {year} {2012})}\BibitemShut {NoStop}%
\bibitem [{\citenamefont {Mandal}\ \emph {et~al.}(2017)\citenamefont {Mandal},
  \citenamefont {Klymko},\ and\ \citenamefont {DeWeese}}]{Mandal2017}%
  \BibitemOpen
  \bibfield  {author} {\bibinfo {author} {\bibfnamefont {Dibyendu}\
  \bibnamefont {Mandal}}, \bibinfo {author} {\bibfnamefont {Katherine}\
  \bibnamefont {Klymko}}, \ and\ \bibinfo {author} {\bibfnamefont {Michael~R.}\
  \bibnamefont {DeWeese}},\ }\bibfield  {title} {\enquote {\bibinfo {title}
  {Entropy production and fluctuation theorems for active matter},}\ }\href
  {\doibase 10.1103/PhysRevLett.119.258001} {\bibfield  {journal} {\bibinfo
  {journal} {Phys. Rev. Lett.}\ }\textbf {\bibinfo {volume} {119}},\ \bibinfo
  {pages} {258001} (\bibinfo {year} {2017})}\BibitemShut {NoStop}%
\bibitem [{\citenamefont {Pietzonka}\ and\ \citenamefont
  {Seifert}(2018)}]{Seifert2018}%
  \BibitemOpen
  \bibfield  {author} {\bibinfo {author} {\bibfnamefont {Patrick}\ \bibnamefont
  {Pietzonka}}\ and\ \bibinfo {author} {\bibfnamefont {Udo}\ \bibnamefont
  {Seifert}},\ }\bibfield  {title} {\enquote {\bibinfo {title} {Entropy
  production of active particles and for particles in active baths},}\ }\href
  {\doibase 10.1088/1751-8121/aa91b9} {\bibfield  {journal} {\bibinfo
  {journal} {J. Phys. A: Math. Theor.}\ }\textbf {\bibinfo {volume} {51}},\
  \bibinfo {pages} {01LT01} (\bibinfo {year} {2018})}\BibitemShut {NoStop}%
\bibitem [{\citenamefont {Shankar}\ and\ \citenamefont
  {Marchetti}(2018)}]{Shankar2018}%
  \BibitemOpen
  \bibfield  {author} {\bibinfo {author} {\bibfnamefont {Suraj}\ \bibnamefont
  {Shankar}}\ and\ \bibinfo {author} {\bibfnamefont {M.~Cristina}\ \bibnamefont
  {Marchetti}},\ }\bibfield  {title} {\enquote {\bibinfo {title} {Hidden
  entropy production and work fluctuations in an ideal active gas},}\ }\href
  {\doibase 10.1103/PhysRevE.98.020604} {\bibfield  {journal} {\bibinfo
  {journal} {Phys. Rev. E}\ }\textbf {\bibinfo {volume} {98}},\ \bibinfo
  {pages} {020604} (\bibinfo {year} {2018})}\BibitemShut {NoStop}%
\bibitem [{\citenamefont {Dabelow}\ \emph {et~al.}(2019)\citenamefont
  {Dabelow}, \citenamefont {Bo},\ and\ \citenamefont {Eichhorn}}]{Bo2019}%
  \BibitemOpen
  \bibfield  {author} {\bibinfo {author} {\bibfnamefont {Lennart}\ \bibnamefont
  {Dabelow}}, \bibinfo {author} {\bibfnamefont {Stefano}\ \bibnamefont {Bo}}, \
  and\ \bibinfo {author} {\bibfnamefont {Ralf}\ \bibnamefont {Eichhorn}},\
  }\bibfield  {title} {\enquote {\bibinfo {title} {Irreversibility in active
  matter systems: Fluctuation theorem and mutual information},}\ }\href
  {\doibase 10.1103/PhysRevX.9.021009} {\bibfield  {journal} {\bibinfo
  {journal} {Phys. Rev. X}\ }\textbf {\bibinfo {volume} {9}},\ \bibinfo {pages}
  {021009} (\bibinfo {year} {2019})}\BibitemShut {NoStop}%
\bibitem [{\citenamefont {B\'enichou}\ \emph {et~al.}(2016)\citenamefont
  {B\'enichou}, \citenamefont {Illien}, \citenamefont {Oshanin}, \citenamefont
  {Sarracino},\ and\ \citenamefont {Voituriez}}]{Voituriez2016}%
  \BibitemOpen
  \bibfield  {author} {\bibinfo {author} {\bibfnamefont {O.}~\bibnamefont
  {B\'enichou}}, \bibinfo {author} {\bibfnamefont {P.}~\bibnamefont {Illien}},
  \bibinfo {author} {\bibfnamefont {G.}~\bibnamefont {Oshanin}}, \bibinfo
  {author} {\bibfnamefont {A.}~\bibnamefont {Sarracino}}, \ and\ \bibinfo
  {author} {\bibfnamefont {R.}~\bibnamefont {Voituriez}},\ }\bibfield  {title}
  {\enquote {\bibinfo {title} {Nonlinear response and emerging nonequilibrium
  microstructures for biased diffusion in confined crowded environments},}\
  }\href {\doibase 10.1103/PhysRevE.93.032128} {\bibfield  {journal} {\bibinfo
  {journal} {Phys. Rev. E}\ }\textbf {\bibinfo {volume} {93}},\ \bibinfo
  {pages} {032128} (\bibinfo {year} {2016})}\BibitemShut {NoStop}%
\bibitem [{\citenamefont {Burkholder}\ and\ \citenamefont
  {Brady}(2017)}]{Brady2017}%
  \BibitemOpen
  \bibfield  {author} {\bibinfo {author} {\bibfnamefont {Eric~W.}\ \bibnamefont
  {Burkholder}}\ and\ \bibinfo {author} {\bibfnamefont {John~F.}\ \bibnamefont
  {Brady}},\ }\bibfield  {title} {\enquote {\bibinfo {title} {Tracer diffusion
  in active suspensions},}\ }\href {\doibase 10.1103/PhysRevE.95.052605}
  {\bibfield  {journal} {\bibinfo  {journal} {Phys. Rev. E}\ }\textbf {\bibinfo
  {volume} {95}},\ \bibinfo {pages} {052605} (\bibinfo {year}
  {2017})}\BibitemShut {NoStop}%
\bibitem [{\citenamefont {Stenhammar}\ \emph {et~al.}(2017)\citenamefont
  {Stenhammar}, \citenamefont {Nardini}, \citenamefont {Nash}, \citenamefont
  {Marenduzzo},\ and\ \citenamefont {Morozov}}]{Stenhammar2017}%
  \BibitemOpen
  \bibfield  {author} {\bibinfo {author} {\bibfnamefont {Joakim}\ \bibnamefont
  {Stenhammar}}, \bibinfo {author} {\bibfnamefont {Cesare}\ \bibnamefont
  {Nardini}}, \bibinfo {author} {\bibfnamefont {Rupert~W.}\ \bibnamefont
  {Nash}}, \bibinfo {author} {\bibfnamefont {Davide}\ \bibnamefont
  {Marenduzzo}}, \ and\ \bibinfo {author} {\bibfnamefont {Alexander}\
  \bibnamefont {Morozov}},\ }\bibfield  {title} {\enquote {\bibinfo {title}
  {Role of correlations in the collective behavior of microswimmer
  suspensions},}\ }\href {\doibase 10.1103/PhysRevLett.119.028005} {\bibfield
  {journal} {\bibinfo  {journal} {Phys. Rev. Lett.}\ }\textbf {\bibinfo
  {volume} {119}},\ \bibinfo {pages} {028005} (\bibinfo {year}
  {2017})}\BibitemShut {NoStop}%
\bibitem [{\citenamefont {Bertrand}\ \emph {et~al.}(2018)\citenamefont
  {Bertrand}, \citenamefont {Zhao}, \citenamefont {B\'enichou}, \citenamefont
  {Tailleur},\ and\ \citenamefont {Voituriez}}]{Tailleur2017}%
  \BibitemOpen
  \bibfield  {author} {\bibinfo {author} {\bibfnamefont {Thibault}\
  \bibnamefont {Bertrand}}, \bibinfo {author} {\bibfnamefont {Yongfeng}\
  \bibnamefont {Zhao}}, \bibinfo {author} {\bibfnamefont {Olivier}\
  \bibnamefont {B\'enichou}}, \bibinfo {author} {\bibfnamefont {Julien}\
  \bibnamefont {Tailleur}}, \ and\ \bibinfo {author} {\bibfnamefont
  {Rapha\"el}\ \bibnamefont {Voituriez}},\ }\bibfield  {title} {\enquote
  {\bibinfo {title} {Optimized diffusion of run-and-tumble particles in crowded
  environments},}\ }\href {\doibase 10.1103/PhysRevLett.120.198103} {\bibfield
  {journal} {\bibinfo  {journal} {Phys. Rev. Lett.}\ }\textbf {\bibinfo
  {volume} {120}},\ \bibinfo {pages} {198103} (\bibinfo {year}
  {2018})}\BibitemShut {NoStop}%
\bibitem [{\citenamefont {Illien}\ \emph {et~al.}(2018)\citenamefont {Illien},
  \citenamefont {B\'enichou}, \citenamefont {Oshanin}, \citenamefont
  {Sarracino},\ and\ \citenamefont {Voituriez}}]{Illien2018}%
  \BibitemOpen
  \bibfield  {author} {\bibinfo {author} {\bibfnamefont {Pierre}\ \bibnamefont
  {Illien}}, \bibinfo {author} {\bibfnamefont {Olivier}\ \bibnamefont
  {B\'enichou}}, \bibinfo {author} {\bibfnamefont {Gleb}\ \bibnamefont
  {Oshanin}}, \bibinfo {author} {\bibfnamefont {Alessandro}\ \bibnamefont
  {Sarracino}}, \ and\ \bibinfo {author} {\bibfnamefont {Rapha\"el}\
  \bibnamefont {Voituriez}},\ }\bibfield  {title} {\enquote {\bibinfo {title}
  {Nonequilibrium fluctuations and enhanced diffusion of a driven particle in a
  dense environment},}\ }\href {\doibase 10.1103/PhysRevLett.120.200606}
  {\bibfield  {journal} {\bibinfo  {journal} {Phys. Rev. Lett.}\ }\textbf
  {\bibinfo {volume} {120}},\ \bibinfo {pages} {200606} (\bibinfo {year}
  {2018})}\BibitemShut {NoStop}%
\bibitem [{\citenamefont {Hansen}\ and\ \citenamefont
  {McDonald}(2013)}]{Hansen2013}%
  \BibitemOpen
  \bibfield  {author} {\bibinfo {author} {\bibfnamefont {Jean-Pierre}\
  \bibnamefont {Hansen}}\ and\ \bibinfo {author} {\bibfnamefont {Ian~R.}\
  \bibnamefont {McDonald}},\ }\href {\doibase
  10.1016/B978-0-12-387032-2.00013-1} {\emph {\bibinfo {title} {Theory of
  Simple Liquids (Fourth Edition)}}}\ (\bibinfo  {publisher} {Academic Press},\
  \bibinfo {address} {Oxford},\ \bibinfo {year} {2013})\BibitemShut {NoStop}%
\bibitem [{\citenamefont {Weeks}\ \emph {et~al.}(1971)\citenamefont {Weeks},
  \citenamefont {Chandler},\ and\ \citenamefont {Andersen}}]{WCA1971}%
  \BibitemOpen
  \bibfield  {author} {\bibinfo {author} {\bibfnamefont {John~D.}\ \bibnamefont
  {Weeks}}, \bibinfo {author} {\bibfnamefont {David}\ \bibnamefont {Chandler}},
  \ and\ \bibinfo {author} {\bibfnamefont {Hans~C.}\ \bibnamefont {Andersen}},\
  }\bibfield  {title} {\enquote {\bibinfo {title} {{Role of Repulsive Forces in
  Determining the Equilibrium Structure of Simple Liquids}},}\ }\href {\doibase
  10.1063/1.1674820} {\bibfield  {journal} {\bibinfo  {journal} {J. Chem.
  Phys.}\ }\textbf {\bibinfo {volume} {54}},\ \bibinfo {pages} {5237--5247}
  (\bibinfo {year} {1971})}\BibitemShut {NoStop}%
\bibitem [{\citenamefont {Lander}\ \emph {et~al.}(2012)\citenamefont {Lander},
  \citenamefont {Mehl}, \citenamefont {Blickle}, \citenamefont {Bechinger},\
  and\ \citenamefont {Seifert}}]{Lander2012}%
  \BibitemOpen
  \bibfield  {author} {\bibinfo {author} {\bibfnamefont {B.}~\bibnamefont
  {Lander}}, \bibinfo {author} {\bibfnamefont {J.}~\bibnamefont {Mehl}},
  \bibinfo {author} {\bibfnamefont {V.}~\bibnamefont {Blickle}}, \bibinfo
  {author} {\bibfnamefont {C.}~\bibnamefont {Bechinger}}, \ and\ \bibinfo
  {author} {\bibfnamefont {U.}~\bibnamefont {Seifert}},\ }\bibfield  {title}
  {\enquote {\bibinfo {title} {Noninvasive measurement of dissipation in
  colloidal systems},}\ }\href {\doibase 10.1103/PhysRevE.86.030401} {\bibfield
   {journal} {\bibinfo  {journal} {Phys. Rev. E}\ }\textbf {\bibinfo {volume}
  {86}},\ \bibinfo {pages} {030401} (\bibinfo {year} {2012})}\BibitemShut
  {NoStop}%
\bibitem [{\citenamefont {Marini Bettolo~Marconi}\ and\ \citenamefont
  {Maggi}(2015)}]{Maggi2015b}%
  \BibitemOpen
  \bibfield  {author} {\bibinfo {author} {\bibfnamefont {Umberto}\ \bibnamefont
  {Marini Bettolo~Marconi}}\ and\ \bibinfo {author} {\bibfnamefont {Claudio}\
  \bibnamefont {Maggi}},\ }\bibfield  {title} {\enquote {\bibinfo {title}
  {Towards a statistical mechanical theory of active fluids},}\ }\href
  {\doibase 10.1039/C5SM01718A} {\bibfield  {journal} {\bibinfo  {journal}
  {Soft Matter}\ }\textbf {\bibinfo {volume} {11}},\ \bibinfo {pages}
  {8768--8781} (\bibinfo {year} {2015})}\BibitemShut {NoStop}%
\bibitem [{\citenamefont {Rein}\ and\ \citenamefont {Speck}(2016)}]{Rein2016}%
  \BibitemOpen
  \bibfield  {author} {\bibinfo {author} {\bibfnamefont {Markus}\ \bibnamefont
  {Rein}}\ and\ \bibinfo {author} {\bibfnamefont {Thomas}\ \bibnamefont
  {Speck}},\ }\bibfield  {title} {\enquote {\bibinfo {title} {Applicability of
  effective pair potentials for active brownian particles},}\ }\href {\doibase
  10.1140/epje/i2016-16084-7} {\bibfield  {journal} {\bibinfo  {journal} {Eur.
  Phys. J. E}\ }\textbf {\bibinfo {volume} {39}},\ \bibinfo {pages} {84}
  (\bibinfo {year} {2016})}\BibitemShut {NoStop}%
\bibitem [{\citenamefont {Wittmann}\ \emph
  {et~al.}(2017{\natexlab{b}})\citenamefont {Wittmann}, \citenamefont {Maggi},
  \citenamefont {Sharma}, \citenamefont {Scacchi}, \citenamefont {Brader},\
  and\ \citenamefont {Marconi}}]{Wittmann2017}%
  \BibitemOpen
  \bibfield  {author} {\bibinfo {author} {\bibfnamefont {Ren{\'{e}}}\
  \bibnamefont {Wittmann}}, \bibinfo {author} {\bibfnamefont {C}~\bibnamefont
  {Maggi}}, \bibinfo {author} {\bibfnamefont {A}~\bibnamefont {Sharma}},
  \bibinfo {author} {\bibfnamefont {A}~\bibnamefont {Scacchi}}, \bibinfo
  {author} {\bibfnamefont {J~M}\ \bibnamefont {Brader}}, \ and\ \bibinfo
  {author} {\bibfnamefont {U~Marini~Bettolo}\ \bibnamefont {Marconi}},\
  }\bibfield  {title} {\enquote {\bibinfo {title} {Effective equilibrium states
  in the colored-noise model for active matter i. pairwise forces in the fox
  and unified colored noise approximations},}\ }\href {\doibase
  10.1088/1742-5468/aa8c1f} {\bibfield  {journal} {\bibinfo  {journal} {J.
  Stat. Mech.}\ }\textbf {\bibinfo {volume} {2017}},\ \bibinfo {pages} {113207}
  (\bibinfo {year} {2017}{\natexlab{b}})}\BibitemShut {NoStop}%
\bibitem [{\citenamefont {Basu}\ \emph {et~al.}(2015)\citenamefont {Basu},
  \citenamefont {Maes},\ and\ \citenamefont {Neto{\v{c}}n{\'{y}}}}]{Maes2015}%
  \BibitemOpen
  \bibfield  {author} {\bibinfo {author} {\bibfnamefont {Urna}\ \bibnamefont
  {Basu}}, \bibinfo {author} {\bibfnamefont {Christian}\ \bibnamefont {Maes}},
  \ and\ \bibinfo {author} {\bibfnamefont {Karel}\ \bibnamefont
  {Neto{\v{c}}n{\'{y}}}},\ }\bibfield  {title} {\enquote {\bibinfo {title}
  {Statistical forces from close-to-equilibrium media},}\ }\href {\doibase
  10.1088/1367-2630/17/11/115006} {\bibfield  {journal} {\bibinfo  {journal}
  {New J. Phys.}\ }\textbf {\bibinfo {volume} {17}},\ \bibinfo {pages} {115006}
  (\bibinfo {year} {2015})}\BibitemShut {NoStop}%
\bibitem [{\citenamefont {Bisker}\ and\ \citenamefont
  {England}(2018)}]{Bisker2018}%
  \BibitemOpen
  \bibfield  {author} {\bibinfo {author} {\bibfnamefont {Gili}\ \bibnamefont
  {Bisker}}\ and\ \bibinfo {author} {\bibfnamefont {Jeremy~L.}\ \bibnamefont
  {England}},\ }\bibfield  {title} {\enquote {\bibinfo {title} {Nonequilibrium
  associative retrieval of multiple stored self-assembly targets},}\ }\href
  {\doibase 10.1073/pnas.1805769115} {\bibfield  {journal} {\bibinfo  {journal}
  {Proc. Natl. Acad. Sci. USA}\ }\textbf {\bibinfo {volume} {115}},\ \bibinfo
  {pages} {E10531--E10538} (\bibinfo {year} {2018})}\BibitemShut {NoStop}%
\bibitem [{\citenamefont {Evans}(2004)}]{Evans2004PRL}%
  \BibitemOpen
  \bibfield  {author} {\bibinfo {author} {\bibfnamefont {R.~M.~L.}\
  \bibnamefont {Evans}},\ }\bibfield  {title} {\enquote {\bibinfo {title}
  {Rules for transition rates in nonequilibrium steady states},}\ }\href
  {\doibase 10.1103/PhysRevLett.92.150601} {\bibfield  {journal} {\bibinfo
  {journal} {Phys. Rev. Lett.}\ }\textbf {\bibinfo {volume} {92}},\ \bibinfo
  {pages} {150601} (\bibinfo {year} {2004})}\BibitemShut {NoStop}%
\bibitem [{\citenamefont {Martin}\ \emph {et~al.}(1973)\citenamefont {Martin},
  \citenamefont {Siggia},\ and\ \citenamefont {Rose}}]{Martin1973}%
  \BibitemOpen
  \bibfield  {author} {\bibinfo {author} {\bibfnamefont {P.~C.}\ \bibnamefont
  {Martin}}, \bibinfo {author} {\bibfnamefont {E.~D.}\ \bibnamefont {Siggia}},
  \ and\ \bibinfo {author} {\bibfnamefont {H.~A.}\ \bibnamefont {Rose}},\
  }\bibfield  {title} {\enquote {\bibinfo {title} {Statistical dynamics of
  classical systems},}\ }\href {\doibase 10.1103/PhysRevA.8.423} {\bibfield
  {journal} {\bibinfo  {journal} {Phys. Rev. A}\ }\textbf {\bibinfo {volume}
  {8}},\ \bibinfo {pages} {423--437} (\bibinfo {year} {1973})}\BibitemShut
  {NoStop}%
\bibitem [{\citenamefont {De~Dominicis}(1975)}]{Dominicis1975}%
  \BibitemOpen
  \bibfield  {author} {\bibinfo {author} {\bibfnamefont {C.}~\bibnamefont
  {De~Dominicis}},\ }\bibfield  {title} {\enquote {\bibinfo {title} {A
  lagrangian version of halperin-hohenberg-ma models for the dynamics of
  critical phenomena},}\ }\href {\doibase 10.1007/BF02785928} {\bibfield
  {journal} {\bibinfo  {journal} {Lettere al Nuovo Cimento (1971-1985)}\
  }\textbf {\bibinfo {volume} {12}},\ \bibinfo {pages} {567--574} (\bibinfo
  {year} {1975})}\BibitemShut {NoStop}%
\bibitem [{\citenamefont {Jack}\ and\ \citenamefont
  {Sollich}(2015)}]{Jack2015}%
  \BibitemOpen
  \bibfield  {author} {\bibinfo {author} {\bibfnamefont {R.~L.}\ \bibnamefont
  {Jack}}\ and\ \bibinfo {author} {\bibfnamefont {P.}~\bibnamefont {Sollich}},\
  }\bibfield  {title} {\enquote {\bibinfo {title} {Effective interactions and
  large deviations in stochastic processes},}\ }\href {\doibase
  10.1140/epjst/e2015-02416-9} {\bibfield  {journal} {\bibinfo  {journal}
  {EPJST (Eur. Phys. J. Special Topics)}\ }\textbf {\bibinfo {volume} {224}},\
  \bibinfo {pages} {2351--2367} (\bibinfo {year} {2015})}\BibitemShut {NoStop}%
\bibitem [{\citenamefont {Popkov}\ \emph {et~al.}(2010)\citenamefont {Popkov},
  \citenamefont {Schütz},\ and\ \citenamefont {Simon}}]{Popkov2010}%
  \BibitemOpen
  \bibfield  {author} {\bibinfo {author} {\bibfnamefont {Vladislav}\
  \bibnamefont {Popkov}}, \bibinfo {author} {\bibfnamefont {Gunter~M}\
  \bibnamefont {Schütz}}, \ and\ \bibinfo {author} {\bibfnamefont {Damien}\
  \bibnamefont {Simon}},\ }\bibfield  {title} {\enquote {\bibinfo {title}
  {{ASEP on a ring conditioned on enhanced flux}},}\ }\href {\doibase
  10.1088/1742-5468/2010/10/P10007} {\bibfield  {journal} {\bibinfo  {journal}
  {J. Stat. Mech.}\ }\textbf {\bibinfo {volume} {2010}},\ \bibinfo {pages}
  {P10007} (\bibinfo {year} {2010})}\BibitemShut {NoStop}%
\bibitem [{\citenamefont {Popkov}\ and\ \citenamefont
  {Sch{\"u}tz}(2011)}]{Popkov2011}%
  \BibitemOpen
  \bibfield  {author} {\bibinfo {author} {\bibfnamefont {V.}~\bibnamefont
  {Popkov}}\ and\ \bibinfo {author} {\bibfnamefont {G.~M.}\ \bibnamefont
  {Sch{\"u}tz}},\ }\bibfield  {title} {\enquote {\bibinfo {title} {{Transition
  Probabilities and Dynamic Structure Function in the ASEP Conditioned on
  Strong Flux}},}\ }\href {\doibase 10.1007/s10955-011-0137-7} {\bibfield
  {journal} {\bibinfo  {journal} {J. Stat. Phys.}\ }\textbf {\bibinfo {volume}
  {142}},\ \bibinfo {pages} {627--639} (\bibinfo {year} {2011})}\BibitemShut
  {NoStop}%
\bibitem [{\citenamefont {Ray}\ \emph {et~al.}(2018{\natexlab{b}})\citenamefont
  {Ray}, \citenamefont {Chan},\ and\ \citenamefont {Limmer}}]{Limmer2018a}%
  \BibitemOpen
  \bibfield  {author} {\bibinfo {author} {\bibfnamefont {Ushnish}\ \bibnamefont
  {Ray}}, \bibinfo {author} {\bibfnamefont {Garnet Kin-Lic}\ \bibnamefont
  {Chan}}, \ and\ \bibinfo {author} {\bibfnamefont {David~T.}\ \bibnamefont
  {Limmer}},\ }\bibfield  {title} {\enquote {\bibinfo {title} {Exact
  fluctuations of nonequilibrium steady states from approximate auxiliary
  dynamics},}\ }\href {\doibase 10.1103/PhysRevLett.120.210602} {\bibfield
  {journal} {\bibinfo  {journal} {Phys. Rev. Lett.}\ }\textbf {\bibinfo
  {volume} {120}},\ \bibinfo {pages} {210602} (\bibinfo {year}
  {2018}{\natexlab{b}})}\BibitemShut {NoStop}%
\bibitem [{\citenamefont {Tiz{\'{o}}n-Escamilla}\ \emph
  {et~al.}(2019)\citenamefont {Tiz{\'{o}}n-Escamilla}, \citenamefont
  {Lecomte},\ and\ \citenamefont {Bertin}}]{Lecomte2019}%
  \BibitemOpen
  \bibfield  {author} {\bibinfo {author} {\bibfnamefont {Nicol{\'{a}}s}\
  \bibnamefont {Tiz{\'{o}}n-Escamilla}}, \bibinfo {author} {\bibfnamefont
  {Vivien}\ \bibnamefont {Lecomte}}, \ and\ \bibinfo {author} {\bibfnamefont
  {Eric}\ \bibnamefont {Bertin}},\ }\bibfield  {title} {\enquote {\bibinfo
  {title} {Effective driven dynamics for one-dimensional conditioned langevin
  processes in the weak-noise limit},}\ }\href {\doibase
  10.1088/1742-5468/aaeda3} {\bibfield  {journal} {\bibinfo  {journal} {J.
  Stat. Mech.}\ }\textbf {\bibinfo {volume} {2019}},\ \bibinfo {pages} {013201}
  (\bibinfo {year} {2019})}\BibitemShut {NoStop}%
\bibitem [{\citenamefont {Proesmans}\ and\ \citenamefont
  {Derrida}(2019)}]{Proesmans2019}%
  \BibitemOpen
  \bibfield  {author} {\bibinfo {author} {\bibfnamefont {Karel}\ \bibnamefont
  {Proesmans}}\ and\ \bibinfo {author} {\bibfnamefont {Bernard}\ \bibnamefont
  {Derrida}},\ }\bibfield  {title} {\enquote {\bibinfo {title} {Large-deviation
  theory for a brownian particle on a ring: a {WKB} approach},}\ }\href
  {\doibase 10.1088/1742-5468/aafa7e} {\bibfield  {journal} {\bibinfo
  {journal} {J. Stat. Mech.}\ }\textbf {\bibinfo {volume} {2019}},\ \bibinfo
  {pages} {023201} (\bibinfo {year} {2019})}\BibitemShut {NoStop}%
\bibitem [{\citenamefont {Majumdar}\ and\ \citenamefont
  {Bray}(2002)}]{Majumdar2002}%
  \BibitemOpen
  \bibfield  {author} {\bibinfo {author} {\bibfnamefont {Satya~N.}\
  \bibnamefont {Majumdar}}\ and\ \bibinfo {author} {\bibfnamefont {Alan~J.}\
  \bibnamefont {Bray}},\ }\bibfield  {title} {\enquote {\bibinfo {title}
  {Large-deviation functions for nonlinear functionals of a gaussian stationary
  markov process},}\ }\href {\doibase 10.1103/PhysRevE.65.051112} {\bibfield
  {journal} {\bibinfo  {journal} {Phys. Rev. E}\ }\textbf {\bibinfo {volume}
  {65}},\ \bibinfo {pages} {051112} (\bibinfo {year} {2002})}\BibitemShut
  {NoStop}%
\bibitem [{\citenamefont {Tsobgni~Nyawo}\ and\ \citenamefont
  {Touchette}(2016)}]{Touchette2016}%
  \BibitemOpen
  \bibfield  {author} {\bibinfo {author} {\bibfnamefont {Pelerine}\
  \bibnamefont {Tsobgni~Nyawo}}\ and\ \bibinfo {author} {\bibfnamefont {Hugo}\
  \bibnamefont {Touchette}},\ }\bibfield  {title} {\enquote {\bibinfo {title}
  {Large deviations of the current for driven periodic diffusions},}\ }\href
  {\doibase 10.1103/PhysRevE.94.032101} {\bibfield  {journal} {\bibinfo
  {journal} {Phys. Rev. E}\ }\textbf {\bibinfo {volume} {94}},\ \bibinfo
  {pages} {032101} (\bibinfo {year} {2016})}\BibitemShut {NoStop}%
\bibitem [{\citenamefont {Nyawo}\ and\ \citenamefont
  {Touchette}(2018)}]{Touchette2018}%
  \BibitemOpen
  \bibfield  {author} {\bibinfo {author} {\bibfnamefont {Pelerine~Tsobgni}\
  \bibnamefont {Nyawo}}\ and\ \bibinfo {author} {\bibfnamefont {Hugo}\
  \bibnamefont {Touchette}},\ }\bibfield  {title} {\enquote {\bibinfo {title}
  {Dynamical phase transition in drifted brownian motion},}\ }\href {\doibase
  10.1103/PhysRevE.98.052103} {\bibfield  {journal} {\bibinfo  {journal} {Phys.
  Rev. E}\ }\textbf {\bibinfo {volume} {98}},\ \bibinfo {pages} {052103}
  (\bibinfo {year} {2018})}\BibitemShut {NoStop}%
\bibitem [{\citenamefont {Fullerton}\ and\ \citenamefont
  {Jack}(2013)}]{Fullerton2013}%
  \BibitemOpen
  \bibfield  {author} {\bibinfo {author} {\bibfnamefont {Christopher~J.}\
  \bibnamefont {Fullerton}}\ and\ \bibinfo {author} {\bibfnamefont {Robert~L.}\
  \bibnamefont {Jack}},\ }\bibfield  {title} {\enquote {\bibinfo {title}
  {Dynamical phase transitions in supercooled liquids: Interpreting
  measurements of dynamical activity},}\ }\href {\doibase 10.1063/1.4808152}
  {\bibfield  {journal} {\bibinfo  {journal} {J. Chem. Phys.}\ }\textbf
  {\bibinfo {volume} {138}},\ \bibinfo {pages} {224506} (\bibinfo {year}
  {2013})}\BibitemShut {NoStop}%
\bibitem [{\citenamefont {Jack}\ \emph {et~al.}(2015)\citenamefont {Jack},
  \citenamefont {Thompson},\ and\ \citenamefont {Sollich}}]{Jack2015b}%
  \BibitemOpen
  \bibfield  {author} {\bibinfo {author} {\bibfnamefont {Robert~L.}\
  \bibnamefont {Jack}}, \bibinfo {author} {\bibfnamefont {Ian~R.}\ \bibnamefont
  {Thompson}}, \ and\ \bibinfo {author} {\bibfnamefont {Peter}\ \bibnamefont
  {Sollich}},\ }\bibfield  {title} {\enquote {\bibinfo {title} {Hyperuniformity
  and phase separation in biased ensembles of trajectories for diffusive
  systems},}\ }\href {\doibase 10.1103/PhysRevLett.114.060601} {\bibfield
  {journal} {\bibinfo  {journal} {Phys. Rev. Lett.}\ }\textbf {\bibinfo
  {volume} {114}},\ \bibinfo {pages} {060601} (\bibinfo {year}
  {2015})}\BibitemShut {NoStop}%
\bibitem [{\citenamefont {{Crosato}}\ \emph {et~al.}(2018)\citenamefont
  {{Crosato}}, \citenamefont {{Prokopenko}},\ and\ \citenamefont
  {{Spinney}}}]{Spinney2018}%
  \BibitemOpen
  \bibfield  {author} {\bibinfo {author} {\bibfnamefont {Emanuele}\
  \bibnamefont {{Crosato}}}, \bibinfo {author} {\bibfnamefont {Mikhail}\
  \bibnamefont {{Prokopenko}}}, \ and\ \bibinfo {author} {\bibfnamefont
  {Richard~E.}\ \bibnamefont {{Spinney}}},\ }\bibfield  {title} {\enquote
  {\bibinfo {title} {{Thermodynamics of emergent structure in active
  matter}},}\ }\href@noop {} {\bibfield  {journal} {\bibinfo  {journal} {ArXiv
  e-prints}\ } (\bibinfo {year} {2018})},\ \Eprint
  {http://arxiv.org/abs/1812.08527} {arXiv:1812.08527} \BibitemShut {NoStop}%
\bibitem [{\citenamefont {Jack}\ and\ \citenamefont
  {Sollich}(2014)}]{Jack2014}%
  \BibitemOpen
  \bibfield  {author} {\bibinfo {author} {\bibfnamefont {Robert~L}\
  \bibnamefont {Jack}}\ and\ \bibinfo {author} {\bibfnamefont {Peter}\
  \bibnamefont {Sollich}},\ }\bibfield  {title} {\enquote {\bibinfo {title}
  {Large deviations of the dynamical activity in the east model: analysing
  structure in biased trajectories},}\ }\href {\doibase
  10.1088/1751-8113/47/1/015003} {\bibfield  {journal} {\bibinfo  {journal} {J.
  Phys. A: Math. Theor.}\ }\textbf {\bibinfo {volume} {47}},\ \bibinfo {pages}
  {015003} (\bibinfo {year} {2014})}\BibitemShut {NoStop}%
\bibitem [{\citenamefont {Brewer}\ \emph
  {et~al.}(2018{\natexlab{b}})\citenamefont {Brewer}, \citenamefont {Clark},
  \citenamefont {Bradford},\ and\ \citenamefont {Jack}}]{Jack2018}%
  \BibitemOpen
  \bibfield  {author} {\bibinfo {author} {\bibfnamefont {Tobias}\ \bibnamefont
  {Brewer}}, \bibinfo {author} {\bibfnamefont {Stephen~R}\ \bibnamefont
  {Clark}}, \bibinfo {author} {\bibfnamefont {Russell}\ \bibnamefont
  {Bradford}}, \ and\ \bibinfo {author} {\bibfnamefont {Robert~L}\ \bibnamefont
  {Jack}},\ }\bibfield  {title} {\enquote {\bibinfo {title} {Efficient
  characterisation of large deviations using population dynamics},}\ }\href
  {\doibase 10.1088/1742-5468/aab3ef} {\bibfield  {journal} {\bibinfo
  {journal} {J. Stat. Mech.}\ }\textbf {\bibinfo {volume} {2018}},\ \bibinfo
  {pages} {053204} (\bibinfo {year} {2018}{\natexlab{b}})}\BibitemShut
  {NoStop}%
\bibitem [{\citenamefont {Ferr{\'e}}\ and\ \citenamefont
  {Touchette}(2018)}]{Ferre2018}%
  \BibitemOpen
  \bibfield  {author} {\bibinfo {author} {\bibfnamefont {Gr{\'e}goire}\
  \bibnamefont {Ferr{\'e}}}\ and\ \bibinfo {author} {\bibfnamefont {Hugo}\
  \bibnamefont {Touchette}},\ }\bibfield  {title} {\enquote {\bibinfo {title}
  {Adaptive sampling of large deviations},}\ }\href {\doibase
  10.1007/s10955-018-2108-8} {\bibfield  {journal} {\bibinfo  {journal} {J.
  Stat. Phys.}\ }\textbf {\bibinfo {volume} {172}},\ \bibinfo {pages}
  {1525--1544} (\bibinfo {year} {2018})}\BibitemShut {NoStop}%
\bibitem [{\citenamefont {Murugan}\ \emph {et~al.}(2015)\citenamefont
  {Murugan}, \citenamefont {Zeravcic}, \citenamefont {Brenner},\ and\
  \citenamefont {Leibler}}]{Murugan2015}%
  \BibitemOpen
  \bibfield  {author} {\bibinfo {author} {\bibfnamefont {Arvind}\ \bibnamefont
  {Murugan}}, \bibinfo {author} {\bibfnamefont {Zorana}\ \bibnamefont
  {Zeravcic}}, \bibinfo {author} {\bibfnamefont {Michael~P.}\ \bibnamefont
  {Brenner}}, \ and\ \bibinfo {author} {\bibfnamefont {Stanislas}\ \bibnamefont
  {Leibler}},\ }\bibfield  {title} {\enquote {\bibinfo {title} {Multifarious
  assembly mixtures: Systems allowing retrieval of diverse stored
  structures},}\ }\href {\doibase 10.1073/pnas.1413941112} {\bibfield
  {journal} {\bibinfo  {journal} {Proc. Natl. Acad. Sci. USA}\ }\textbf
  {\bibinfo {volume} {112}},\ \bibinfo {pages} {54--59} (\bibinfo {year}
  {2015})}\BibitemShut {NoStop}%
\bibitem [{\citenamefont {Stern}\ \emph {et~al.}(2017)\citenamefont {Stern},
  \citenamefont {Pinson},\ and\ \citenamefont {Murugan}}]{Murugan2017b}%
  \BibitemOpen
  \bibfield  {author} {\bibinfo {author} {\bibfnamefont {Menachem}\
  \bibnamefont {Stern}}, \bibinfo {author} {\bibfnamefont {Matthew~B.}\
  \bibnamefont {Pinson}}, \ and\ \bibinfo {author} {\bibfnamefont {Arvind}\
  \bibnamefont {Murugan}},\ }\bibfield  {title} {\enquote {\bibinfo {title}
  {The complexity of folding self-folding origami},}\ }\href {\doibase
  10.1103/PhysRevX.7.041070} {\bibfield  {journal} {\bibinfo  {journal} {Phys.
  Rev. X}\ }\textbf {\bibinfo {volume} {7}},\ \bibinfo {pages} {041070}
  (\bibinfo {year} {2017})}\BibitemShut {NoStop}%
\bibitem [{\citenamefont {Joshi}\ \emph {et~al.}(2019)\citenamefont {Joshi},
  \citenamefont {Putzig}, \citenamefont {Baskaran},\ and\ \citenamefont
  {Hagan}}]{Joshi2017}%
  \BibitemOpen
  \bibfield  {author} {\bibinfo {author} {\bibfnamefont {Abhijeet}\
  \bibnamefont {Joshi}}, \bibinfo {author} {\bibfnamefont {Elias}\ \bibnamefont
  {Putzig}}, \bibinfo {author} {\bibfnamefont {Aparna}\ \bibnamefont
  {Baskaran}}, \ and\ \bibinfo {author} {\bibfnamefont {Michael~F.}\
  \bibnamefont {Hagan}},\ }\bibfield  {title} {\enquote {\bibinfo {title} {{The
  interplay between activity and filament flexibility determines the emergent
  properties of active nematics}},}\ }\href {\doibase 10.1039/C8SM02202J}
  {\bibfield  {journal} {\bibinfo  {journal} {Soft Matter}\ }\textbf {\bibinfo
  {volume} {15}},\ \bibinfo {pages} {94--101} (\bibinfo {year}
  {2019})}\BibitemShut {NoStop}%
\bibitem [{\citenamefont {Nguyen}\ \emph {et~al.}(2014)\citenamefont {Nguyen},
  \citenamefont {Klotsa}, \citenamefont {Engel},\ and\ \citenamefont
  {Glotzer}}]{Nguyen2014b}%
  \BibitemOpen
  \bibfield  {author} {\bibinfo {author} {\bibfnamefont {Nguyen H.~P.}\
  \bibnamefont {Nguyen}}, \bibinfo {author} {\bibfnamefont {Daphne}\
  \bibnamefont {Klotsa}}, \bibinfo {author} {\bibfnamefont {Michael}\
  \bibnamefont {Engel}}, \ and\ \bibinfo {author} {\bibfnamefont {Sharon~C.}\
  \bibnamefont {Glotzer}},\ }\bibfield  {title} {\enquote {\bibinfo {title}
  {Emergent collective phenomena in a mixture of hard shapes through active
  rotation},}\ }\href {\doibase 10.1103/PhysRevLett.112.075701} {\bibfield
  {journal} {\bibinfo  {journal} {Phys. Rev. Lett.}\ }\textbf {\bibinfo
  {volume} {112}},\ \bibinfo {pages} {075701} (\bibinfo {year}
  {2014})}\BibitemShut {NoStop}%
\end{thebibliography}%

\end{document}